\documentclass[onecolumn]{IEEEtran}
\usepackage{cite}
\usepackage{amsmath,amssymb,amsfonts}
\usepackage[ruled]{algorithm2e}
    \SetKwInOut{Parameter}{Parameter}
\usepackage{dsfont}
\usepackage{textcomp}
\usepackage{makecell}
\usepackage[inline]{enumitem}
\usepackage[utf8]{inputenc}
\usepackage[english]{babel}
\usepackage{amsthm}
\usepackage[bb=boondox]{mathalfa}
\usepackage{mathtools}
\usepackage{amssymb}
\usepackage{float}
\usepackage[pdftex]{graphicx}
\usepackage{tikz}
\usepackage{tkz-euclide}
\usepackage{tabto}
\usetikzlibrary{positioning}
\graphicspath{{./pdf/}{./jpeg/}{./png}}
\DeclareGraphicsExtensions{.pdf,.jpeg,.png}

\definecolor{phiGreen}{RGB}{0,128,32}

\def\BibTeX{{\rm B\kern-.05em{\sc i\kern-.025em b}\kern-.08em
    T\kern-.1667em\lower.7ex\hbox{E}\kern-.125emX}}

\theoremstyle{definition}

\theoremstyle{remark}

\title{Optimizing Secrecy Codes Using Gradient Descent}

\author{\IEEEauthorblockN{David Hunn}, \IEEEmembership{Student Member, IEEE}, and \IEEEauthorblockN{Willie K. Harrison}
\IEEEmembership{Senior Member, IEEE}
\thanks{This work was funded by the US National Science Foundation: Grant Award Number \#1910812.}}

\begin{document}

\maketitle

\begin{abstract}
Recent theoretical developments in coset coding theory have provided continuous-valued functions which give the equivocation and maximum likelihood (ML) decoding probability of coset secrecy codes. In this work, we develop a method for incorporating these functions, along with a complex set of constraints, into a gradient descent optimization algorithm. This algorithm employs a movement cost function and trigonometric update step to ensure that the continuous-valued code definition vector ultimately reaches a value which yields a realizable coset code. This algorithm is used to produce coset codes with blocklength up to a few thousand. These codes were compared against published codes, including both short-blocklength and capacity-achieving constructions. For most code sizes, codes generated using gradient descent outperformed all others, especially capacity-achieving constructions, which performed significantly worse than randomly-generated codes at short blocklength. 
\end{abstract}



\section{Introduction}
\label{sec:introduction}

Secrecy coding refers to the coding of information for transmission over a communication channel with the goal of keeping the information transmitted from being revealed to an eavesdropper. In contrast to encryption, secrecy coding may be implemented without the use of a secret key. The aim of secrecy coding is to limit the eavesdropper's knowledge using information-theoretic properties of coding and data transmission.  

\subsection{Binary Erasure Wiretap Channel}
\begin{figure}[H]
    \centering
    \scriptsize
    \begin{tikzpicture}
        \node at (-0.1,0) [rectangle, minimum height = 0.4cm, anchor = north, node font=\bfseries, inner sep=0.05cm, outer sep=0] (S) {Sender}; 
        \node at (1.5,0) [rectangle, fill=blue!20, draw=blue, minimum height = 0.6cm, minimum width = 1.1cm] (En) {Encoder}; 
        \node at (3.9,0) [rectangle, align=center, fill=green!20, draw=green, minimum height = 0.6cm, minimum width = 1.1cm] (NL) {Noiseless\\Channel};
        \node at (6.3,0) [rectangle, fill=blue!20, draw=blue, minimum height = 0.6cm, minimum width = 1.1cm] (De) {Decoder}; 
        \node at (7.7,0) [rectangle, minimum height = 0.4cm, minimum width = 1.1cm, align=left, anchor = north, node font=\bfseries] (LR) {Legitimate\\Receiver}; 
        \node at (3.9,-1.3) [rectangle, align=center, fill=green!20, draw=green, minimum height = 0.6cm, minimum width = 1.1cm] (BEC) {Binary Erasure\\Channel (BEC)};
        \node at (6.6,-1.3) [rectangle, minimum height = 0.4cm, minimum width = 1.1cm, align=left, anchor = north, node font=\bfseries] (Eve) {Eavesdropper};
        \draw[->] (En) -- (NL) node[pos=0.5, anchor=south]{\!$X\!$ ($n\!$ bits)}; 
        \draw[->] (NL) -- (De) node[pos=0.5, anchor=south]{\!$X\!$ ($n\!$ bits)};; 
        \draw[->] (S.north) + (-0.15cm,0) -- (En) node[pos=0.5, anchor=south]{\!$M\!$ ($k\!$ bits)};; 
        \draw[->] (De) -- (8.0cm,0) node[pos=0.5, anchor=south]{\!$M\!$ ($k\!$ bits)};
        \draw[->] (En.east)+(0.3,0) |- (BEC.west) ;
        \draw[->] (BEC) -- (6.8cm,-1.3) node[pos=0.95, anchor=south]{$Z\!$ ($n\!$ symbols, $Z_i \!\in\! \{0,\!1,\!?\}$)};
    \end{tikzpicture}

    \caption{Binary erasure wiretap channel.}
    \label{fig:wiretap_channel}
\end{figure}
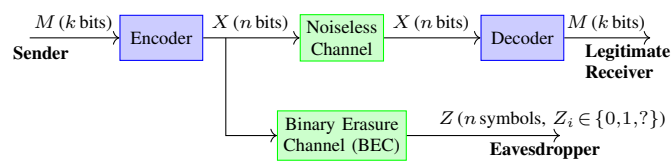

Introduced by Wyner in~\cite{Wyner1975}, the first and most widely studied model for communication with an eavesdropper is called the \emph{wiretap channel} and comprises a transmitter, ``Alice'', a legitimate receiver, ``Bob'', who receives the communication through a main channel, and an eavesdropper, ``Eve'', who receives the communication through an eavesdropper's channel. The primary wiretap channel studied in this work is the binary erasure wiretap channel (BEWC) and is shown in Fig.~\ref{fig:wiretap_channel}. In this channel, a message is encoded and sent by Alice. It is then received via a noiseless channel by a legitimate receiver and received via a binary erasure channel (BEC) by an eavesdropper. Arguably the simplest nontrivial wiretap channel, the BEWC was among the first channel models employed to analyze the performance of wiretap codes. Because of its simplicity, the BEWC is a promising candidate for study with the aim of identifying codes which perform well and which could also be applied to other binary-input wiretap channels. In spite of the extent of the research on the BEWC, however, the problem of designing good secrecy codes over this channel remains open. 

\subsection{Coset Coding}
Secrecy coding for the BEWC involves encoding the original $k$-bit message $m$ (a realization of the random variable $M$) to an $n$-bit codeword $x$ (a realization of $X$). This may be done by creating a base $(n,\kappa)$ linear block code $\mathcal{C}$ defined by generator matrix $G$, with $\kappa = n-k$. Each of the possible realizations $m$ of $M$ is then assigned to one of the cosets of $\mathcal{C}$. The codeword $x$ is then formed by selecting an element at random from the chosen coset. This form of secrecy coding is called \emph{coset coding}. It is implemented in practice by defining an auxiliary generator matrix $G'$ of size $k \times n$ comprised of rows linearly independent to the rows of $G$. A $\kappa$-bit random auxiliary message $m'$ is then generated in order to select a random coset element. The auxiliary message $m'$ is then appended to $m$, and the resulting vector is multiplied by the $n \times n$ matrix $G^*$ formed by vertically concatenating $G'$ with $G$. The final codeword $x$ is given by 
\begin{equation}
    x = \begin{bmatrix} m & m' \end{bmatrix} \begin{bmatrix} G' \\ G \end{bmatrix}\mathrm{.}
\end{equation}

\subsection{Coset Code Performance}
A number of metrics have been used to quantify the effectiveness of secrecy codes under a variety of different circumstances. These metrics include measures of information leakage under average and worst-case conditions, channel-dependent and channel-independent metrics, and requirements for various levels of asymptotic secrecy. Several of these metrics are discussed in the following sections. 

\subsubsection{Information Leakage Measures}
Leakage of information to an eavesdropper is typically measured using information-theoretic metrics on the eavesdropper's observation $Z$. The most common such metric is the mutual information between the message and the eavesdropper's observation, $I(M;Z)$. This metric is referred to as the equivocation loss, with the eavesdropper message entropy $H(M \!\! \mid \!\! Z) = H(M) - I(M;Z)$ being referred to as the equivocation. The equivocation loss is also equal to the Kullback-Leibler divergence between the joint distribution $p_{MZ}(m,z)$ of $M$ and $Z$ and the product of the marginal distributions $p_M(m)$ of $M$ and $p_Z(z)$ of $Z$. That is, 
\begin{equation}
    \label{eqn:equivocationDefinition}
    I(M;Z) = \mathds{D}(p_{MZ},p_M p_Z)\mathrm{.}
\end{equation}
Other information-theoretic secrecy metrics include norms on the difference between joint and marginal distributions~\cite{Bloch2021OverviewInformationTheoreticSecurity, Bloch2013StrongSecrecyResolvability}, 
\begin{equation}
    \label{eqn:normDefinition}
    \| p_{MZ},p_M p_Z\|_\beta = \!\! \left(\!\!\!\!\!\!\!\!\!\! \sum_{\;\;\;\;\;\;\;\;\substack{m\in \{0,1\}^{k}, \\ z\in\{0,1,?\}^n}}{\!\!\!\!\!\!\!\!\!\!\left|p_{MZ}(m,z)-p_M(m)p_Z(z)\right|^\beta}\!\!\right)^{\frac{1}{\beta}}\!\!\!\!\mathrm{,}
\end{equation}
with the total variation distance being half the 1-norm, 
\begin{equation}
    \label{eqn:oneNorm}
    \mathds{V}(p_{MZ},p_M p_Z) = \frac{1}{2} \!\!\!\!\!\!\!\!\!\!\!\! \sum_{\;\;\;\;\;\;\;\;\substack{m\in \{0,1\}^{k}, \\ z\in\{0,1,?\}^n}}{\!\!\!\!\!\!\!\!\!\!\!\!\left|p_{MZ}(m,z)-p_M(m)p_Z(z)\right|}\mathrm{.}
\end{equation}
The $\chi^2$ divergence is another metric which is particularly significant for the BEWC, as it was shown in~\cite{hunn2024subspace} to relate closely to an eavesdropper's maximum likelihood (ML) decoding probability. It is given by 
\begin{equation}
    \label{eqn:def_x2}
    \chi^2(p_{MZ},p_M p_Z) =  \!\!\!\!\!\!\!\!\!\!\sum_{\;\;\;\;\substack{m\in \{0,1\}^{k}, \\ z\in\{0,1,?\}^n}}{\!\! \left( \frac{p_{MZ}(m,z)}{p_M(m)p_Z(z)}-1\right)^2p_M(m) p_Z(z)} \mathrm{.}
\end{equation}

Some works have employed non-information-theoretic secrecy metrics, such as the eavesdropper's error rate with maximum likelihood decoding~\cite{Klinc2011MLDecoding} or with an efficiently-implementable decoder~\cite{Baldi2012EfficientSuboptimalDecoder, Harrison2018AnalysisShortSecrecyCodes}, but these measures are considered less rigorous than message entropy~\cite{Bloch2021OverviewInformationTheoreticSecurity, Harrison2018AnalysisShortSecrecyCodes}. (This contrasts with the case of error correcting codes in which conditional message entropy, though of some significance, is overshadowed by practical decoding requirements) 

\subsubsection{Equivocation Loss}
\label{sec:MessageEquivocationDefinition}
Equivocation is the most widely-studied measure of secrecy code performance, but calculating equivocation (and therefore equivocation loss) for a specific code is nontrivial, even over a channel as simple as the BEWC. The equivocation can clearly be calculated as an expectation over all possible values of $z$, 
\begin{equation}
    \label{eqn:EquivocationGeneralExpectation}
    H(M \mid Z) = \sum_{z \in \{0,1,?\}^n}{\mathrm{Pr}(Z=z) \cdot H(M \mid Z=z)}\mathrm{.}
\end{equation}

An important result presented in \cite{Pfister2017QuantifyingEquivocation} is that for uniformly distributed $M$, the equivocation may be calculated based on the set, denoted $r(z)$, of revealed bit positions of the eavesdropper's observation $z$, as 
\begin{equation}
    \label{eqn:EquivocationPfister}
    H(M \! \mid \! Z=z) = H(M) - \lvert r(z) \rvert + \mathrm{rank}(G_{r(z)})\mathrm{,}
\end{equation}
where $G_{r(z)}$ is the submatrix of $G$ formed by concatenating the columns of $G$ which are indexed by the set $r(z)$. It is assumed throughout this work that the assumption required by \eqref{eqn:EquivocationPfister} holds--- namely, that the message is uniformly distributed across the $2^k$ possibilities and, therefore, $H(M) = k$. 
One of the implications of \eqref{eqn:EquivocationPfister} is that the eavesdropper's message equivocation may be calculated via a matrix rank calculation for each of the $2^n$ possible erasure patterns $r(z)$, as 
\begin{equation}
    \label{eqn:EquivocationTotalPfister}
    H(M \! \mid \! Z) = \!\!\!\!\!\!\!\!\sum_{r \in P(\{1 \dots n\})}  \epsilon^{\lvert r \rvert} (1-\epsilon)^{\lvert r \rvert} \left(k - \lvert r \rvert + \mathrm{rank}(G_{r})\right)\mathrm{,}
\end{equation}
where $P(\cdot)$ is the power set function. Because $\mathcal{O}(n^2)$ calculations over $GF_2^{\kappa}$ are required to calculate the rank of a matrix with a size up to $\kappa \times n$ ~\cite{Bard2009GF2Operations}, the total complexity of an equivocation calculation by this method is $\mathcal{O}(n^2 2^n)$. 
Using \eqref{eqn:EquivocationTotalPfister}, it is practical to calculate a code's equivocation for a given $\epsilon$ up to a blocklength of a few tens. Beyond this blocklength, researchers have generally resorted to approximate equivocation values generated using, e.g., Monte Carlo simulations, as studied in ~\cite{Pfister2017QuantifyingEquivocation} or asymptotic methods, as in ~\cite{Al-Hassan2013BestKnownLinearCodes}. 

\subsubsection{Asymptotic Secrecy}
Equivocation loss is also at the heart of the most commonly used criteria for asymptotic security, \emph{weak secrecy} and \emph{strong secrecy}~\cite{Bloch2008ArbitraryWiretapChannelsMetrics}. Weak secrecy requires that mutual information per codeword symbol approach zero with increasing blocklength. That is, 
\begin{equation}
    \label{eqn:WeakSecrecy}
    \lim_{n \to \infty} \frac{1}{n} I(M ; Z) = 0\mathrm{.}
\end{equation}
Strong secrecy, on the other hand, imposes a similar restriction on the total, rather than per-symbol, mutual information: 
\begin{equation}
    \label{eqn:StrongSecrecy}
    \lim_{n \to \infty} I(M ; Z) = 0\mathrm{.}
\end{equation}

\subsubsection{Achievability Gap}
\label{sec:achievabilityGap}
It can be seen by inspection of \eqref{eqn:EquivocationTotalPfister} that for a given coset code, equivocation depends on $G$, a property of the code, and $\epsilon$, a property of the channel. The authors of ~\cite{Pfister2017QuantifyingEquivocation} recommend removing the channel dependence by specifying a metric called the achievability gap, denoted $A_g$, equal to 
\begin{equation}
    \label{eqn:AchievabilityGap}
    A_g = \frac{\kappa}{n} - \left. \frac{H(M \mid Z)}{n}\right|_{\epsilon = \frac{k}{n}} \mathrm{.}
\end{equation}
The achievability gap thus represents the per-symbol equivocation loss evaluated at $\epsilon = k / n$ and gives the maximum difference between per-symbol performance of the finite-blocklength code and that of the ideal infinite-blocklength case. In accordance with this rationale, code performance will be measured at an erasure probability of $\epsilon = k/n$ throughout this work.

\subsection{Finite Blocklength Code Construction}
\label{sec:finiteBlocklengthConstructions}
Since the introduction of the wiretap channel in 1975, Research into code design for wiretap channels, including the BEWC, has focused primarily on capacity-achieving constructions which provide security guarantees in the limit of infinite blocklength. The purpose of these capacity-achieving code constructions is not to provide good performance at short blocklength. Nevertheless, most of the short-blocklength secrecy codes evaluated in the literature are constructed by simply using capacity-achieving constructions to generate short-blocklength codes~\cite{Thangaraj2007WiretapLDPC, Mahdavifar2011PolarCapacityAchieving, Herfeh2021FinitePolarReedMullerComparison}. Remarkably few code constructions have been developed specifically for use at finite blocklength, and those that have been proposed are often not suitable to produce coset codes for the BEWC at practical blocklength. These code construction techniques are discussed below in the context of finite-blocklength coset codes over the BEWC. 
\subsubsection{Capacity-Based Codes}
A number of schemes have been proposed for generating codes which provide secrecy guarantees over the BEWC in the limit of large blocklength. In his introduction of the wiretap channel, Wyner showed that weak secrecy can be achieved using random coding arguments. Since that time, a number of structured wiretap codes have also been proposed which provide security guarantees. These codes frequently make use of a sequence of codes $\mathcal{C}_n$ which approach capacity over the eavesdropper's channel with increasing $n$. A number of capacity-achieving codes have been studied over the BEWC, including LDPC codes and their duals~\cite{Thangaraj2007WiretapLDPC}, polar codes~\cite{Mahdavifar2011PolarCapacityAchieving, Taleb2021FiniteBlocklengthPolarWiretapConiderations}, and Reed-Muller codes~\cite{Herfeh2021FinitePolarReedMullerComparison}. Such constructions can be proven to achieve weak secrecy using an argument laid out aptly in Theorem 1 of ~\cite{Thangaraj2007WiretapLDPC}. It has been shown, however, that these constructions do not, and in many cases cannot, achieve strong secrecy~\cite{Bloch2015ErrorControlForSecrecy}. 
\subsubsection{Constrained Binning Codes}
More recently, alternative wiretap code design techniques have been devised based on constrained binning techniques. These techniques, though more complex than the capacity-based codes and requiring adaptation to produce coset codes, are capable of achieving strong secrecy. Binning techniques are typically based on principles of channel resolvability or privacy amplification. Codes designed on the principle of channel resolvability have been devised in which the base code is a polar code~\cite{Mahdavifar2011PolarCapacityAchieving} or the dual of an LDPC code~\cite{Subramanian2011StrongSecrecyLDPC, Bloch2015ErrorControlForSecrecy}, while codes designed on the principle of channel resolvability have primarily been constructed based on polar codes~\cite{Chou2016PolarRandomBinning, Bloch2015ErrorControlForSecrecy}. 
\subsubsection{Finite-Blocklength-Specific Constructions}
Code designs intended for performance at short blocklength have typically been either brute force searches for best codes~\cite{Al-Hassan2014NewBestCodes}, which provide optimal performance but which are limited to very short blocklength, and autoencoders~\cite{Besser2020Autoencoders, Rana_Chou2021ShortBlocklengthDeepLearning}, which produce promising results but which cannot readily be applied to explicit design of coset codes. One code design technique useful for moderate-blocklength coset codes was developed by Al-Hassan, Ahmed, and Tomlinson in \cite{Al-Hassan2013BestKnownLinearCodes}. Their technique was developed for the binary symmetric wiretap channel (BSWC) but may be adapted to the BEWC.

\subsubsection{Subspace Exclusion Codes}
\label{sec:SEC}
Of particular interest to this study are a class of codes called \emph{subspace exclusion codes}. Introduced in~\cite{hunn2024subspace}, these codes are defined by a generator matrix of dimension $\kappa$ which includes all possible columns except those that lie within a particular subspace of dimension $u$. (The choice of a particular dimension-$u$ subspace does not affect the code performance.) These codes were proven to be optimal for their size in terms of $\chi^2$ divergence. Subspace exclusion codes with $u=\kappa-1$, equivalent to extended Hadamard codes, were also proven to be locally optimal in terms of equivocation loss. 

\subsection{Finite Blocklength Performance Bounds}
\label{sec:bounds}

Although the secrecy capacity of the BEWC has been known since its introduction, the secrecy performance at finite blocklength remained unclear. It was not until Hayashi applied the channel resolvability technique to secrecy coding that useful bounds were identified for finite-blocklength performance~\cite{Hayashi2006OriginalBounds}. Following the introduction of the concept of channel dispersion techniques by Polyansky \emph{et al.}, these methods were applied to secrecy coding to produce significantly improved performance bounds~\cite{Polyansky2010}. The best achievability bound currently known for the BEWC was presented in~\cite{Yang2016FiniteBlocklengthBoundsConference} and is given by 
\begin{equation}
    \label{eqn:achievabilityBound}
    \mathds{V}(p_{MZ},p_M p_Z) < Q\left((\epsilon - \frac{k}{n}) \sqrt{\frac{n}{\epsilon - \epsilon^2}}\right) \mathrm{,} 
\end{equation}
where $Q(\cdot)$ is the Gaussian tail probability function. The best known (nonasymptotic) converse bound for the BEWC was presented in~\cite{Yang2019NonasymptoticLimits} and is given by  
\begin{equation}
    \label{eqn:converseBound}
    \begin{split}
    \mathds{V}(&p_{MZ},p_M p_Z) \geq E_{2^\kappa}(p_{YZ},p_Y p_Z) = \sum_{i=0}^{k}{\left( \binom{n}{i} \epsilon^{i} (1-\epsilon)^{n-i} \cdot (1-2^{i-k}) \right)} \mathrm{.}
    \end{split}
\end{equation}


\subsection{Continuously Valued Performance Functions}
\label{sec:continuouslyValuedFunctions}
Calculating the exact value of an information-theoretic security metric for a particular coset code typically requires considering each of the $2^n$ revealed bit patterns $r(z)$ which could occur in the eavesdropper's observation $Z$, then performing a calculation on the submatrix $G_{r(z)}$ formed by the columns of $G$ indexed by $r(z)$. This approach clearly does not permit application of techniques such as gradient descent. Recently, however, several functions were identified which take a continuously-valued vector as input and which output correct security metric values whenever the input vector defines a real coset code. These functions are described in more detail below. 

\subsubsection{Continuous Code Specification}
Rather than defining a coset code by its generator matrix $G$, the functions defined in~\cite{hunn2024subspace} make use of a code definition vector $q$, such that the $i^{\mathrm{th}}$ element $q_i$ of $q$ represents the fraction of the columns of $G$ equal to $\nu(i)$, where $\nu(i)$ gives the $i^{\mathrm{th}}$ binary vector. This would ordinarily produce a $q$ with indices in the range $[\![0,2^{\kappa}-1]\!]$. It has been proven in~\cite{Harrison2018DualRelationships}, however, that the presence of the all-zero column in $G$ is always detrimental to code performance. We therefore define $q$ with indices in the range $[\![1,2^{\kappa}-1]\!]$ with the assumption that the all-zero column is not present. 

To be considered valid, $q$ must satisfy a nonnegativity constraint, 
\begin{equation}
    \label{eqn:QConstraintNonnegative}
    q_i \geq 0 \;\;\forall\; i \in [\![1,2^\kappa-1]\!]\mathrm{,}
\end{equation}
as well as a unit-sum constraint,
\begin{equation}
    \label{eqn:QConstraintTotal}
    \sum_{i=1}^{2^\kappa-1}q_i = 1 \mathrm{.}
\end{equation}
With these definitions and constraints, every valid generator matrix $G$ has a corresponding $q$. In order to find a valid generator matrix $G$ from a valid $q$, however, it is also necessary for $q$ to satisfy  
\begin{equation}
    \label{eqn:QConstraintRealizable}
    n q_i \in \mathbb{N}, \;\; \forall \;i \in [\![1,2^\kappa-1]\!] \mathrm{,}
\end{equation}
where $n$ is an integer which also represents the blocklength for the realized code. 


\subsubsection{Code Metric Functions}
As identified in~\cite{hunn2024subspace}, for a $\kappa$-dimensional coset code defined by $q$ with blocklength $n$ over a BEWC with erasure probability $\epsilon$, the equivocation loss, denoted $l(n,\epsilon,q)$, is given by 
\begin{equation}
    \label{eqn:expectedEquivocationEpsilon}
    l(n,\epsilon,q) = n (1-\epsilon) - \kappa + \!\! \sum_{\delta = 1 }^{\kappa}{\left( K_{\delta} \!\!\! \sum_{S \in \Xi(W,\kappa-\delta)} \!\!\!\!\!\!  \phi(S,n,\epsilon,q)\right)}\mathrm{,} 
\end{equation}
where $\Xi(W,d)$ is a function which returns the set of all $d$-dimensional subspaces of the code space $W$, $\phi(S,n,\epsilon,q)$ is given by 
\begin{equation}
    \label{eqn:def_phi}
    \phi(S,n,\epsilon,q) = \epsilon^{n(1-\zeta(S,q))}\mathrm{,}
\end{equation}
$\zeta(S,q)$ is equal to
\begin{equation}
    \label{eqn:def_zeta}
    \zeta(S,q) = \sum_{i:\nu(i) \in S}{q_i}\mathrm{,}
\end{equation}
and $K_\delta$ is a series of constants given by 
\begin{equation}                    
    \label{eqn:defKdelta}
    K_\delta = \prod_{i=1}^{\delta-1} {(1-2^{i})}\mathrm{.}
\end{equation}
It was also shown in~\cite{hunn2024subspace} that for a given $q$, $\epsilon$, and $n$, the $\chi^2$ divergence between the joint and marginal message/observation distributions, denoted $\lambda(n,\epsilon,q)$, is given by 
\begin{equation}
    \label{eqn:x2_lambdaFinal}
    \lambda(n,\epsilon,q) = (2-\epsilon)^{n}2^{-\kappa} \left( 1+ \!\! \sum_{S \in \Xi(W,\kappa-1)} \!\!\!\!\!\!  \varphi(S,n,\epsilon,q) \right) -1 \mathrm{,} 
\end{equation}
where 
\begin{equation}
    \label{eqn:x2_def_varphi}
    \varphi(S,n,\epsilon,q) = \left(\frac{\epsilon}{2-\epsilon}\right)^{n(1-\zeta(S,q))} \mathrm{.} 
\end{equation}

\subsubsection{Code Metric Derivatives}
It is clear by inspection of \eqref{eqn:expectedEquivocationEpsilon} and \eqref{eqn:x2_def_varphi} that these functions are continuous and continuously differentiable. The gradient $\nabla l(n,\epsilon,q)$ may then be calculated from \eqref{eqn:expectedEquivocationEpsilon}, \eqref{eqn:def_phi}, and \eqref{eqn:def_zeta}, and its elements are given by 
\begin{equation}
    \label{eqn:ufc_nabla1}
    \begin{split}
    \nabla l(n,\epsilon,q)_{i} &= -n \ln(\epsilon) \sum_{\delta=1}^{\kappa}{\left( K_{\delta} \!\!\!\!\!\!\!\!\!\!\!\!\!\!\!\!\!\!\!\! \sum_{\;\;\;\;\;\;\;\;\;\;\;\; S: S \in \Xi(W,\kappa-\delta), \nu(i) \in S}{\!\!\!\!\!\!\!\!\!\!\!\!\!\!\!\!\!\!\!\! e^{n \ln(\epsilon)(1-\zeta(S))}} \right)} \mathrm{.} 
    \end{split}
\end{equation}
Similarly, $\nabla \lambda(n,\epsilon,q)$ may be calculated from \eqref{eqn:x2_lambdaFinal}, \eqref{eqn:x2_def_varphi}, and \eqref{eqn:def_zeta}, and its elements are given by  
\begin{equation}
    \label{eqn:nablaLambda}
    \begin{split}
    \nabla \lambda(n,\epsilon,q)_{i} &= -n \ln\left(\frac{\epsilon}{2-\epsilon}\right) \left(\frac{\epsilon}{2}\right)^{n}   \!\!\!\!\!\!\!\!\!\!\!\!\!\!\!\!\!\!\!\! \sum_{\;\;\;\;\;\;\;\;\;\;\;\; S: S \in \Xi(W,\kappa-1), \nu(i) \in S}{\!\!\!\!\!\!\!\!\!\!\!\!\!\!\!\!\!\!\!\! e^{n \ln\left(\frac{\epsilon}{2-\epsilon}\right)(1-\zeta(S))}}  \mathrm{.} 
    \end{split}
\end{equation}

\section{Gradient Descent Methods}
\label{sec:GradDescMethods}

Although the gradient $\nabla l(\epsilon,n,q)$ is easy to calculate, the problem of finding a good realizable coset code via gradient descent involves several complications. First, the elements of $q$ must always satisfy total constraint \eqref{eqn:QConstraintTotal}. This may be ensured by computing the constraint-compliant movement vector $\dot{q}$ which is the projection of the gradient $\nabla l(\epsilon,n,q)$ onto the allowable space for the vector $q$. Thus $\dot{q}$ must satisfy 
\begin{equation}
    \label{eqn:xiConstraintTotal}
    \sum_{i}{\dot{q}_i} = 0 \mathrm{.} 
\end{equation}
When this constraint is applied, another complication becomes apparent: In every known starting condition, $q$ tends toward a code defined by uniform $q_i = 1/(2^{\kappa}-1)$ for all $i > 0$. We call this the \emph{uniform fraction code} and represent it as $\bar{q}$.

Although this appears to be the optimal $q$ for any $n,\kappa$, it does not define a realizable code unless $n$ is an integer multiple of $2^{\kappa}-1$. To deal with this difficulty, the natural solution is to force $q$ away from $\bar{q}$ by adding either a hard constraint on the distance from $\bar{q}$ or by adding a centrifugal penalty term to the gradient. These methods are adequate to move $q$ away from the uniform fraction code, but the next problem is that some of the elements of $q$ may become negative in order to satisfy the radial constraint. It then becomes necessary to enforce the nonnegativity constraint \eqref{eqn:QConstraintNonnegative}. If we accept the additional constraint that in the final code, provided $n \leq 2^{\kappa}-1$, no column should appear more than once in $G$, we can also apply an upper bound of 
\begin{equation}
    \label{eqn:QConstraintUpper}
    q_i \leq \frac{1}{n} \;\;\forall\; i \in [1..2^\kappa-1] 
\end{equation}
to each element of $q$. Combining these constraints, as the distance of $q$ from $\bar{q}$ increases, each $q_i$ will eventually reach zero or $\frac{1}{n}$, ensuring a realizable code. 

The upper and lower bounds on the elements of $q$ could be applied by calculating the dimension-reduced gradient formed by finding a linear projection which would zero out all the elements of $\dot{q}$ which would otherwise violate \eqref{eqn:QConstraintNonnegative} or \eqref{eqn:QConstraintUpper}. This approach, however, has some significant drawbacks. For instance, it requires collision detection at each step to determine whether any of the $q_i$ would violate upper or lower bounds. Additionally, for any $q_i$ equal to the upper or lower bounds, the dimension-reduced gradient must be evaluated after each element is zeroed out to determine whether the dimension-reduced gradient requires successive elements to be zeroed out. If, alternatively, gradient elements are permanently zeroed out when they reach one of the limits, some of the elements of $q_i$ might be prevented from otherwise allowable movement which would reduce the objective function. 

To overcome these drawbacks, a novel approach to enforcing the upper and lower bounds on $q_i$ is presented here. In this approach, a diagonal movement cost matrix $\tilde{Q}$ is defined which assigns a unit cost $\tilde{Q}_{i,i}$ to each $\dot{q}_i$, given by 
\begin{equation}
    \label{eqn:GradDesc_def_qtilde}
    \tilde{Q}_{i,i} = \frac{1}{\sqrt{q_i\left({1}/{n}-q_i\right)}} \mathrm{.}
\end{equation}
The cost of movement in a particular direction $\dot{q}$ is then simply $\lvert \tilde{Q} \dot{q}\rvert $. 
Using this definition, imposing a movement cost limit effectively prevents any $q_i$ from exceeding the upper or lower bounds because the movement cost becomes infinite once either bound has been reached. In practice, even with discrete steps of a finite step size, it is possible to use the cost limit to ensure the bounds are not exceeded. To see how, observe that $\dot{q}$ may be expressed in terms of another vector $\grave{q}$ defined by $\grave{q} = \tilde{Q}\dot{q}$. In this case, fixing the movement cost is identical to fixing $\lvert \grave{q} \rvert$. Then to take a movement step of cost $s^2$ in the direction defined by a unit-magnitude $\grave{q}$, instead of calculating $\dot{q}$ and incrementing each $q_i$ linearly in the direction of $\dot{q}_i$, we may identify a function $f(q_i,x)$ of $q_i$ and $x$, where $x$ represents a movement intention for element $q_i$. This function should have the following properties: 
\begin{equation}
    \label{eqn:GradDesc_fPropertyEquality}
    f(q_i,0) = q_i \mathrm{;}
\end{equation}
\begin{equation}
    \label{eqn:GradDesc_fPropertyDerivative}
    \left. \frac{\delta f(q_i,x)}{\delta x} \right|_{x = 0} = \frac{1}{\tilde{Q}_{i,i}} = \sqrt{q_i(1/n-q_i)} \mathrm{;}
\end{equation}
\begin{equation}
    \label{eqn:GradDesc_fPropertyRange}
    f(q_i,x) \in [0,\tfrac{1}{n}] \; \forall \; q_i \in [0,\tfrac{1}{n}] \mathrm{.}
\end{equation}
Then we may update each $q_i$ to its new value $q'_i$ using 
\begin{equation}
    \label{eqn:GradDesc_qGraveUpdate}
    q'_{i} = f(q_i,s\grave{q}_i) \mathrm{.}
\end{equation}
This ensures that the behavior of the update approaches that of the linear method for small $s$ and that the new $q_i$ never exceeds the required bounds. The function that satisfies requirements \eqref{eqn:GradDesc_fPropertyEquality}, \eqref{eqn:GradDesc_fPropertyDerivative}, and \eqref{eqn:GradDesc_fPropertyRange} is 
\begin{equation}
    \label{eqn:GradDesc_f_cos}
    f(q_i,x) = \tfrac{1}{2n}(1+\mathrm{cos}(x + \mathrm{cos}^{-1}(2nq_i-1)))\mathrm{.} 
\end{equation}

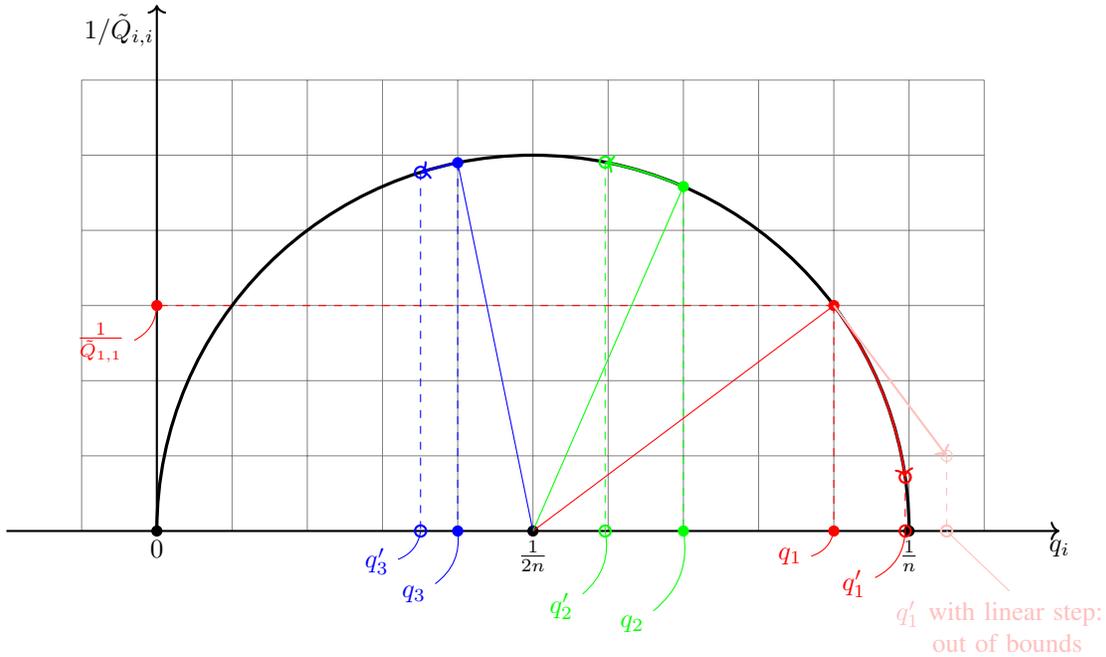
\begin{figure}[!h]
    \centering
    \hspace{0.5cm}
    \begin{tikzpicture}
    \draw[step=1cm,gray,very thin] (-1,0) grid (11,6);
    \draw[->,thick] (-2,0) -- (12,0);
    \draw[->,thick] (0,0) -- (0,7);
    \draw[very thick] (10,0) arc (0:180:5cm);
    \tkzDefPoint(0,0){ll};
    \tkzDefPoint(5,0){c};
    \tkzDefPoint(10,0){ul};
    \node at (c)[circle, fill, inner sep=1.5pt]{};
    \node at (ll)[circle, fill, inner sep=1.5pt]{};
    \node at (ul)[circle, fill, inner sep=1.5pt]{};
    \tkzLabelPoint (c) {$\tfrac{1}{2n}$};
    \tkzLabelPoint (ul) {$\tfrac{1}{n}$};
    \tkzLabelPoint (ll) {$0$};
    \tkzLabelPoint (12,0) {$q_i$};
    \tkzLabelPoint (-0.5,7) {$1/\tilde{Q}_{i,i}$};
    \tkzDefPoint(9,3){q1};
    \tkzDefPoint(9,0){q1x};
    \tkzDefPoint(0,3){q1y};
    \tkzDefPoint(9.949,0.715){q1new};
    \tkzDefPoint(9.949,0){q1newx};
    \tkzDefPoint(0,0.715){q1newy};
    \tkzDefPoint(10.5,1){q1lin};
    \tkzDefPoint(10.5,0){q1linx};
    \tkzDefPoint(0,1){q1liny};
    \tkzDefPoint(7,4.583){q2};
    \tkzDefPoint(7,0){q2x};
    \tkzDefPoint(0,4.583){q2y};
    \tkzDefPoint(5.960,4.907){q2new};
    \tkzDefPoint(5.960,0){q2newx};
    \tkzDefPoint(0,4.907){q2newy};
    \tkzDefPoint(4,4.899){q3};
    \tkzDefPoint(4,0){q3x};
    \tkzDefPoint(0,4.899){q3y};
    \tkzDefPoint(3.506,4.772){q3new};
    \tkzDefPoint(3.506,0){q3newx};
    \tkzDefPoint(0,4.772){q3newy};
    \node at (q1)[circle, fill, red, inner sep=1.5pt]{};
    \draw[red] (q1) -- (c);
    \node at (q1x)[circle,fill,red,thick,inner sep=1.5pt]{};
    \node at (q1y)[circle,fill,red,thick,inner sep=1.5pt]{};
    \draw[dashed, red] (q1) -- (q1x);
    \draw[dashed, red] (q1) -- (q1y);
    \node at (q1new)[circle,draw,red,thick,inner sep=1.5pt]{};
    \draw[<-,red,thick,rotate around={8.222:(5,0)}] (10,0) arc (0:28.648:5cm);
    \node at (q1lin)[circle,draw,pink,inner sep=1.5pt]{};
    \draw[<-,pink,thick] (q1lin) -- (q1);
    \node at (q1newx)[circle,draw,red,thick,inner sep=1.5pt]{};
    \draw[dashed, red] (q1new) -- (q1newx);
    \node at (q1linx)[circle,draw,pink,thick,inner sep=1.5pt]{};
    \draw[dashed, pink] (q1lin) -- (q1linx);
    \node[red, below left=0.1cm and 0.3cm of q1x] (q1xl) {$q_1$}
        edge[red, bend right] (q1x);
    \node[red, below left=0.4cm and 0.4cm of q1newx] (q1newxl) {$q'_1$}
        edge[red, bend right] (q1newx);
    \node[pink, below right=0.8cm and -0.8cm of q1linx, label={[align=left]}, text width = 4cm] (q1linxl) {$q'_1$ with linear step:\\\;\;\;\;\;out of bounds}
        edge[pink] (q1linx);
    \node[red, below left=0.1cm and 0.3cm of q1y] (q1yl) {$\tfrac{1}{\tilde{Q}_{1,1}}$}
        edge[red, bend right] (q1y);
    \node at (q2)[circle, fill, green, inner sep=1.5pt]{};
    \draw[green] (q2) -- (c);
    \node at (q2x)[circle,fill,green,thick,inner sep=1.5pt]{};
    \draw[dashed, green] (q2) -- (q2x);
    \node at (q2new)[circle,draw,green,thick,inner sep=1.5pt]{};
    \draw[->,green,thick,rotate around={66.421:(5,0)}] (10,0) arc (0:12.503:5cm);
    \node at (q2newx)[circle,draw,green,thick,inner sep=1.5pt]{};
    \draw[dashed, green] (q2new) -- (q2newx);
    \node[green, below left=1.0cm and 0.4cm of q2x] (q2xl) {$q_2$}
        edge[green, bend right] (q2x);
    \node[green, below left=0.7cm and 0.3cm of q2newx] (q2newxl) {$q'_2$}
        edge[green, bend right] (q2newx);
    \node at (q3)[circle, fill, blue, inner sep=1.5pt]{};
    \draw[blue] (q3) -- (c);
    \node at (q3x)[circle,fill,blue,thick,inner sep=1.5pt]{};
    \draw[dashed, blue] (q3) -- (q3x);
    \node at (q3new)[circle,draw,blue,thick,inner sep=1.5pt]{};
    \draw[->,blue,thick,rotate around={101.537:(5,0)}] (10,0) arc (0:5.848:5cm);
    \node at (q3newx)[circle,draw,blue,thick,inner sep=1.5pt]{};
    \draw[dashed, blue] (q3new) -- (q3newx);
    \node[blue, below left=0.6cm and 0.3cm of q3x] (q3xl) {$q_3$}
        edge[blue, bend right] (q3x);
    \node[blue, below left=0.1cm and 0.3cm of q3newx] (q3newxl) {$q'_3$}
        edge[blue, bend right] (q3newx);
    \end{tikzpicture}
    \caption{Geometric interpretation of the boundary-compliant update procedure. }
    \label{fig:tikz_VisualAnalogy1}
\end{figure}

The function defined in \eqref{eqn:GradDesc_f_cos}, together with the update procedure given in \eqref{eqn:GradDesc_qGraveUpdate}, permits a natural geometric interpretation. This interpretation is shown in Fig. \ref{fig:tikz_VisualAnalogy1}. In this interpretation, each element $q_i$ of $q$ is mapped to a point on the upper semicircle centered at $\tfrac{1}{2n}$ with radius $\tfrac{1}{2n}$. The base movement vector elements $\grave{q}_i$ then correspond to the relative rotation magnitudes of these points around the semicircle. (Note that only the top semicircle is considered, so if a point would rotate into the lower semicircle, it may instead be considered to have been reflected across the axis back to the top semicircle.) The elements of the movement vector $\dot{q}_i$ correspond to the initial movement rate of the semicircle points in the horizontal ($q_i$) direction, while the vertical axis represents the movement efficiency $\tfrac{1}{\tilde{Q}_{i,i}}$--- that is, the inverse of the movement cost $\tilde{Q}_{i,i}$. 

\begin{figure}[!h]
    \centering
    \hspace{0.5cm}
    \begin{tikzpicture}
    \draw[step=1cm,gray,very thin] (-1,0) grid (11,6);
    \draw[->,thick] (-2,0) -- (12,0);
    \draw[->,thick] (0,0) -- (0,7);
    \draw[very thick] (10,0) arc (0:180:5cm);
    \tkzDefPoint(0,0){ll};
    \tkzDefPoint(5,0){c};
    \tkzDefPoint(10,0){ul};
    \node at (c)[circle, fill, inner sep=1.5pt]{};
    \node at (ll)[circle, fill, inner sep=1.5pt]{};
    \node at (ul)[circle, fill, inner sep=1.5pt]{};
    \tkzLabelPoint (c) {$\tfrac{1}{2n}$};
    \tkzLabelPoint (ul) {$\tfrac{1}{n}$};
    \tkzLabelPoint (ll) {$0$};
    \tkzLabelPoint (12,0) {$q_i$};
    \tkzLabelPoint (-0.5,7) {$1/\tilde{Q}_{i,i}$};
    \tkzDefPoint(9.949,0.715){q1};
    \tkzDefPoint(9.949,0){q1x};
    \tkzDefPoint(0,0.715){q1y};
    \tkzDefPoint(9.981,0.432){q1new};
    \tkzDefPoint(9.981,0){q1newx};
    \tkzDefPoint(0,0.432){q1newy};
    \tkzDefPoint(5.960,4.907){q2};
    \tkzDefPoint(5.960,0){q2x};
    \tkzDefPoint(0,4.907){q2y};
    \tkzDefPoint(6.238,4.844){q2new};
    \tkzDefPoint(6.238,0){q2newx};
    \tkzDefPoint(0,4.844){q2newy};
    \tkzDefPoint(3.506,4.772){q3};
    \tkzDefPoint(3.506,0){q3x};
    \tkzDefPoint(0,4.772){q3y};
    \tkzDefPoint(3.780,4.849){q3new};
    \tkzDefPoint(3.780,0){q3newx};
    \tkzDefPoint(0,4.849){q3newy};
    \tkzDefPoint(6.472,3.465){com1};
    \tkzDefPoint(6.472,0){com1x};
    \tkzDefPoint(0,3.465){com1y};
    \tkzDefPoint(6.667,3.376){com2};
    \tkzDefPoint(6.667,0){com2x};
    \tkzDefPoint(0,3.376){com2y};

    \tkzDefPoint(6.580,3.510){comAngle};
    
    \node at (q1)[circle, fill, red, inner sep=1.5pt]{};
    \node at (q1x)[circle,fill,red,thick,inner sep=1.5pt]{};
    \draw[dashed, red] (q1) -- (q1x);
    \node at (q1new)[circle,draw,red,thick,inner sep=1.5pt]{};
    \draw[<-,red,thick,rotate around={4.956:(5,0)}] (10,0) arc (0:3.266:5cm);
    \node at (q1newx)[circle,draw,red,thick,inner sep=1.5pt]{};
    \draw[dashed, red] (q1new) -- (q1newx);
    \node[red, below left=0.1cm and 0.6cm of q1x] (q1xl) {$q_1$}
        edge[red, bend right] (q1x);
    \node[red, below left=0.5cm and 0.2cm of q1newx] (q1newxl) {$q'_1$}
        edge[red, bend right] (q1newx);
    \node at (q2)[circle, fill, green, inner sep=1.5pt]{};
    \node at (q2x)[circle,fill,green,thick,inner sep=1.5pt]{};
    \draw[dashed, green] (q2) -- (q2x);
    \node at (q2new)[circle,draw,green,thick,inner sep=1.5pt]{};
    \draw[<-,green,thick,rotate around={75.658:(5,0)}] (10,0) arc (0:3.266:5cm);
    \node at (q2newx)[circle,draw,green,thick,inner sep=1.5pt]{};
    \draw[dashed, green] (q2new) -- (q2newx);
    \node[green, below left=0.8cm and 0.2cm of q2x] (q2xl) {$q_2$}
        edge[green, bend right] (q2x);
    \node[green, below left=0.7cm and -0.21cm of q2newx] (q2newxl) {$q'_2$}
        edge[green, bend right] (q2newx);
    \node at (q3)[circle, fill, blue, inner sep=1.5pt]{};
    \node at (q3x)[circle,fill,blue,thick,inner sep=1.5pt]{};
    \draw[dashed, blue] (q3) -- (q3x);
    \node at (q3new)[circle,draw,blue,thick,inner sep=1.5pt]{};
    \draw[<-,blue,thick,rotate around={104.119:(5,0)}] (10,0) arc (0:3.266:5cm);
    \node at (q3newx)[circle,draw,blue,thick,inner sep=1.5pt]{};
    \draw[dashed, blue] (q3new) -- (q3newx);
    \node[blue, below left=0.1cm and 0.4cm of q3x] (q3xl) {$q_3$}
        edge[blue, bend right] (q3x);
    \node[blue, below left=0.5cm and 0.3cm of q3newx] (q3newxl) {$q'_3$}
        edge[blue, bend right] (q3newx);
    \draw[dashed, black] (6.667,0) -- (6.667,6);
    \node at (com1) [circle,fill,black,thick,inner sep=1.5pt]{};
    \draw[black] (c) -- (com1);
    \node at (com2) [circle,draw, black,thick,inner sep=1.5pt]{};
    \draw[<-,black,thick,rotate around={63.713:(5,0)}] (8.765,0) arc (0:3.266:5cm);
    \draw[black, dashed] (c) -- (com2);
    \node at (com2x) [circle,draw, black,thick,inner sep=1.5pt]{};
    \node at (com1x) [circle,fill,black,thick,inner sep=1.5pt]{};
    \draw[black, dashed] (com1) -- (com1x);
    \node[black, align=center, font=\tiny, below right=0.9cm and 0.8cm of com1x] (com1xl) {Current \\ mean}
        edge[black, bend left] (com1x);
    \node[black, align=center, font=\tiny, below right=0.1cm and 0.5cm of com2x] (com2xl) {Required \\ mean}
        edge[black, bend left] (com2x);
    \node[black, align=center, font=\tiny, above left=0.1 and 0.36 of com1] (com1l) {Centroid}
        edge[black, bend right] (com1);
    \node[black, align=center, font=\tiny, above right=-0.05 and 0.22 of com2] (com2l) {Update \\ angle, $\omega$}
        edge[black, bend right] (comAngle);
    \end{tikzpicture}
    \caption{Geometric representation of the boundary-compliant mean adjustment procedure. }
    \label{fig:tikz_VisualAnalogy2}
\end{figure}
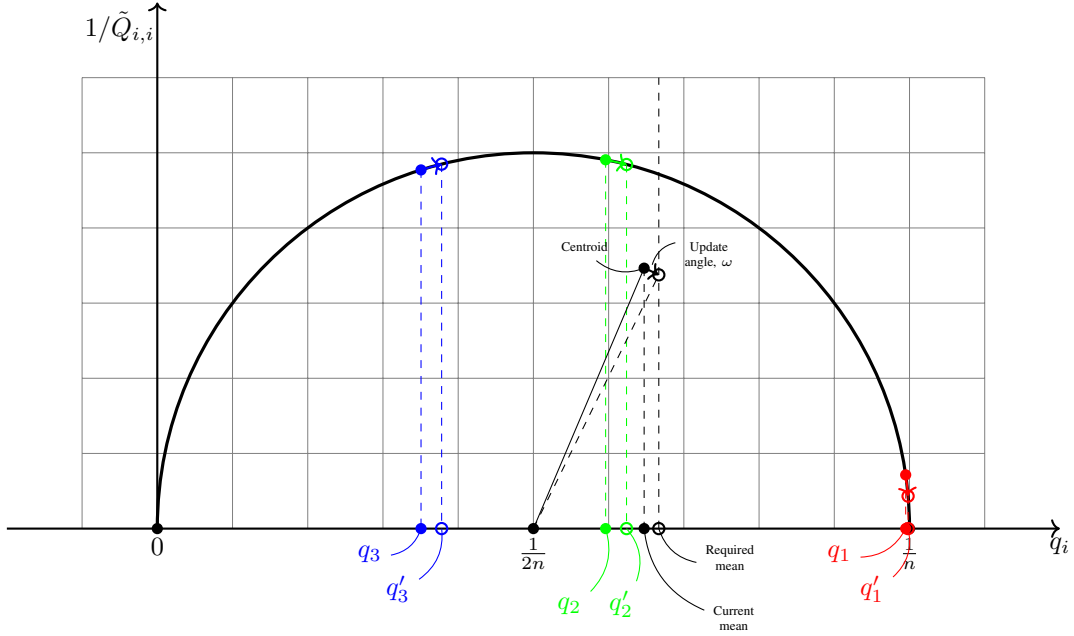

Although the update procedure described above ensures that \eqref{eqn:QConstraintNonnegative} and \eqref{eqn:QConstraintUpper} are satisfied, the nonlinear character of the update function $f(\cdot)$ means that \eqref{eqn:QConstraintTotal} might not be satisfied, even if \eqref{eqn:xiConstraintTotal} is satisfied. For that reason, after each update using $f(\cdot)$, it is necessary to adjust the $q_i$ to bring their mean into compliance. Fortunately, the geometric interpretation delineated in Fig. \ref{fig:tikz_VisualAnalogy1} suggests a natural procedure for updating the mean without violating the upper or lower bound requirements. Consider the semicircle points illustrated in Fig. \ref{fig:tikz_VisualAnalogy1}. If we construct the centroid of these points, the horizontal coordinate of this centroid equals the mean of the $q_i$, which we require to equal $\tfrac{1}{n}$. This centroid may be rotated by an angle $\omega$ about the semicircle's center by rotating each of the points by the same $\omega$. Thus to ensure that \eqref{eqn:QConstraintTotal} is satisfied, we calculate the rotation $\omega$ that would bring the centroid to the correct horizontal position, then apply this rotation to each $q_i$ using the function $f(\cdot)$. More precisely, $\omega$ and $q'_i$ are given by 
\begin{equation}
    \label{eqn:GradDesc_alpha}
    \begin{split}
    \omega &= \mathrm{cos}^{-1} \left(\frac{ \sqrt{\left( \frac{\sum_{i=1}^{2^\kappa-1}{q_i}}{2^\kappa-1}-\frac{1}{2n}\right)^2 + \left(\frac{\sum_{i=1}^{2^\kappa-1}{\sqrt{q_i(1/n - q_i)}}}{2^\kappa-1}\right)^2}}{\frac{1}{2^\kappa-1} - \frac{1}{2n}}\right) - \mathrm{tan}^{-1}\left( \frac{\frac{\sum_{i=1}^{2^\kappa-1}{\sqrt{q_i(1/n - q_i)}}}{2^\kappa-1}}{\frac{\sum_{i=1}^{2^\kappa-1}{q_i}}{2^\kappa-1}-\frac{1}{2n}}\right) \\
    &= \mathrm{cos}^{-1} \left(\frac{ \sqrt{\left( 2n \sum_{i=1}^{2^\kappa-1}{q_i}-(2^\kappa-1)\right)^2 + \left(2n \sum_{i=1}^{2^\kappa-1}{\sqrt{q_i(1/n - q_i)}}\right)^2}}{2n - (2^\kappa-1)}\right) \\
    &\;\;\;\;\;\;\;\; - \mathrm{tan}^{-1}\left( \frac{2n \sum_{i=1}^{2^\kappa-1}{\sqrt{q_i(1/n - q_i)}}}{2n \sum_{i=1}^{2^\kappa-1}{q_i}-(2^\kappa-1)}\right)
    \end{split}
\end{equation}
and 
\begin{equation}
    \label{eqn:GradDesc_SafeAdjustMean}
    q_i = f(q_i,\omega) \mathrm{.} 
\end{equation}    
This procedure is illustrated in Fig. \ref{fig:tikz_VisualAnalogy2}. 

With suitable procedures for updating $q$ based on a given step size and unit vector $\grave{q}$, the next step is to identify the appropriate $\grave{q}$ at each step. Because the change in $q$ must satisfy two objectives- descending the gradient and increasing the distance from the uniform fraction code- $\grave{q}$ may be considered as a linear combination of two vectors, $\grave{q}_{\mathrm{g}}$, which advances $q$ down the gradient, and $\grave{q}_{\mathrm{m}}$, which increases the distance from the uniform fraction code: 
\begin{equation}
    \label{eqn:GradDesc_graveqLinearCombo}
    \grave{q} = c_{\mathrm{g}} \grave{q}_{\mathrm{g}} + c_{\mathrm{m}} \grave{q}_{\mathrm{m}} \mathrm{.}
\end{equation}

The gradient-descending vector $\grave{q}_{\mathrm{g}}$ is required to have the following properties: 1. The elements of its corresponding $\dot{q}_{\mathrm{g}}$ must sum to zero; 2. Its corresponding $\dot{q}_{\mathrm{g}}$ must have zero inner product with the radius vector $q-\bar{q}$; 3. It must have unit magnitude; and 4. Its corresponding $\dot{q}_{\mathrm{g}}$ has minimal inner product with the gradient $\nabla l(\epsilon,n,q)$, subject to the previous constraints. Such a $\grave{q}$ will yield a movement direction which minimizes the objective function $l(\epsilon,n,q)$ subject to constraints \eqref{eqn:QConstraintTotal}, \eqref{eqn:QConstraintNonnegative}, and \eqref{eqn:QConstraintUpper} and while maintaining the distance from the uniform fraction code. To find such a $\grave{q}_{\mathrm{g}}$, we use the method of Lagrange multipliers. The Lagrangian function for this vector, denoted $\mathcal{L}_{\mathrm{g}}$, is defined by 
\begin{equation}
    \label{eqn:GradDesc_Stage1Lagrangian}
    \begin{split}        
    &\mathcal{L}_{\mathrm{g}}(\grave{q}_{\mathrm{g}},\xi_1,\xi_1,\xi_3) = \nabla l(\epsilon,n,q) \tilde{Q}^{-1} \grave{q}_{\mathrm{g}} + \xi_1 \mathds{1}^{\intercal} \tilde{Q}^{-1} \grave{q}_{\mathrm{g}} + \xi_2 (q-\bar{q})^{\intercal} \tilde{Q}^{-1} \grave{q}_{\mathrm{g}} + \xi_3 (\grave{q}_{\mathrm{g}}^{\intercal} \grave{q}_{\mathrm{g}} - 1)  \mathrm{.} 
    \end{split}
\end{equation}
Then the desired $\grave{q}_{\mathrm{g}}$ must satisfy 
\begin{equation}
    \label{eqn:GradDesc_QgraveConstraintTotal}
    \mathds{1}^{\intercal} \tilde{Q}^{-1} \grave{q}_{\mathrm{g}} = 0 \mathrm{,}
\end{equation}
\begin{equation}
    \label{eqn:GradDesc_QgraveConstraintRadius}
    (q-\bar{q})^{\intercal} \tilde{Q}^{-1} \grave{q}_{\mathrm{g}} = 0 \mathrm{,}
\end{equation}
and 
\begin{equation}
    \label{eqn:GradDesc_QgraveConstraintUnit}
    \grave{q}_{\mathrm{g}}^{\intercal} \grave{q}_{\mathrm{g}} = 1 \mathrm{,}
\end{equation}
as well as 
\begin{equation}
    \label{eqn:GradDesc_Lagrangian1}
    \mathbb{0} = \tilde{Q}^{-1} \nabla l(\epsilon,n,q)^{\intercal} + \xi_1 \tilde{Q}^{-1}\mathds{1} + \xi_2 \tilde{Q}^{-1}(q-\bar{q}) + 2\xi_3  \grave{q}_{\mathrm{g}} \mathrm{.} 
\end{equation}
Using the substitutions $\xi_1' = -\tfrac{\xi_1}{2\xi_3}$, $\xi_2' = - \tfrac{\xi_2}{2\xi_3}$, and $\xi_3' = -\tfrac{1}{2\xi_3}$, \eqref{eqn:GradDesc_Lagrangian1} becomes 
\begin{equation}
    \label{eqn:GradDesc_Lagrangian2}
    \begin{split}
    \grave{q}_{\mathrm{g}} &= \xi_3' \tilde{Q}^{-1} \nabla l(\epsilon,n,q)^{\intercal} + \xi_1' \tilde{Q}^{-1}\mathds{1} + \xi_2' \tilde{Q}^{-1}(q-\bar{q}) \mathrm{.} 
    \end{split}
\end{equation}
Then substituting between \eqref{eqn:GradDesc_QgraveConstraintTotal},\eqref{eqn:GradDesc_QgraveConstraintUnit}, and \eqref{eqn:GradDesc_Lagrangian1}  yields 
\begin{equation}
    \label{eqn:GradDesc_Lagrangian3}
    0 = \xi_3' \mathds{1}^{\intercal} \tilde{Q}^{-1} \tilde{Q}^{-1} \nabla l(\epsilon,n,q)^{\intercal} + \xi_1' \mathds{1}^{\intercal} \tilde{Q}^{-1} \tilde{Q}^{-1}\mathds{1} + \xi_2' \mathds{1}^{\intercal} \tilde{Q}^{-1} \tilde{Q}^{-1}(q-\bar{q}) \mathrm{,} 
\end{equation}
\begin{equation}
    \label{eqn:GradDesc_Lagrangian4}
    \xi_1' = - \xi_3' \frac{\mathds{1}^{\intercal} \tilde{Q}^{-2} \nabla l(\epsilon,n,q)^{\intercal}}{\mathds{1}^{\intercal} \tilde{Q}^{-2}\mathds{1}} - \xi_2'\frac{\mathds{1}^{\intercal} \tilde{Q}^{-2}(q-\bar{q})}{\mathds{1}^{\intercal} \tilde{Q}^{-2}\mathds{1}} \mathrm{,} 
\end{equation}
\begin{equation}
    \label{eqn:GradDesc_Lagrangian5}
    \begin{split}
    \grave{q}_{\mathrm{g}} &= \xi_3' \tilde{Q}^{-1} \nabla l(\epsilon,n,q)^{\intercal} + \xi_2' \tilde{Q}^{-1}(q-\bar{q}) \\
    &\;\;\;\;\;\;\;\; - \xi_3' \frac{\tilde{Q}^{-1}\mathds{1} \mathds{1}^{\intercal} \tilde{Q}^{-2} \nabla l(\epsilon,n,q)^{\intercal}}{\mathds{1}^{\intercal} \tilde{Q}^{-2}\mathds{1}} - \xi_2' \frac{\tilde{Q}^{-1}\mathds{1} \mathds{1}^{\intercal} \tilde{Q}^{-2}(q-\bar{q})}{\mathds{1}^{\intercal} \tilde{Q}^{-2}\mathds{1}} \\
    &= \left(I - \frac{\tilde{Q}^{-1}\mathds{1} \mathds{1}^{\intercal} \tilde{Q}^{-1}}{\mathds{1}^{\intercal} \tilde{Q}^{-2}\mathds{1}}\right) \left( \xi_3' \tilde{Q}^{-1} \nabla l(\epsilon,n,q)^{\intercal} + \xi_2' \tilde{Q}^{-1}(q-\bar{q}) \right) \mathrm{,} 
    \end{split}
\end{equation}
and with the substitutions $P_{\mathds{1}} = \left(I - \tfrac{\tilde{Q}^{-1} \mathds{1} \mathds{1}^{\intercal} \tilde{Q}^{-1}}{\mathds{1}^{\intercal} \tilde{Q}^{-2}\mathds{1}}\right)$ and $\tilde{r} = P_{\mathds{1}} \tilde{Q}^{-1} (q-\bar{q})$, 
\begin{equation}
    \label{eqn:GradDesc_Lagrangian6}
    \begin{split}
    0 &= (q-\bar{q})^{\intercal} \tilde{Q}^{-1} P_{\mathds{1}} \left( \xi_3' \tilde{Q}^{-1} \nabla l(\epsilon,n,q)^{\intercal} + \xi_2' \tilde{Q}^{-1}(q-\bar{q}) \right) \mathrm{,} 
    \end{split}
\end{equation}
\begin{equation}
    \label{eqn:GradDesc_Lagrangian7}
    \begin{split}
    \xi_2' &= -\xi_3' \frac{(q-\bar{q})^{\intercal} \tilde{Q}^{-1} P_{\mathds{1}} \tilde{Q}^{-1} \nabla l(\epsilon,n,q)^{\intercal}}{(q-\bar{q})^{\intercal} \tilde{Q}^{-1} P_{\mathds{1}} \tilde{Q}^{-1}(q-\bar{q})} \\
    &= -\xi_3' \frac{\tilde{r}^{\intercal} P_{\mathds{1}} \tilde{Q}^{-1} \nabla l(\epsilon,n,q)^{\intercal}}{\tilde{r}^{\intercal} \tilde{r}}\mathrm{,} 
    \end{split}
\end{equation}
\begin{equation}
    \label{eqn:GradDesc_Lagrangian8}
    \begin{split}
    \grave{q}_{\mathrm{g}} &= \xi_3' P_{\mathds{1}}\tilde{Q}^{-1} \nabla l(\epsilon,n,q)^{\intercal} - \xi_3' \frac{\tilde{r} \tilde{r}^{\intercal} P_{\mathds{1}} \tilde{Q}^{-1} \nabla l(\epsilon,n,q)^{\intercal}}{\tilde{r}^{\intercal} \tilde{r}} \\
    &= \xi_3'\left(I - \frac{\tilde{r} \tilde{r}^{\intercal}}{\tilde{r}^{\intercal} \tilde{r}} \right) P_{\mathds{1}} \tilde{Q}^{-1} \nabla l(\epsilon,n,q)^{\intercal} \\
    &= \frac{\left(I - \frac{\tilde{r} \tilde{r}^{\intercal}}{\tilde{r}^{\intercal} \tilde{r}} \right) P_{\mathds{1}} \tilde{Q}^{-1} \nabla l(\epsilon,n,q)^{\intercal}}{\lvert \left(I - \frac{\tilde{r} \tilde{r}^{\intercal}}{\tilde{r}^{\intercal} \tilde{r}} \right) P_{\mathds{1}} \tilde{Q}^{-1} \nabla l(\epsilon,n,q)^{\intercal} \rvert}\mathrm{.} 
    \end{split}
\end{equation}
Although $\grave{q}_{\mathrm{g}}$ is updated at each step of the gradient descent, it may not be necessary to recalculate the gradient $\nabla l(\epsilon,n,q)$ at each step, as the gradient generally changes relatively slowly. Reducing the frequency of the gradient calculation significantly accelerates the gradient descent performance, as the gradient calculation is the most computationally expensive portion of the calculation of $\grave{q}_{\mathrm{g}}$. In practice, we define a constant $n_{\mathrm{g}}$ and recalculate $\nabla l(\epsilon,n,q)$ once for every $n_{\mathrm{g}}$ descent steps. 

The radial movement vector $\grave{q}_{\mathrm{m}}$ is similarly the legitimate movement vector that increases the distance from the uniform fraction code with the greatest cost efficiency. It is required to have the following properties: 1. The elements of its corresponding $\dot{q}_{\mathrm{m}}$ must sum to zero; 2. It must have unit magnitude; and 4. Its corresponding $\dot{q}_{\mathrm{m}}$ has maximal inner product with the negative radial vector $(\bar{q}-q)$, subject to the previous constraints. We  again use the method of Lagrange multipliers to find such a $\grave{q}_{\mathrm{m}}$. The Lagrangian function for this vector, denoted $\mathcal{L}_{\mathrm{m}}$, is defined by 
\begin{equation}
    \label{eqn:GradDesc_Stage2Lagrangian}
    \begin{split}        
    &\mathcal{L}_{\mathrm{m}}(\grave{q}_{\mathrm{m}},\xi_1,\xi_2) = (q-\bar{q})^{\intercal} \tilde{Q}^{-1} \grave{q}_{\mathrm{m}} + \xi_1 \mathds{1}^{\intercal} \tilde{Q}^{-1} \grave{q}_{\mathrm{m}} + \xi_2 (\grave{q}_{\mathrm{m}}^{\intercal} \grave{q}_{\mathrm{m}} - 1)  \mathrm{,} 
    \end{split}
\end{equation}
and $\grave{q}_{\mathrm{m}}$ must satisfy 
\begin{equation}
    \label{eqn:GradDesc_QgraveMTotal}
    \mathds{1}^{\intercal} \tilde{Q}^{-1} \grave{q}_{\mathrm{m}} = 0 \mathrm{,}
\end{equation}
\begin{equation}
    \label{eqn:GradDesc_QgraveMUnit}
    \grave{q}_{\mathrm{m}}^{\intercal} \grave{q}_{\mathrm{m}} = 1 \mathrm{,}
\end{equation}
and 
\begin{equation}
    \label{eqn:GradDesc_Lagrangian10}
    \mathbb{0} = \tilde{Q}^{-1} (q-\bar{q}) + \xi_1 \tilde{Q}^{-1}\mathds{1} + 2\xi_2  \grave{q}_{\mathrm{m}} \mathrm{.} 
\end{equation}
Using the substitutions $\xi_1' = -\tfrac{\xi_1}{2\xi_2}$ and $\xi_2' = - \tfrac{1}{2\xi_2}$, \eqref{eqn:GradDesc_Lagrangian11} becomes 
\begin{equation}
    \label{eqn:GradDesc_Lagrangian11}
    \begin{split}
    \grave{q}_{\mathrm{m}} &= \xi_2' \tilde{Q}^{-1} (q-\bar{q}) + \xi_1' \tilde{Q}^{-1}\mathds{1} \mathrm{.} 
    \end{split}
\end{equation}
Substituting between \eqref{eqn:GradDesc_QgraveMTotal}, \eqref{eqn:GradDesc_QgraveMUnit}, and \eqref{eqn:GradDesc_Lagrangian11} yields
\begin{equation}
    \label{eqn:GradDesc_Lagrangian12}
    \begin{split}
    0 &= \xi_2' \mathds{1}^{\intercal} \tilde{Q}^{-1} \tilde{Q}^{-1} (q-\bar{q}) + \xi_1' \mathds{1}^{\intercal} \tilde{Q}^{-1} \tilde{Q}^{-1}\mathds{1} \mathrm{,} 
    \end{split}
\end{equation}
\begin{equation}
    \label{eqn:GradDesc_Lagrangian13}
    \begin{split}
    \xi_1' &= - \xi_2' \frac{\mathds{1}^{\intercal} \tilde{Q}^{-1} \tilde{Q}^{-1} (q-\bar{q})}{\mathds{1}^{\intercal} \tilde{Q}^{-1} \tilde{Q}^{-1}\mathds{1}} \mathrm{,} 
    \end{split}
\end{equation}
\begin{equation}
    \label{eqn:GradDesc_Lagrangian14}
    \begin{split}
    \grave{q}_{\mathrm{m}} &= \xi_2' \tilde{Q}^{-1} (q-\bar{q}) - \xi_2' \frac{\mathds{1}^{\intercal} \tilde{Q}^{-1} \tilde{Q}^{-1} (q-\bar{q})}{\mathds{1}^{\intercal} \tilde{Q}^{-1} \tilde{Q}^{-1}\mathds{1}} \tilde{Q}^{-1}\mathds{1} \\
    &= \xi_2' \tilde{Q}^{-1} (q-\bar{q}) - \xi_2' \frac{\tilde{Q}^{-1}\mathds{1} \mathds{1}^{\intercal} \tilde{Q}^{-1} }{\mathds{1}^{\intercal} \tilde{Q}^{-1} \tilde{Q}^{-1}\mathds{1}} \tilde{Q}^{-1} (q-\bar{q}) = \xi_2' P_{\mathds{1}} \tilde{Q}^{-1} (q-\bar{q}) \mathrm{,} 
    \end{split}
\end{equation}
\begin{equation}
    \label{eqn:GradDesc_Lagrangian15}
    1 = \xi_2'^2 (q-\bar{q})^{\intercal} \tilde{Q}^{-1} P_{\mathds{1}} \tilde{Q}^{-1} (q-\bar{q}) \mathrm{,}
\end{equation} 
and 
\begin{equation}
    \label{eqn:GradDesc_Lagrangian16}
    \begin{split}
    \grave{q}_{\mathrm{m}} &=  \frac{P_{\mathds{1}} \tilde{Q}^{-1} (q-\bar{q})}{\lvert P_{\mathds{1}} \tilde{Q}^{-1} (q-\bar{q}) \rvert} \mathrm{.} 
    \end{split}
\end{equation}

The choice of $c_{\mathrm{g}}$ and $c_{\mathrm{m}}$ in \eqref{eqn:GradDesc_graveqLinearCombo} depends on several factors. We desire a constant total step size $s$, but the magnitude of the movement down the gradient should not exceed the distance likely to be required to reach a local minimum. We would also like movement down the gradient to be made when possible, with movement outward from $\bar{q}$ prioritized when $q$ has reached a local minimum with respect to the gradient. To this end, we define the value of movement in the $\grave{q}_{\mathrm{m}}$ direction as the inner product of the corresponding $\dot{q}_{\mathrm{m}}$ with a unit vector in the radial direction, while the value of movement in the $\grave{q}_{\mathrm{g}}$ direction is equal to a weighting factor, denoted $k_\mathrm{g}$, times the inner product of $\dot{q}_{\mathrm{g}}$ with the gradient $\nabla l(\epsilon,n,q)$. With these definitions, the optimal value per unit cost is achieved by selecting $c_{\mathrm{g}}$ and $c_{\mathrm{m}}$ in proportion to the value of $\grave{q}_{\mathrm{g}}$ and $\grave{q}_{\mathrm{m}}$, respectively. Then $\grave{q}$ is given by 
\begin{equation}
    \label{eqn:GradDesc_QGrave_kg}
    \begin{split}
    \grave{q} = \frac{-k_{\mathrm{g}} \nabla l(\epsilon,n,q) \tfrac{ \tilde{Q}^{-1} \grave{q}_{\mathrm{g}} }{\lvert \tilde{Q}^{-1} \grave{q}_{\mathrm{g}} \rvert} \grave{q}_{\mathrm{g}} + \tfrac{(q-\bar{q})^{\intercal} \tilde{Q}^{-1} \grave{q}_{\mathrm{m}} }{\lvert q-\bar{q} \rvert \lvert \tilde{Q}^{-1} \grave{q}_{\mathrm{m}} \rvert} \grave{q}_{\mathrm{m}}} {\sqrt{ \left(k_{\mathrm{g}} \nabla l(\epsilon,n,q) \tfrac{ \tilde{Q}^{-1} \grave{q}_{\mathrm{g}} }{\lvert \tilde{Q}^{-1} \grave{q}_{\mathrm{g}} \rvert}\right)^2 + \left(\tfrac{(q-\bar{q})^{\intercal} \tilde{Q}^{-1} \grave{q}_{\mathrm{m}} }{\lvert q-\bar{q} \rvert \lvert \tilde{Q}^{-1} \grave{q}_{\mathrm{m}} \rvert}  \right)^2 } } \mathrm{.}
    \end{split}
\end{equation}

The choice of $k_{\mathrm{g}}$ has the effect of determining how quickly the distance from the uniform fraction code increases. Too low a value will not prevent the distance from increasing when there is a strong gradient to descent, but too high a value will cause oscillations around local minima with a large enough amplitude to permanently stop outward movement from the uniform fraction code. To select an appropriate value for $k_{\mathrm{g}}$, the magnitude of undesirable fluctuations in $\grave{q}_{\mathrm{g}}$ may be estimated and set to a fixed fraction of the value $\tfrac{(q-\bar{q})^{\intercal} \tilde{Q}^{-1} \grave{q}_{\mathrm{m}} }{\lvert q-\bar{q} \rvert \lvert \tilde{Q}^{-1} \grave{q}_{\mathrm{m}} \rvert}$ of movement in the $\grave{q}_{\mathrm{m}}$ direction. The value of the undesirable components of $\grave{q}_{\mathrm{g}}$ is estimated by calculating an exponentially weighted moving variance on the quantity $\nabla l(\epsilon,n,q) \tfrac{ \tilde{Q}^{-1} \grave{q}_{\mathrm{g}} }{\lvert \tilde{Q}^{-1} \grave{q}_{\mathrm{g}} \rvert} \grave{q}_{\mathrm{g}}$. This moving variance, denoted $\mathds{W}_{\alpha} \left(\nabla l(\epsilon,n,q) \tfrac{ \tilde{Q}^{-1} \grave{q}_{\mathrm{g}} }{\lvert \tilde{Q}^{-1} \grave{q}_{\mathrm{g}} \rvert} \grave{q}_{\mathrm{g}} \right)$ for exponent $\alpha$, is then used to calculate the ratio $\tau$ of the undesirable fluctuations to the $\grave{q}_{\mathrm{m}}$ value: 
\begin{equation}
    \label{eqn:GradDesc_tauDef}
    \tau = k_{\mathrm{g}} \frac{\sqrt{\mathds{W}_{\alpha} \left(\nabla l(\epsilon,n,q) \tfrac{ \tilde{Q}^{-1} \grave{q}_{\mathrm{g}} }{\lvert \tilde{Q}^{-1} \grave{q}_{\mathrm{g}} \rvert} \grave{q}_{\mathrm{g}} \right)}}{\tfrac{(q-\bar{q})^{\intercal} \tilde{Q}^{-1} \grave{q}_{\mathrm{m}} }{\lvert q-\bar{q} \rvert \lvert \tilde{Q}^{-1} \grave{q}_{\mathrm{m}} \rvert}} \mathrm{.} 
\end{equation}
The value of $\tau$ is then set to a target $\tau_{\mathrm{t}}$ by adjusting $k_{\mathrm{g}}$ according to a proportional controller law with gain $\alpha$. 

With the suitable $\grave{q}$ and $f(q_i,x)$, the only remaining decisions necessary to perform gradient descent optimization are the choice of initial $q$, step size $s$, gradient recalculation period $n_{\mathrm{g}}$, and the gain exponent $\alpha$ and target $\tau_{\mathrm{t}}$ used for updating $k_{\mathrm{g}}$. The final pseudocode for gradient optimization is shown in Algorithm \ref{alg:gradDesc}.  

\begin{algorithm}
    \caption{Full Gradient Descent Algorithm}
    \label{alg:gradDesc}
    \KwIn{$\kappa$}
    \KwIn{$n$}
    \KwIn{$\epsilon$}
    \Parameter{$s \text{, step size}$}
    \Parameter{$n_{\mathrm{g}} \text{, gradient recalculation period}$}
    \Parameter{$\alpha \text{, gain exponent}$}
    \Parameter{$\tau_{\mathrm{t}} \text{, fluctuation ratio target}$}
    \Parameter{$\sigma \text{, random offset standard deviation}$}
    $k_{\mathrm{g}} \gets 1$
    
    $q \gets \bar{q}$
    
    $\mathbb{g} \gets \mathbb{0}$
    
    $\mathds{W} \gets 1$

    \While{$\lvert \, q \, \rvert^2 < 1/n$}
    {
        //Add random offset \\        
        $r \gets N(0,\sigma)$ \tabto{4cm}//Normally distributed random vector \\        
        $q \gets f(q,r)$ \tabto{4cm}//Update q \\
        $q \gets f(q,\omega(q))$ \tabto{4cm}//Safe adjust mean \\
        //Calculate gradient using subspace decomposition \\        
        $g \gets \nabla l(\epsilon,n,q)$\\       
        \For{$n \gets 1$ \KwTo $n_{\mathrm{g}}$}
        {
            $\tilde{q}_{i:i \in [1 .. 2^\kappa-1]} \gets \frac{n}{\sqrt{n q_i(1 - n q_i)}}$ \\
            $\tilde{Q} \gets \mathrm{diag}(\tilde{q})$ \\
            $P_{\mathds{1}} \gets \left(I - \tfrac{\tilde{Q}^{-1} \mathds{1} \mathds{1}^{\intercal} \tilde{Q}^{-1}}{\mathds{1}^{\intercal} \tilde{Q}^{-2}\mathds{1}}\right)$ \\
            $\tilde{r} \gets P_{\mathds{1}} \tilde{Q}^{-1} (q-\bar{q})$ \\
            $\grave{q}_{\mathrm{m}} \gets \frac{\tilde{r}}{\lvert \tilde{r} \rvert}$ \\
            $\grave{q}_{\mathrm{g}} \gets \left( I - \frac{\tilde{r} \tilde{r}^{\intercal}}{\tilde{r}^{\intercal} \tilde{r}} \right) P_{\mathds{1}} \tilde{Q}^{-1} g^{\intercal}$ \\
            $\grave{q}_{\mathrm{g}} \gets \frac{\grave{q}_{\mathrm{g}}}{\lvert \grave{q}_{\mathrm{g}} \rvert}$ \\
            \uIf{$g \tilde{Q}^{-1} \grave{q}_{\mathrm{g}} > 0$} 
            {
                $\grave{q}_{\mathrm{g}} \gets -\grave{q}_{\mathrm{g}}$ \tabto{4cm} //To make sure $\grave{q}_{\mathrm{g}}$ represents a min, not a max
            }
            $\grave{q} \gets -k_{\mathrm{g}} g \tfrac{ \tilde{Q}^{-1} \grave{q}_{\mathrm{g}} }{\lvert \tilde{Q}^{-1} \grave{q}_{\mathrm{g}} \rvert} \grave{q}_{\mathrm{g}} + \tfrac{(q-\bar{q})^{\intercal} \tilde{Q}^{-1} \grave{q}_{\mathrm{m}} }{\lvert q-\bar{q} \rvert \lvert \tilde{Q}^{-1} \grave{q}_{\mathrm{m}} \rvert} \grave{q}_{\mathrm{m}}$ \\
            $\grave{q} \gets \frac{\grave{q}}{\lvert \grave{q} \rvert}$ \\
            $q \gets f(q,\frac{s \grave{q}}{\lvert \tilde{Q}^{-1} \grave{q} \rvert})$ \tabto{4cm}//Update q \\
            $q \gets f(q,\omega(q))$ \tabto{4cm}//Safe adjust mean 
        }
        //Update $k_{\mathrm{g}}$ \\
        $\mathds{W} \gets (1-\alpha) \mathds{W} + \alpha (1-\alpha) \left| \tilde{Q}^{-1} \grave{q}_{\mathrm{g}} g \frac{\tilde{Q}^{-1} \grave{q}_{\mathrm{g}}}{\lvert \tilde{Q}^{-1} \grave{q}_{\mathrm{g}} \rvert} - \mathbb{g} \right|^2$ \\
        $\mathbb{g} \gets (1-\alpha) \mathbb{g} + \alpha \tilde{Q}^{-1} \grave{q}_{\mathrm{g}} g \frac{\tilde{Q}^{-1} \grave{q}_{\mathrm{g}}}{\lvert \tilde{Q}^{-1} \grave{q}_{\mathrm{g}} \rvert}$ \\
        $k_{\mathrm{g}} \gets  k_{\mathrm{g}} \cdot (\frac{\grave{q}_{\mathrm{m}}^{\intercal} \tilde{Q}^{-1} (q-\bar{q})}{k_{\mathrm{g}} \sqrt{\mathds{W}} \lvert q-\bar{q} \rvert})^{\alpha}$
    }    
\end{algorithm}
\newpage

\section{Comparison Methods}
\label{sec:comparisonMethods}

To evaluate the effectiveness of the methods described in this work, Algorithm \ref{alg:gradDesc} was implemented and executed for a variety of code sizes. Both equivocation loss and $\chi^2$ divergence were used as the function to be optimized. The techniques described in the literature were also implemented for similar code sizes and used to generate codes for comparison. As a baseline, we also generated a sample of random codes of each size. Additionally, the limits described in the literature and summarized in \ref{sec:bounds} were calculated for each code size and presented for comparison. The details of each of these computations are given below. 

\subsection{Gradient Descent Execution}
Algorithm \ref{alg:gradDesc} was used to generate codes which minimize both equivocation loss and $\chi^2$ divergence. For codes minimizing equivocation loss, dimensions in the range $\kappa=[\![8,10]\!]$ were considered, while for $\chi^2$ divergence, the range of dimensions included $\kappa=[\![8,12]\!]$. For each code dimension, a total of sixteen blocklengths were considered, starting with $n=2^{\kappa-4}$, in intervals of $2^{\kappa-4}$, up to $n=2^{\kappa} - 2^{\kappa-4}$. In all cases, the erasure probability was set to $\epsilon = k/n = (n-\kappa)/n$. The remaining parameters for Algorithm \ref{alg:gradDesc} were as follows: $s=0.0001$; $n_g=25$; $\alpha=0.5$; $\tau_t=1$; $\sigma=0.000001$.

\subsection{LDPC-Based Secrecy Codes}
As reported in \cite{Subramanian2011StrongSecrecyLDPC}, the duals of LDPC codes achieve strong secrecy in the limit of large blocklength. Generating such a code is ordinarily simple, as it only requires creation of a randomly-generated low-density generator matrix. For the code sizes used in the present application, however, the process is complicated by the fact that the blocklengths are often very large compared with the code dimension. This implies that it is not practical to maintain a constant distribution of row weights as blocklength increases (or the generator matrix would contain many all-zero columns). For this reason, the code is instead defined by starting with a column weight of two, then adding all weight-two columns to the generator matrix. Next, all weight-three columns are considered. If $n$ does not permit all weight-three columns, then a random selection of these columns is included. If $n$ is sufficiently large, all of them are added, and the process is repeated for weight-four columns, etc., until the required blocklength is reached. This process is repeated for each code size to produce the required LDPC-based secrecy codes.  

\subsection{BKLC-Based Secrecy Codes}
The authors of \cite{Al-Hassan2013BestKnownLinearCodes} describe a procedure for generating secrecy codes incrementally by sequentially adding columns to a generator matrix. In this scheme, each new column is selected to optimize the performance of the code over the binary symmetric channel (BSC). The initial generator matrix is formed from a best known linear code (BKLC)--- a code that is the best for its size in terms of minimum distance among all published code constructions. This technique has the advantage that at each step, the message-observation mutual information $I(M;Z)$ is simply a function of the new column and the current code's syndrome probabilities. 

A BKLC-based code was generated for each of the required code sizes. The initial BKLC was identified using the Magma BKLC database \cite{Magma}. Because this technique optimizes for a BSC, a bit error probability $p$ is required to calculate the syndrome probabilities. The value of $p$ was selected, using a rationale similar to that described in \ref{sec:achievabilityGap}, to set the secrecy capacity of the corresponding binary symmetric wiretap channel (BSWC) equal to the desired code rate $k/n$. The expression relating $p$ and $\epsilon$ is thus 
\begin{equation}
    \label{eqn:BSC_epsilon_relation}
    \epsilon = - p \cdot \log_2(p) - (1-p)\log_2(1-p) \mathrm{.} 
\end{equation}

\subsection{Random Secrecy Codes}
For each of the required code sizes, a sample of 256 random codes were generated. These codes were generated by selecting at random $n$ of the $2^{\kappa}-1$ possible nonzero generator matrix columns (with uniform distribution and without replacement). Each random sample was then used to estimate the mean and standard deviation of the underlying distribution of random codes. This permits the performance of other code constructions, as well as applicable limits, to be expressed in terms of deviation from random code performance. 

\subsection{Limits on Code Metrics}
In addition to other secrecy code constructions, we use known limits on code performance metrics to evaluate the performance of codes generated using Algorithm \ref{alg:gradDesc}. The limits listed in \ref{sec:bounds} apply to total variation distance $\mathds{V}(p_{MZ},p_M p_Z)$, but over the BEWC, the total variation distance may be related to both the $\chi^2$ divergence $\chi^2(p_{MZ},p_M p_Z)$ and to the equivocation loss (which is also equal to the Kullback-Leibler divergence $\mathds{D}(p_{MZ},p_M p_Z)$). These relations exist because all three of these metrics fall into the category of $f$-divergences~\cite{SilveyFDivergence}, and they may therefore be expressed as 
\begin{equation}
    \label{eqn:def_f-divergence}
    \mathds{X}(p_{MZ},p_M p_Z) = \sum_{m,z}{f_{\mathds{X}}\left( \frac{p_{MZ}(m,z)}{p_M(m) p_Z(z)} \right) p_M(m) p_Z(z)} = \sum_{z}{p_Z(z) \sum_{m}{f_{\mathds{X}}\left( \frac{p_{MZ}(m,z)}{p_M(m) p_Z(z)} \right) p_M(m) }}\mathrm{.} 
\end{equation}
Here, $\mathds{X} \in \{\mathds{V}, \mathds{D}, \chi^2  \}$, and the $f_{\mathds{X}}$ are known as the kernels of the divergences and are given by 
\begin{equation}
    \label{eqn:kernel_V}
    f_{\mathds{V}}(a) = \frac{1}{2}|a-1| \mathrm{,}
\end{equation}
\begin{equation}
    \label{eqn:kernel_D}
    f_{\mathds{D}}(a) = a \log_2(a) \mathrm{,}
\end{equation}
and 
\begin{equation}
    \label{eqn:kernel_x2}
    f_{\chi^2}(a) = a^2-1 \mathrm{.}
\end{equation}
Considering \eqref{eqn:def_f-divergence}, we may note that for a given $z$, the equivocation loss $H(M)-H(M|Z=z)$ is in the range $[\![0 , k]\!]$. Furthermore, if a given $z$ results in $b$ bits of equivocation loss, then there are $2^{k-b}$ possible messages $m$ which are consistent with $z$. Thus the value of the fraction in \eqref{eqn:def_f-divergence} is zero for $2^{k}-2^{k-b}$ values of $m$ and $2^b$ for the remaining $2^{k-b}$ values of $m$. Using these observations, we may express the inner sum in \eqref{eqn:def_f-divergence} as 
\begin{equation}
    \label{eqn:innerSum_X}
    \sum_{m}{f_{\mathds{X}}\left( \frac{p_{MZ}(m,z)}{p_M(m) p_Z(z)} \right) p_M(m) } = (1-2^{-b})f_{\mathds{X}}(0) + 2^{-b}f_{\mathds{X}}(2^b) \mathrm{.} 
\end{equation}
Substituting in the different metrics, we obtain 
\begin{equation}
    \label{eqn:innerSum_V}
    \sum_{m}{f_{\mathds{V}}\left( \frac{p_{MZ}(m,z)}{p_M(m) p_Z(z)} \right) p_M(m) } = 1-2^{-b} \mathrm{,} 
\end{equation}
\begin{equation}
    \label{eqn:innerSum_D}
    \sum_{m}{f_{\mathds{D}}\left( \frac{p_{MZ}(m,z)}{p_M(m) p_Z(z)} \right) p_M(m) } = b \mathrm{,} 
\end{equation}
and
\begin{equation}
    \label{eqn:innerSum_x2}
    \sum_{m}{f_{\chi^2}\left( \frac{p_{MZ}(m,z)}{p_M(m) p_Z(z)} \right) p_M(m) } = 2^{b}-1 \mathrm{.} 
\end{equation}
Next, we may take $b$ in the range $[\![1,k]\!]$ (because the sum is always zero for $b=0$) to find ranges for the ratios of the sums in \eqref{eqn:innerSum_V}, \eqref{eqn:innerSum_D}, and \eqref{eqn:innerSum_x2}. This yields (using the expressions $S_{\mathds{V}}$, $S_{\mathds{D}}$, and $S_{\chi^2}$ as shorthand for the sums in \eqref{eqn:innerSum_V}, \eqref{eqn:innerSum_D}, and \eqref{eqn:innerSum_x2}, respectively) 
\begin{equation}
    \label{eqn:innerRatio_VD}
    \frac{S_{\mathds{V}}}{S_{\mathds{D}}} = \frac{1-2^{-b}}{b} \in \left[\frac{1-2^{-k}}{k},\frac{1}{2}\right] \mathrm{,}
\end{equation}
\begin{equation}
    \label{eqn:innerRatio_Dx2}
    \frac{S_{\mathds{D}}}{S_{\chi^2}} = \frac{b}{2^{b}-1} \in \left[\frac{k}{2^{k}-1},1\right] \mathrm{,}
\end{equation}
and 
\begin{equation}
    \label{eqn:innerRatio_x2V}
    \frac{S_{\chi^2}}{S_{\mathds{V}}} = 2^b \in \left[2,2^{k}\right] \mathrm{.}
\end{equation}
Substituting these expressions back into \eqref{eqn:def_f-divergence}, we may bound the $\chi^2$ divergence and the equivocation loss by constant factors of the total variation distance. This gives
\begin{equation}
    \label{eqn:tvd_x2_relation}
    2\mathds{V}(p_{MZ},p_M p_Z) \leq \chi^2(p_{MZ},p_M p_Z) \leq 2^k \mathds{V}(p_{MZ},p_M p_Z)
\end{equation}
and
\begin{equation}
    \label{eqn:tvd_equiv_relation}
    2\mathds{V}(p_{MZ},p_M p_Z) \leq \mathds{D}(p_{MZ},p_M p_Z) \leq \frac{k}{1-2^{-k}} \mathds{V}(p_{MZ},p_M p_Z) \mathrm{.} 
\end{equation}

Combining the limits expressed in \eqref{eqn:achievabilityBound} and \eqref{eqn:converseBound} with relations \eqref{eqn:tvd_x2_relation} and \eqref{eqn:tvd_equiv_relation}, we find that the optimal equivocation loss and $\chi^2$ divergence are bounded by
\begin{equation}
    \label{eqn:finalLimit_EqLoss}
    2\sum_{i=0}^{k}{\left( \binom{n}{i} \epsilon^{i} (1-\epsilon)^{n-i} \cdot (1-2^{i-k}) \right)} \leq \mathds{D}(p_{MZ},p_M p_Z) \leq \frac{k}{1-2^{-k}}  Q\left((\epsilon - \frac{k}{n}) \sqrt{\frac{n}{\epsilon - \epsilon^2}}\right)
\end{equation}
and 
\begin{equation}
    \label{eqn:finalLimit_x2}
    2\sum_{i=0}^{k}{\left( \binom{n}{i} \epsilon^{i} (1-\epsilon)^{n-i} \cdot (1-2^{i-k}) \right)} \leq \chi^2(p_{MZ},p_M p_Z) \leq 2^{k}  Q\left((\epsilon - \frac{k}{n}) \sqrt{\frac{n}{\epsilon - \epsilon^2}}\right) \mathrm{.} 
\end{equation}

In addition to bounds based on total variation distance, at least one converse bound is known for the $\chi^2$ divergence~\cite{hunn2024subspace}. It is found by setting $q$ to $\bar{q}$ in \eqref{eqn:x2_lambdaFinal} and is given by 
\begin{equation}
    \label{eqn:boundDirectX2}
    \lambda(n,\epsilon,q) \geq (2-\epsilon)^{n}2^{-\kappa} \left( 1+ \!\! (2^{\kappa}-1)\left( \frac{\epsilon}{2-\epsilon} \right)^{\frac{n 2^{\kappa-1}}{2^{\kappa}-1}} \right) -1 \mathrm{.} 
\end{equation}



\section{Results}
Each of the code generation techniques described in \ref{sec:finiteBlocklengthConstructions} was executed for the code sizes specified, and the resulting final codes were recorded. The codes were then evaluated in terms of equivocation loss and $\chi^2$ divergence, and the results were compared. The limits specified in \ref{sec:bounds} were also compared. The results are shown graphically below and are given in tabular form in Appendix \ref{apx:EqLossResultsTabular} (equivocation loss) and Appendix \ref{apx:X2ResultsTabular} ($\chi^2$ divergence). The final codes generated by Algorithm \ref{alg:gradDesc} are also listed in the text file accompanying this paper. During the execution of Algorithm \ref{alg:gradDesc}, the value of $q$ was also recorded periodically as the algorithm progressed. These $q$ values define a path from $\bar{q}$ to the final realizable code. These paths are shown for selected code sizes below to illustrate the complexity of the dynamics of the gradient descent. 

\subsection{Gradient Descent Path}
\begin{figure}
    \centering
    \includegraphics[scale=1.0]{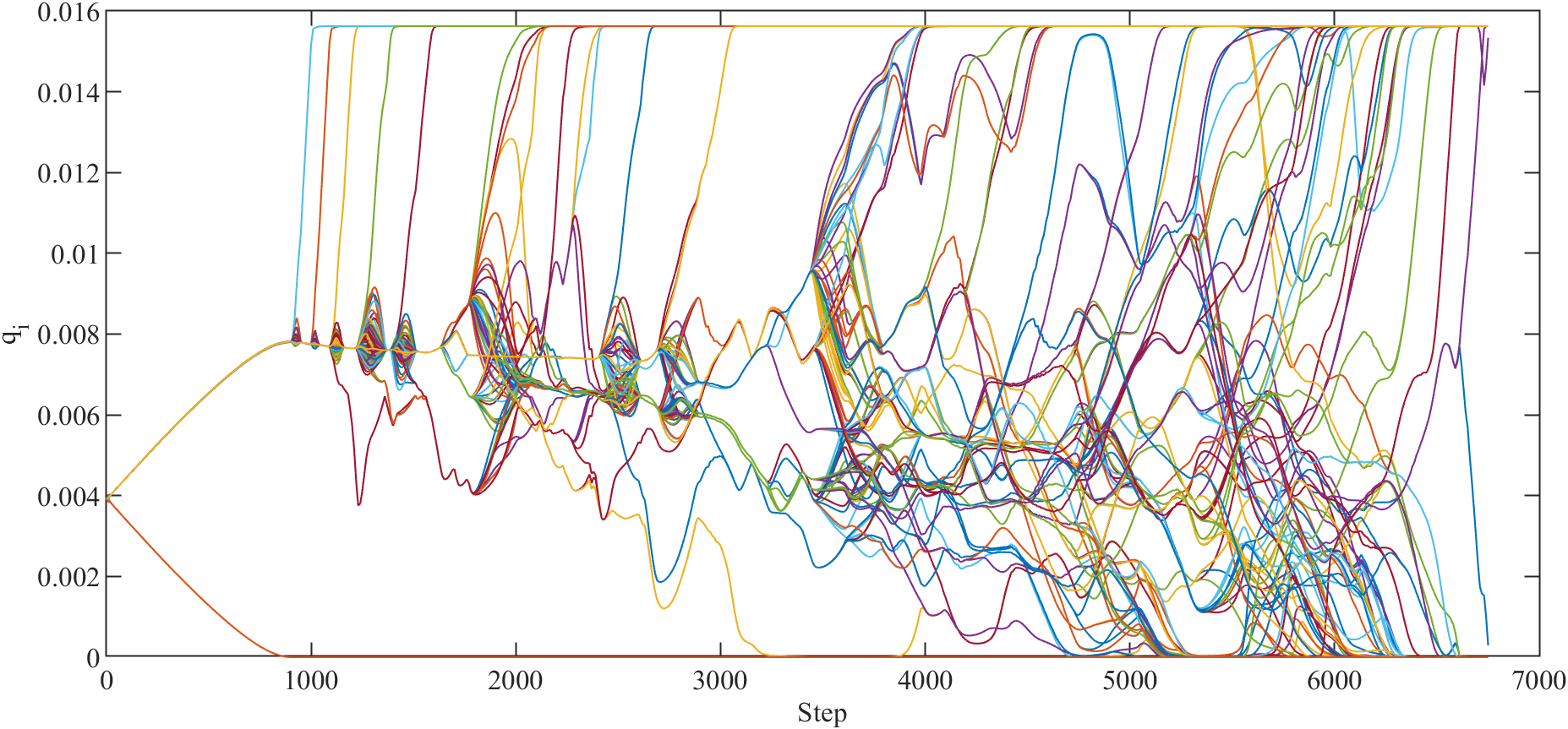}
    \caption{Gradient descent path minimizing equivocation loss for $n=64$ and $\kappa=8$.}
    \label{fig:pathEqLoss_8_64}
\end{figure}

\begin{figure}
    \centering
    \includegraphics[scale=1.0]{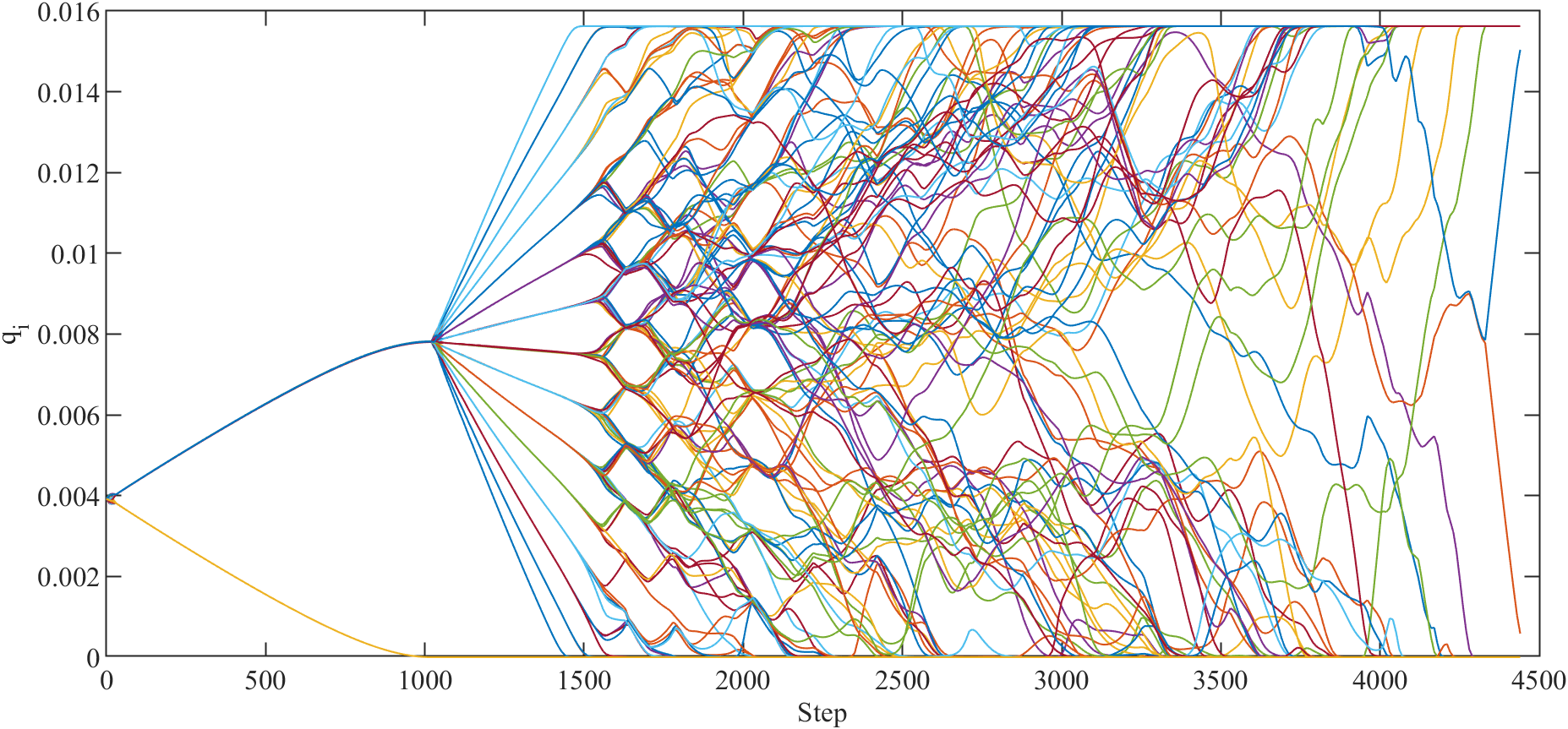}
    \caption{Gradient descent path minimizing $\chi^2$ divergence for $n=64$ and $\kappa=8$.}
    \label{fig:pathX2_8_64}
\end{figure}

\begin{figure}
    \centering
    \includegraphics[scale=1.0]{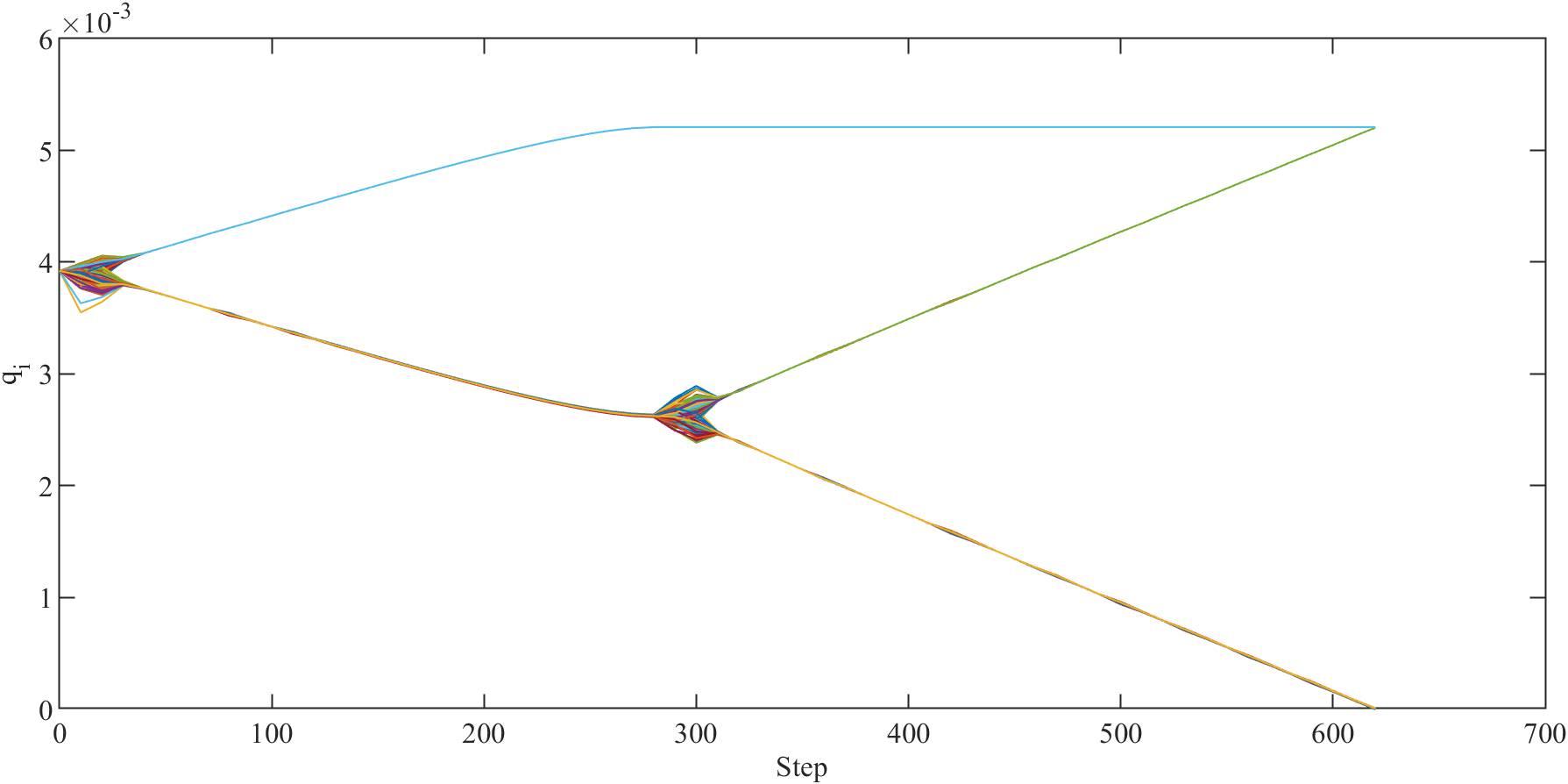}
    \caption{Gradient descent path minimizing equivocation loss for $n=192$ and $\kappa=8$.}
    \label{fig:pathEqLoss_8_192}
\end{figure}

\begin{figure}
    \centering
    \includegraphics[scale=1.0]{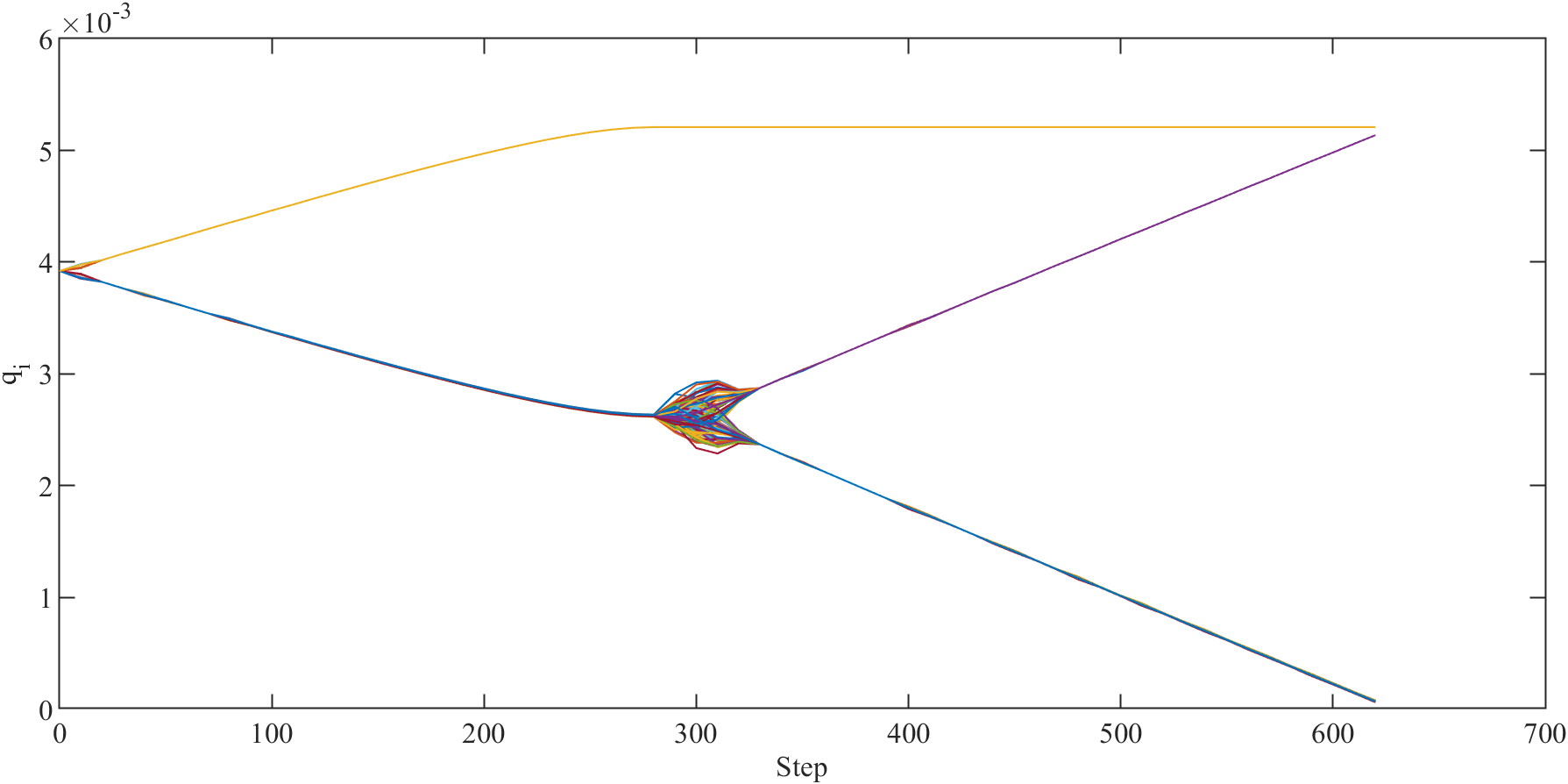}
    \caption{Gradient descent path minimizing $\chi^2$ divergence for $n=192$ and $\kappa=8$.}
    \label{fig:pathX2_8_1924}
\end{figure}

The path followed by the $q$ vector as Algorithm \ref{alg:gradDesc} progressed are shown for code size (64,8) for equivocation loss in Fig. \ref{fig:pathEqLoss_8_64} and for $\chi^2$ divergence in Fig. \ref{fig:pathX2_8_64}. In these figures, each trace represents one element of $q$. The examples shown in Figs. \ref{fig:pathEqLoss_8_64} and \ref{fig:pathX2_8_64} are characteristic of the gradient descent paths for all code sizes except those for which a subspace exclusion code is possible, i.e., those for which $2^{\kappa}-n$ is an integer power of two. A representative example of the gradient descent for such a code size is shown for the (192,8) case in Fig. \ref{fig:pathEqLoss_8_192} (equivocation loss) and in Fig. \ref{fig:pathX2_8_1924} ($\chi^2$ divergence). 

\newpage

\subsection{Equivocation Loss Performance}
\begin{figure}
    \centering
    \includegraphics[trim= 0 0.25cm 0 0.85cm, scale=1.0]{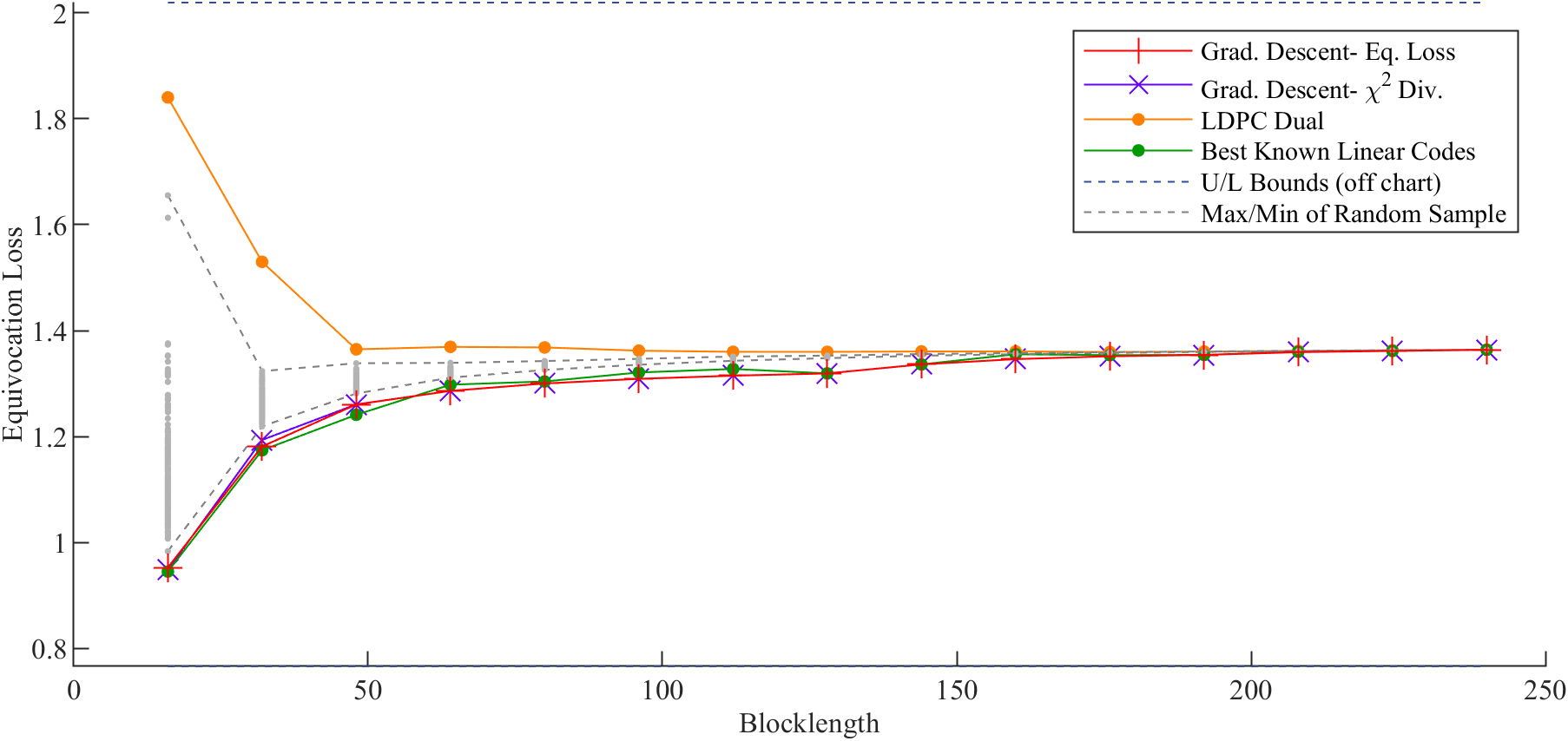}
    \caption{Equivocation Loss for Codes of Dimension $\kappa=8$.}
    \label{fig:eqLossNonscaled8}
\end{figure}

\begin{figure}
    \centering
    \includegraphics[trim= 0 0.25cm 0 0.85cm, scale=1.0]{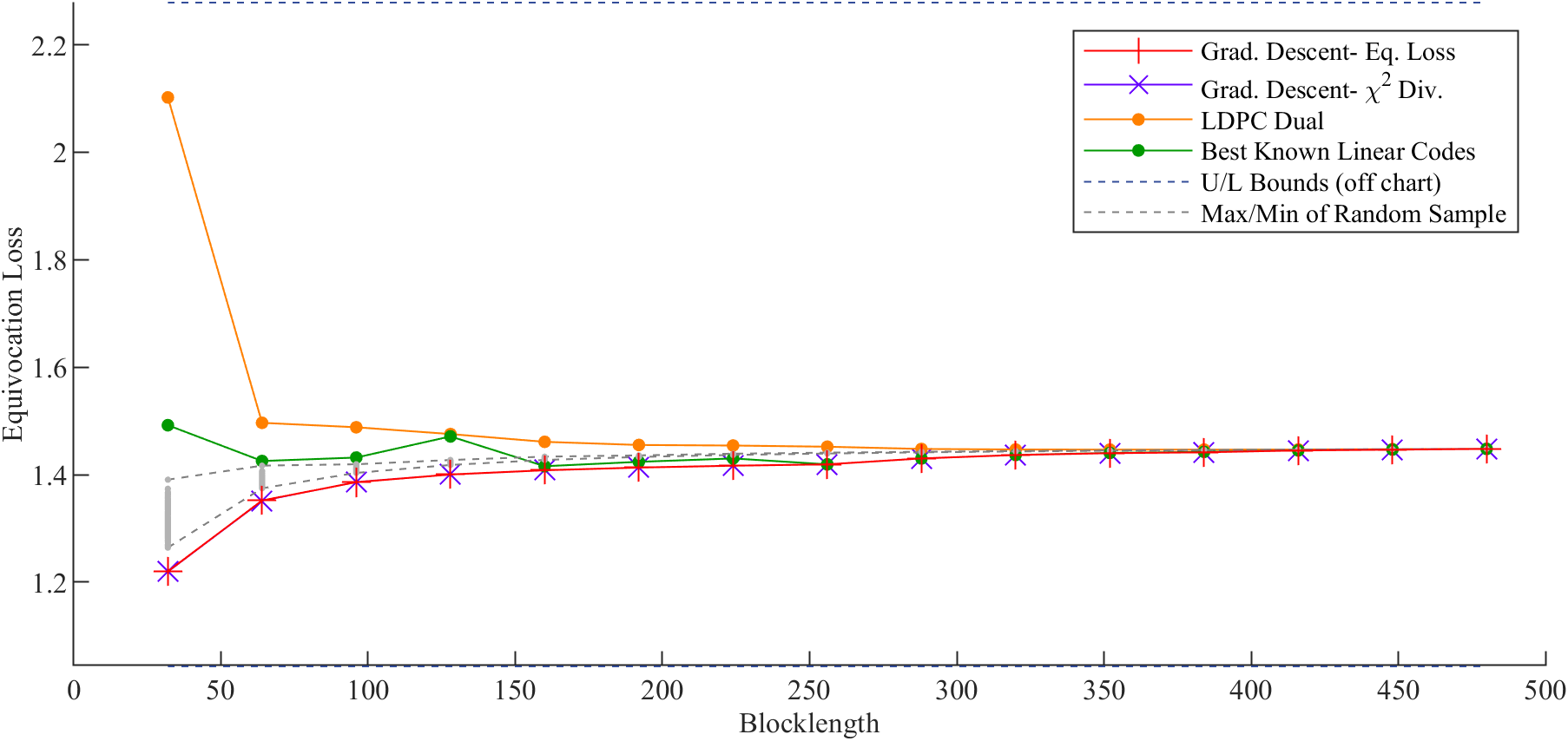}
    \caption{Equivocation Loss for Codes of Dimension $\kappa=9$.}
    \label{fig:eqLossNonscaled9}
\end{figure}

\begin{figure}
    \centering
    \includegraphics[trim= 0 0.25cm 0 0.85cm, scale=1.0]{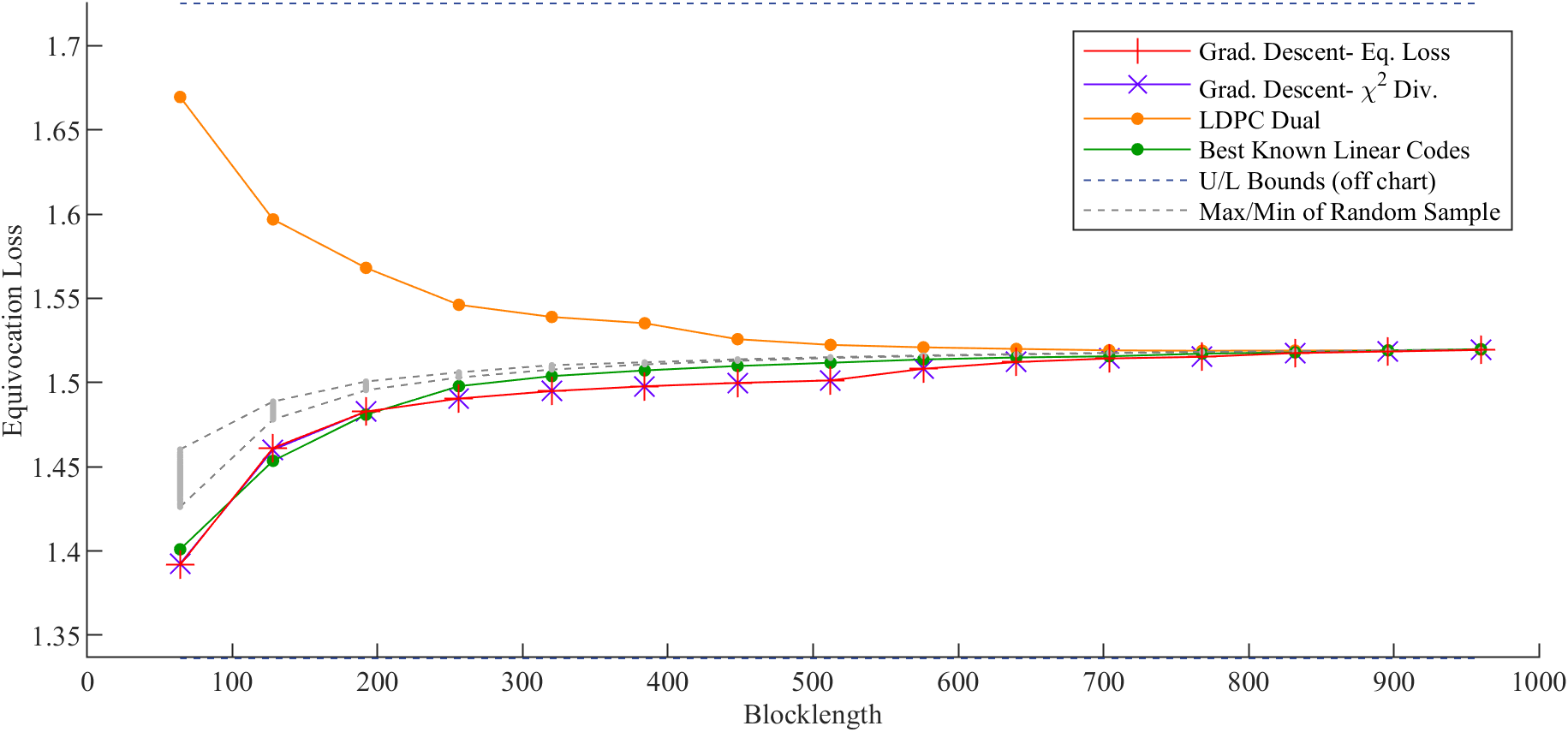}
    \caption{Equivocation Loss for Codes of Dimension $\kappa=10$.}
    \label{fig:eqLossNonscaled10}
\end{figure}

\begin{figure}
    \centering
    \includegraphics[trim= 0 0.25cm 0 0.85cm, scale=1.0]{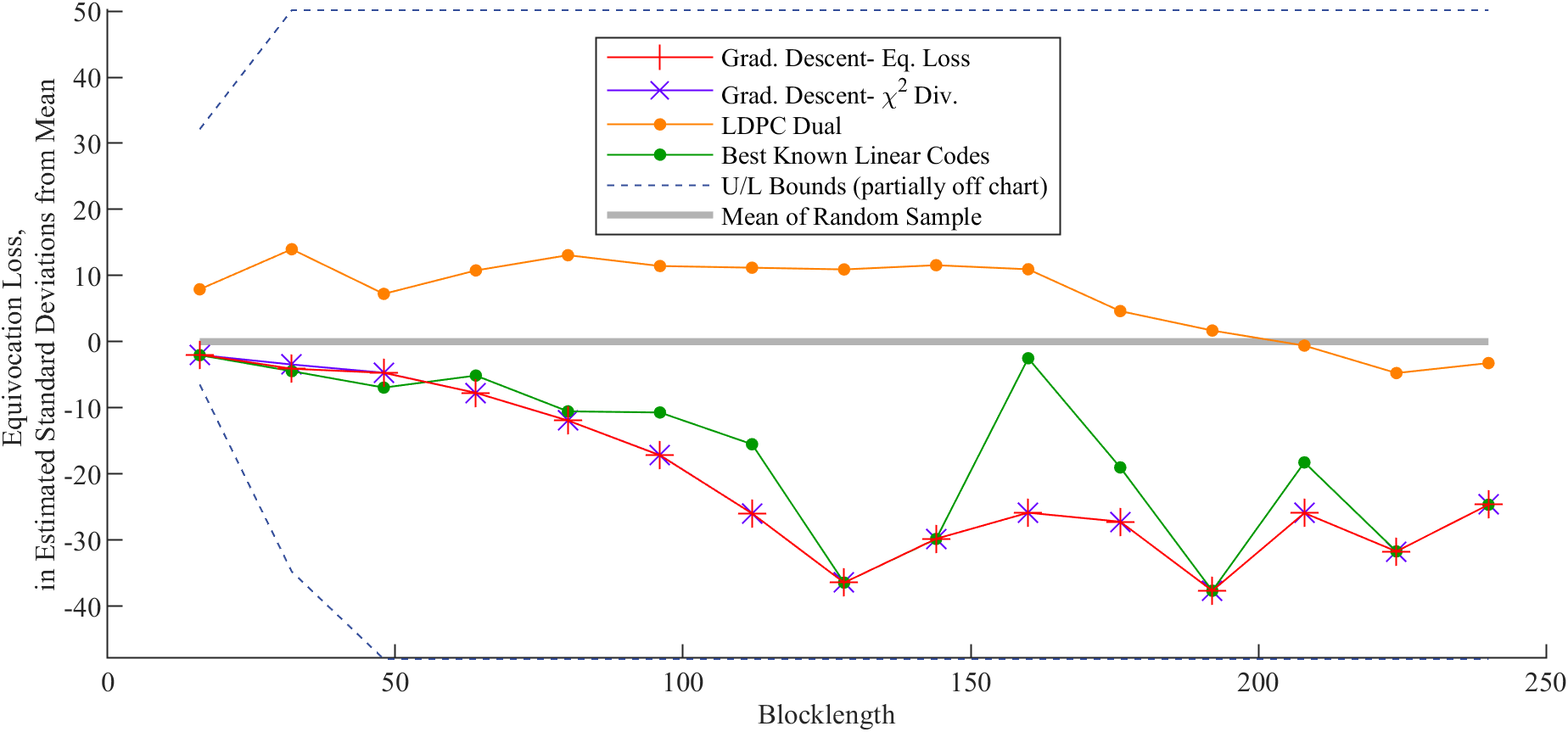}
    \caption{Equivocation Loss for Codes of Dimension $\kappa=8$, Relative to Random Sample Performance.}
    \label{fig:eqLossScaled8}
\end{figure}

\begin{figure}
    \centering
    \includegraphics[trim= 0 0.25cm 0 0.85cm, scale=1.0]{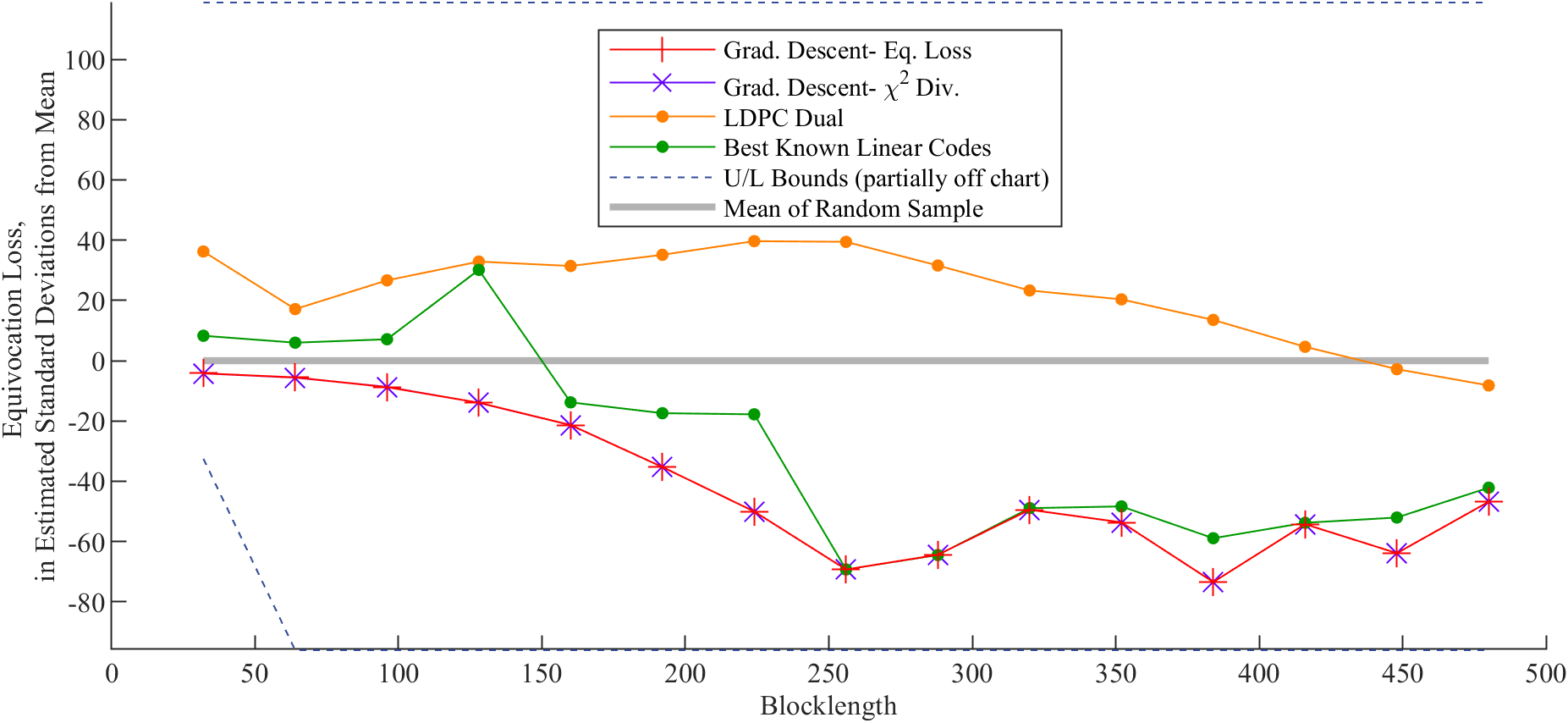}
    \caption{Equivocation Loss for Codes of Dimension $\kappa=9$, Relative to Random Sample Performance.}
    \label{fig:eqLossScaled9}
\end{figure}

\begin{figure}
    \centering
    \includegraphics[trim= 0 0.25cm 0 0.85cm, scale=1.0]{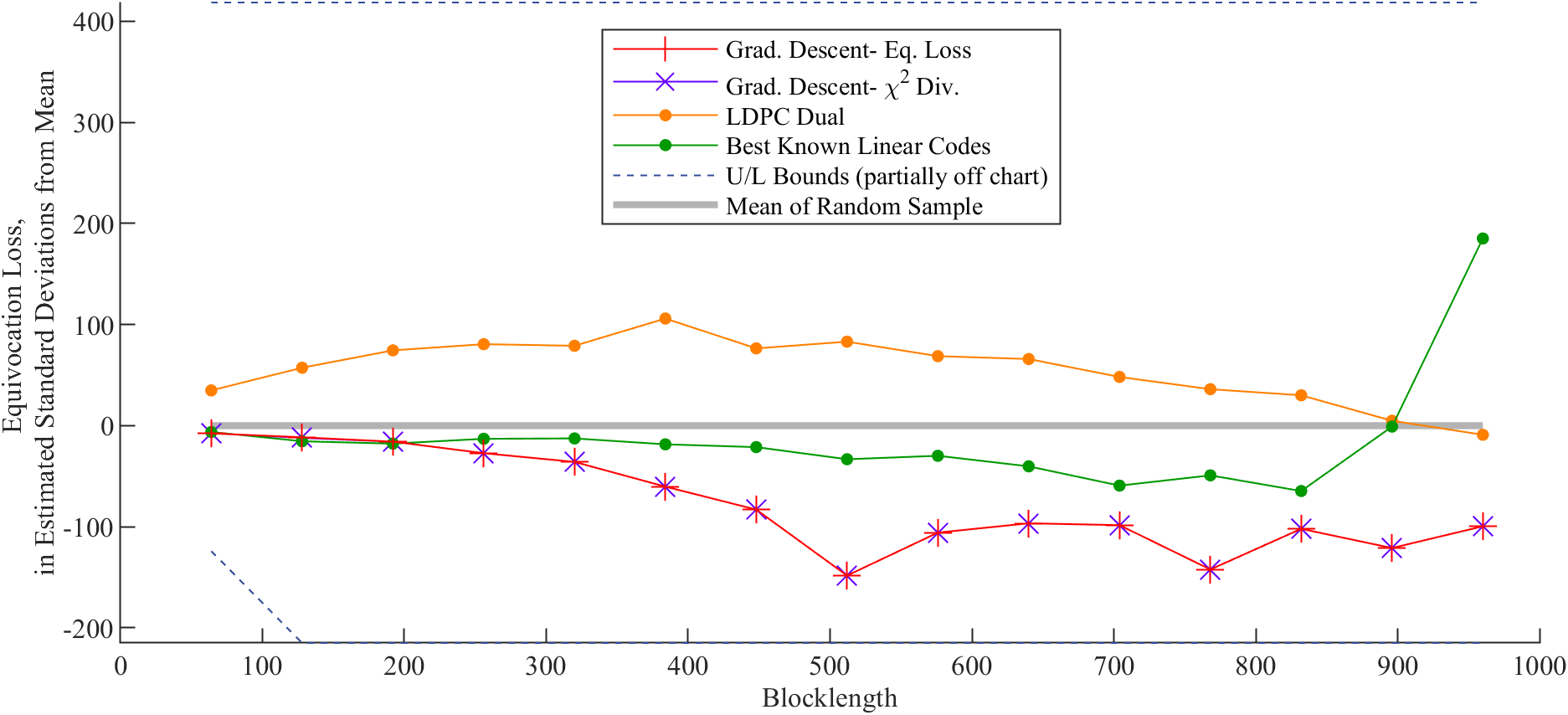}
    \caption{Equivocation Loss for Codes of Dimension $\kappa=10$, Relative to Random Sample Performance.}
    \label{fig:eqLossScaled10}
\end{figure}

The results of gradient descent optimization in terms of equivocation loss, as well as limits and comparison codes, are shown for dimension $\kappa=8$, $\kappa=9$, and $\kappa=10$ in Fig. \ref{fig:eqLossNonscaled8}, Fig. \ref{fig:eqLossNonscaled9}, and Fig. \ref{fig:eqLossNonscaled10}, respectively. In Figs. \ref{fig:eqLossScaled8}, \ref{fig:eqLossScaled9}, and \ref{fig:eqLossScaled10}, the same results are shown in terms of the estimated standard deviation of the random samples. 

\clearpage

\subsection{$\chi^2$ Divergence Performance}
\begin{figure}[htpb]
    \centering
    \includegraphics[trim= 0 0.25cm 0 -0.25cm, scale=1.0]{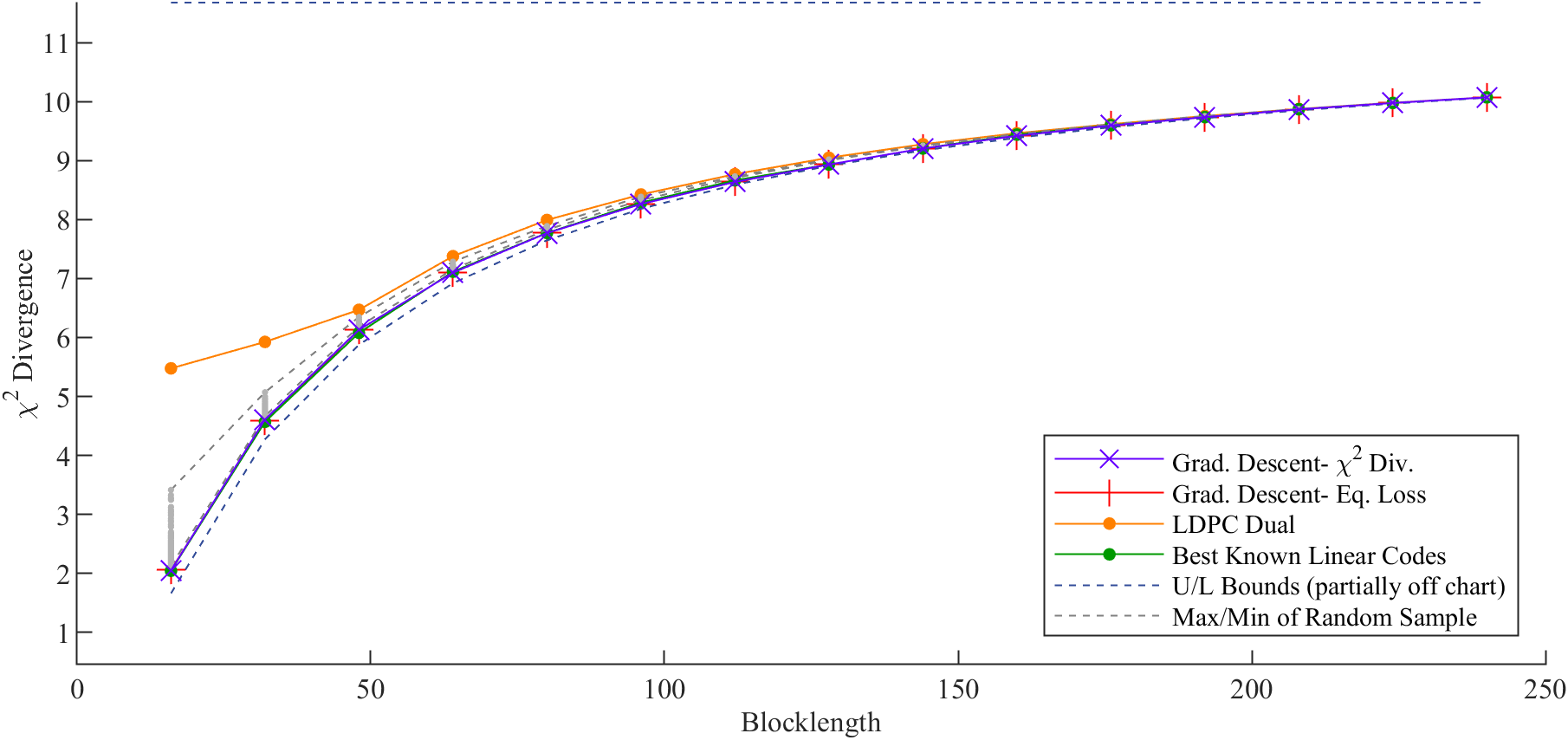}
    \caption{$\chi^2$ Divergence for Codes of Dimension $\kappa=8$.}
    \label{fig:x2Nonscaled8}
\end{figure}

\begin{figure}[htpb]
    \centering
    \includegraphics[trim= 0 0.25cm 0 0.25cm, scale=1.0]{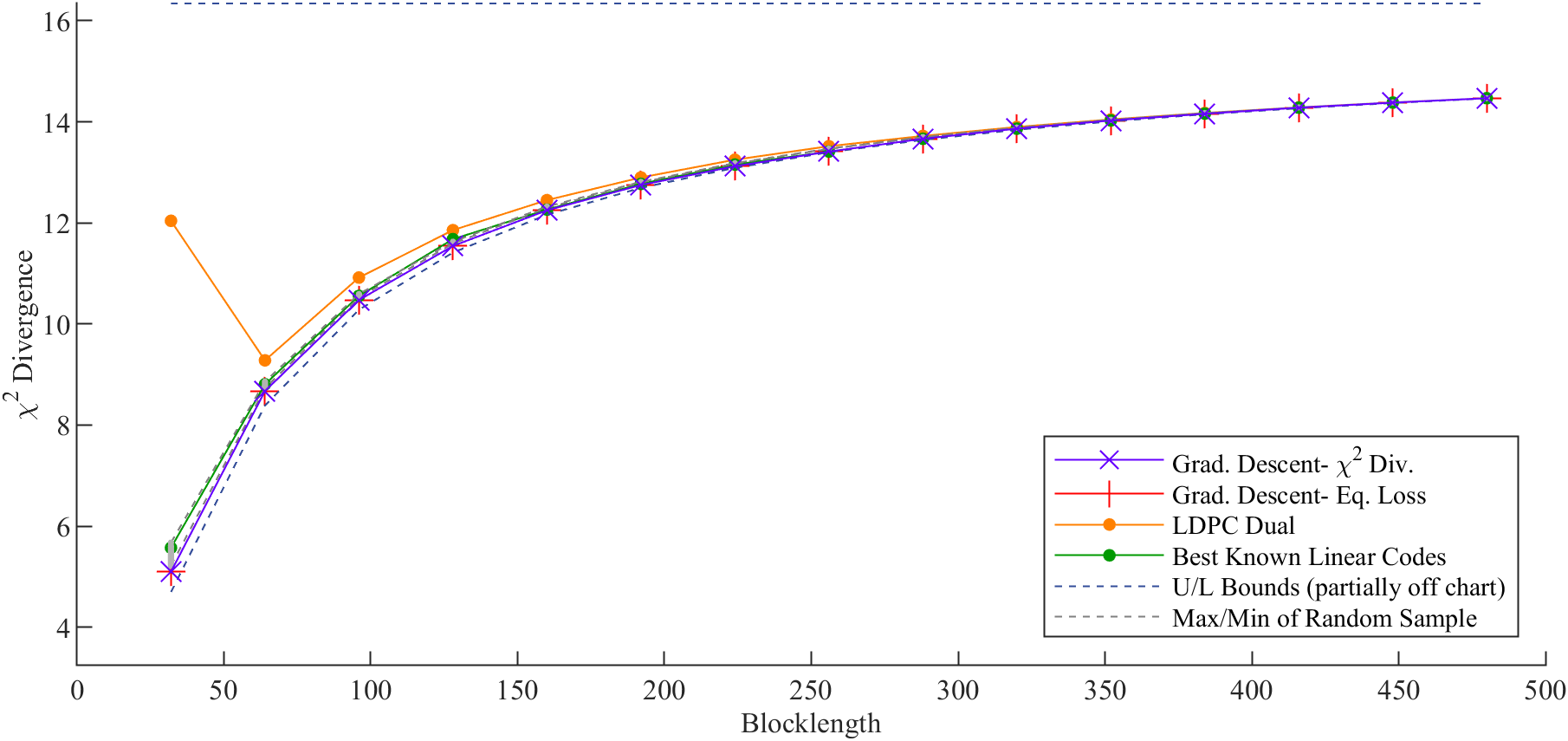}
    \caption{$\chi^2$ Divergence for Codes of Dimension $\kappa=9$.}
    \label{fig:x2Nonscaled9}
\end{figure}

\begin{figure}
    \centering
    \includegraphics[trim= 0 0.25cm 0 0.85cm, scale=1.0]{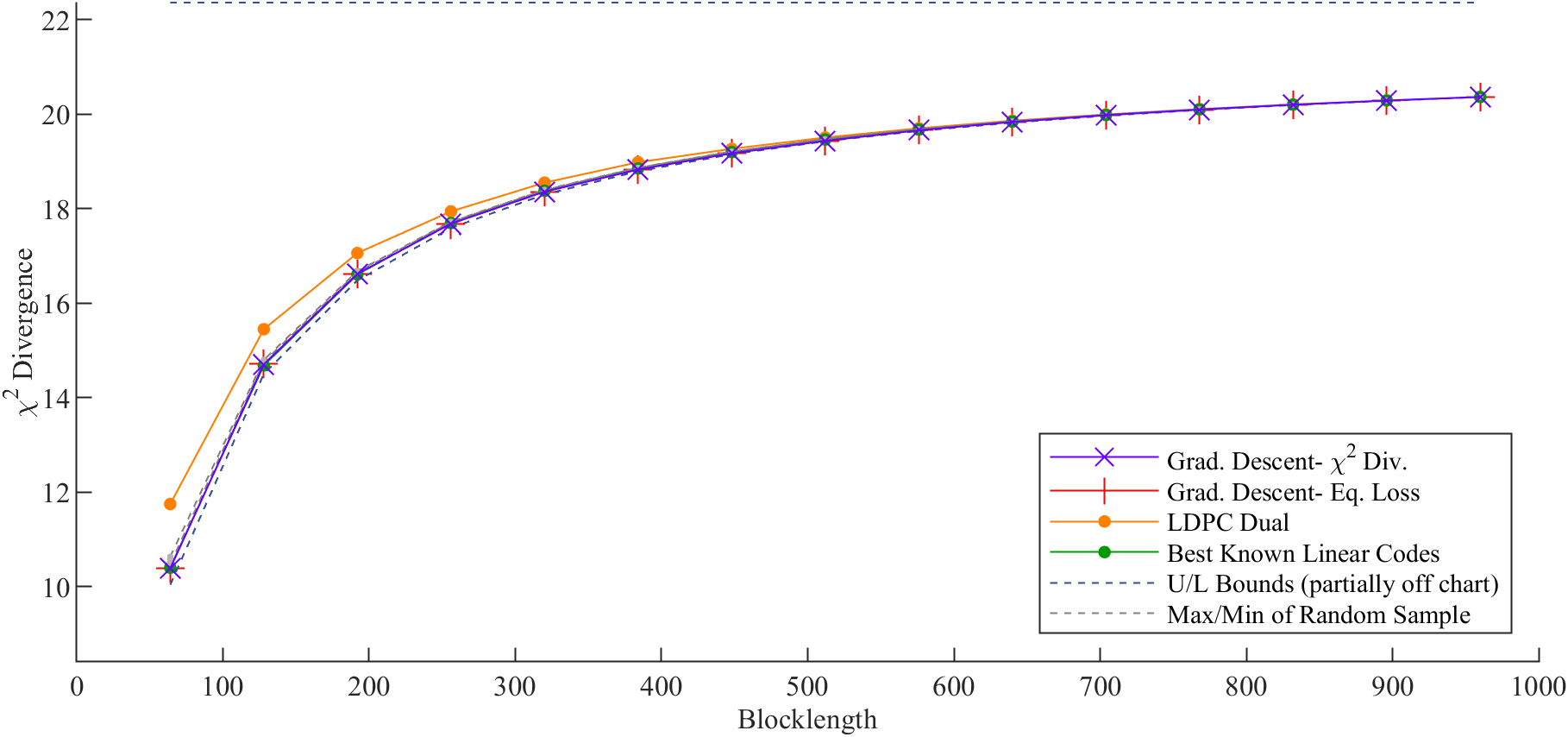}
    \caption{$\chi^2$ Divergence for Codes of Dimension $\kappa=10$.}
    \label{fig:x2Nonscaled10}
\end{figure}

\begin{figure}
    \centering
    \includegraphics[trim= 0 0.25cm 0 0.85cm, scale=1.0]{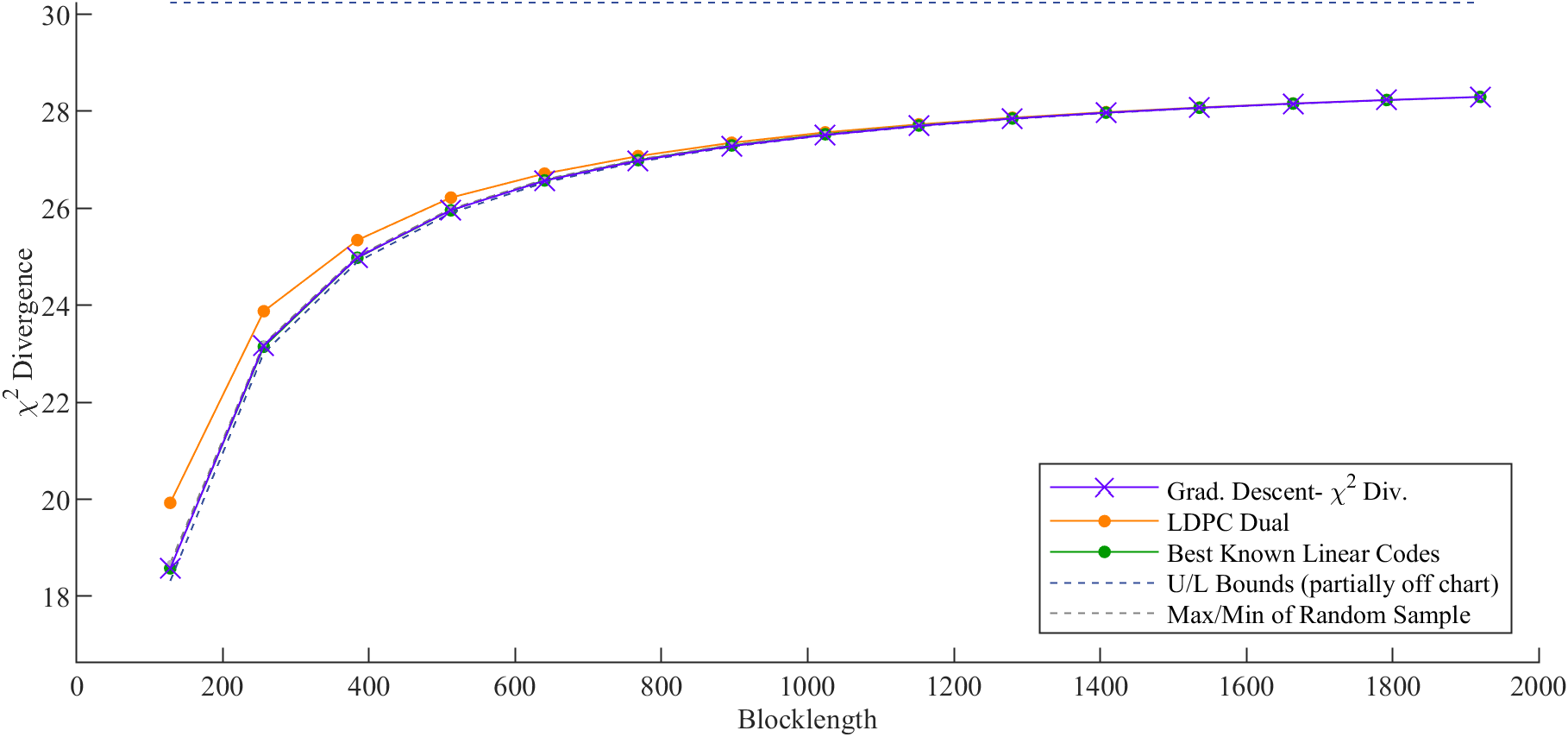}
    \caption{$\chi^2$ Divergence for Codes of Dimension $\kappa=11$.}
    \label{fig:x2Nonscaled11}
\end{figure}

\begin{figure}
    \centering
    \includegraphics[trim= 0 0.25cm 0 0.85cm, scale=1.0]{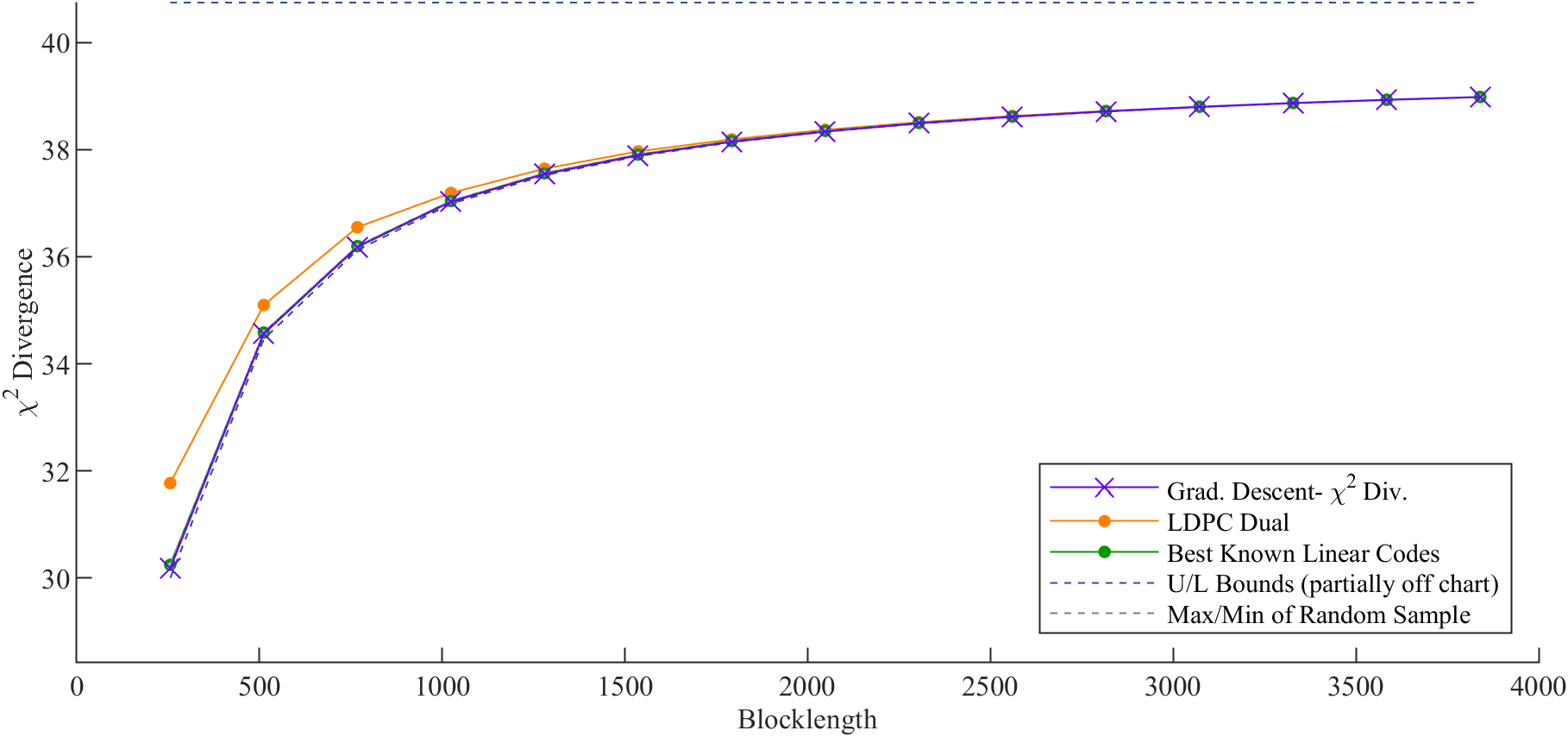}
    \caption{$\chi^2$ Divergence for Codes of Dimension $\kappa=12$.}
    \label{fig:x2Nonscaled12}
\end{figure}

\begin{figure}
    \centering
    \includegraphics[trim= 0 0.25cm 0 0.85cm, scale=1.0]{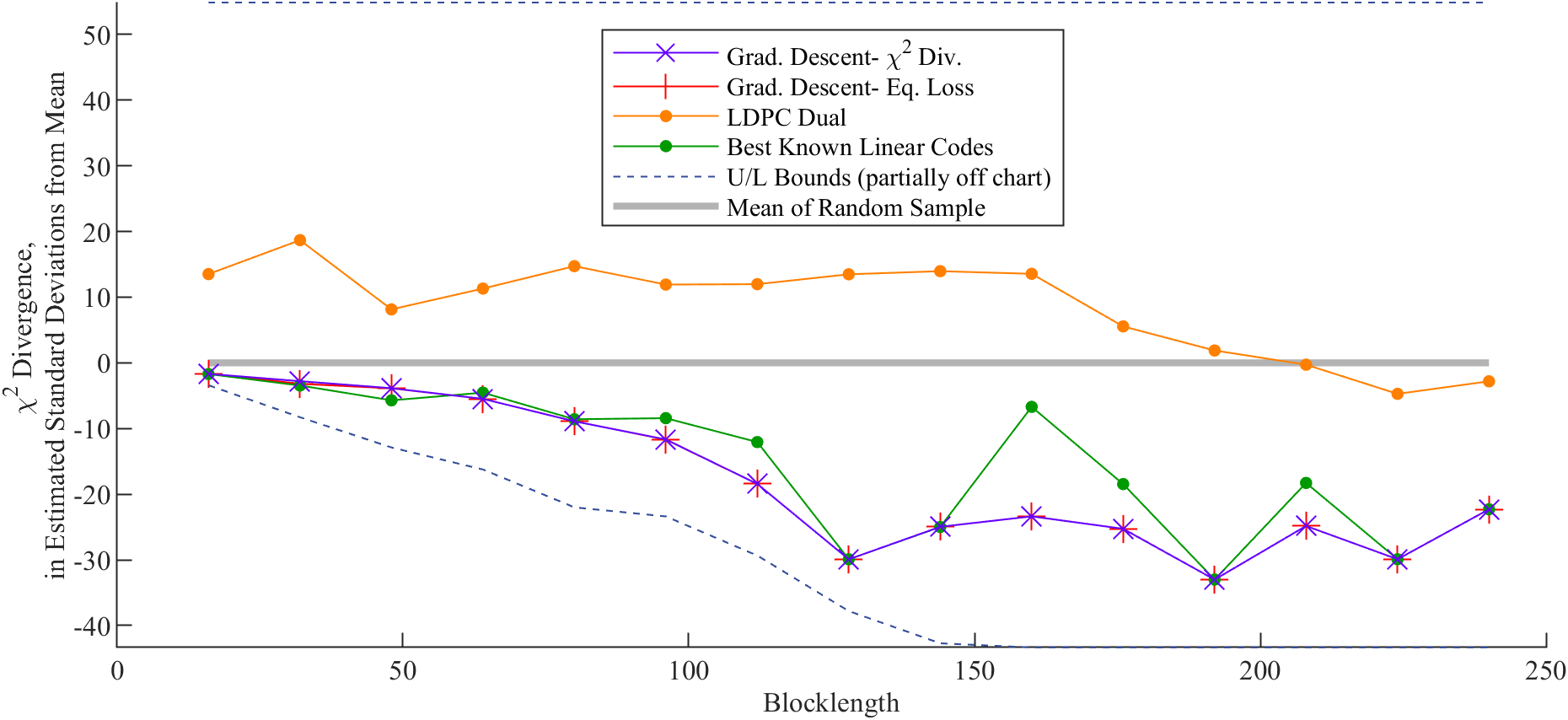}
    \caption{$\chi^2$ Divergence for Codes of Dimension $\kappa=8$, Relative to Random Sample Performance.}
    \label{fig:x2Scaled8}
\end{figure}

\begin{figure}
    \centering
    \includegraphics[trim= 0 0.25cm 0 0.85cm, scale=1.0]{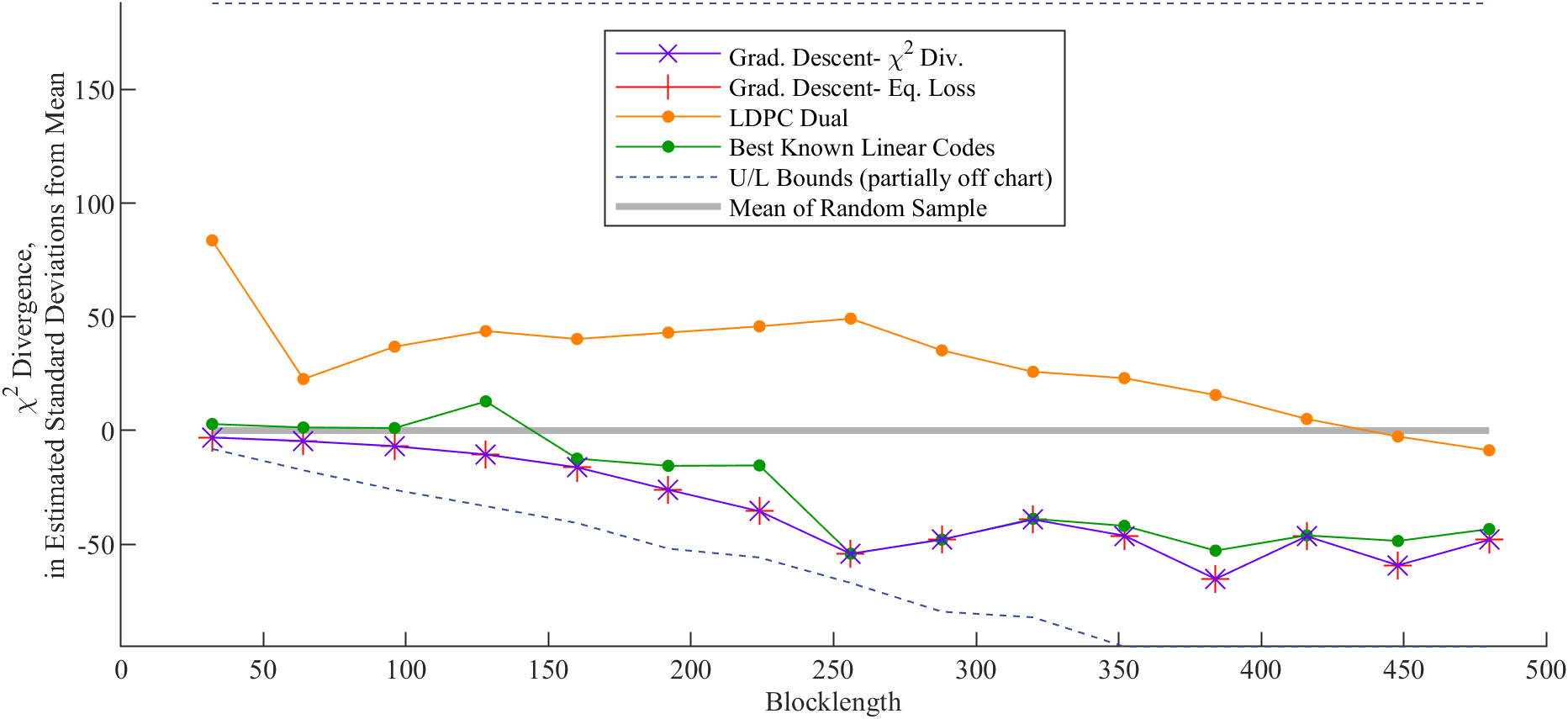}
    \caption{$\chi^2$ Divergence for Codes of Dimension $\kappa=9$, Relative to Random Sample Performance.}
    \label{fig:x2Scaled9}
\end{figure}

\begin{figure}
    \centering
    \includegraphics[trim= 0 0.25cm 0 0.85cm, scale=1.0]{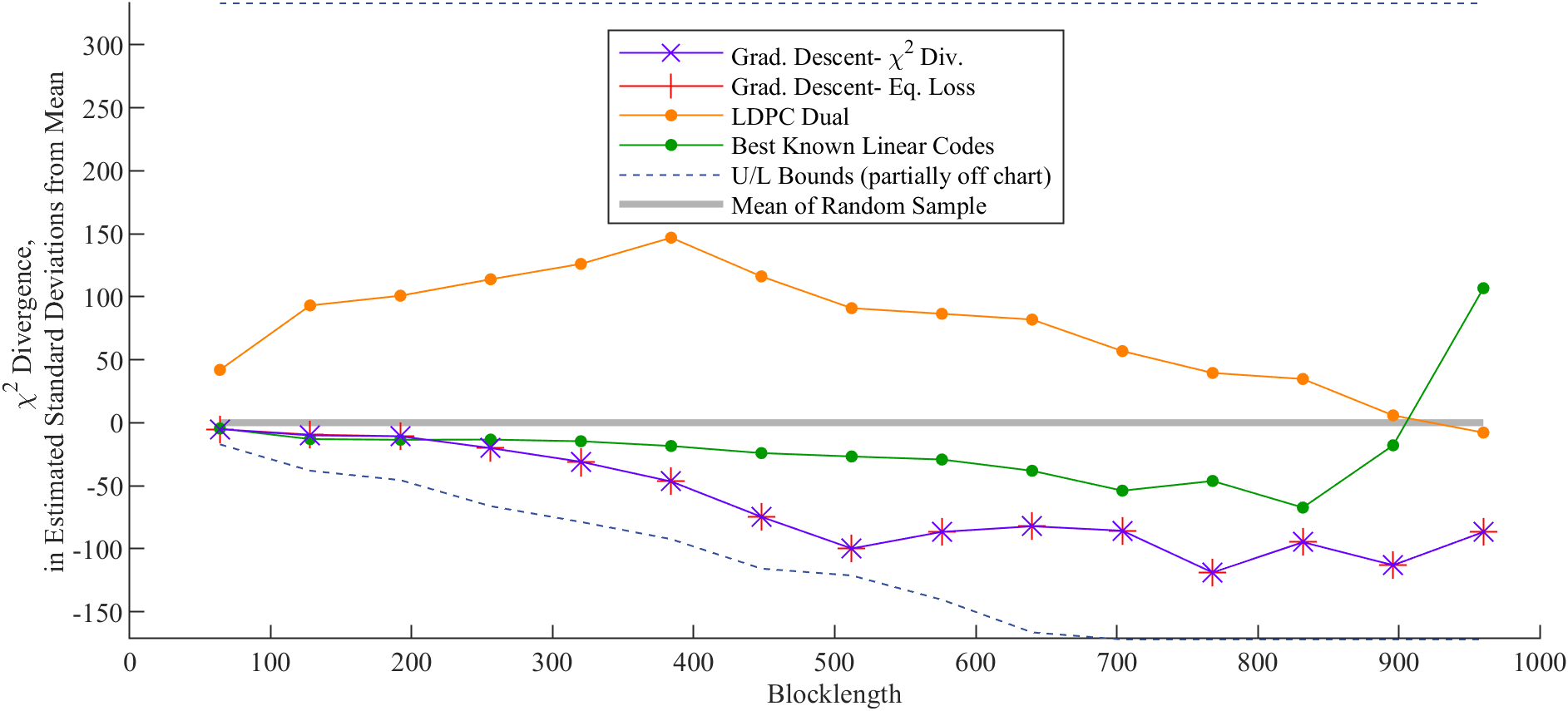}
    \caption{$\chi^2$ Divergence for Codes of Dimension $\kappa=10$, Relative to Random Sample Performance.}
    \label{fig:x2Scaled10}
\end{figure}

\begin{figure}
    \centering
    \includegraphics[trim= 0 0.25cm 0 0.25cm, scale=1.0]{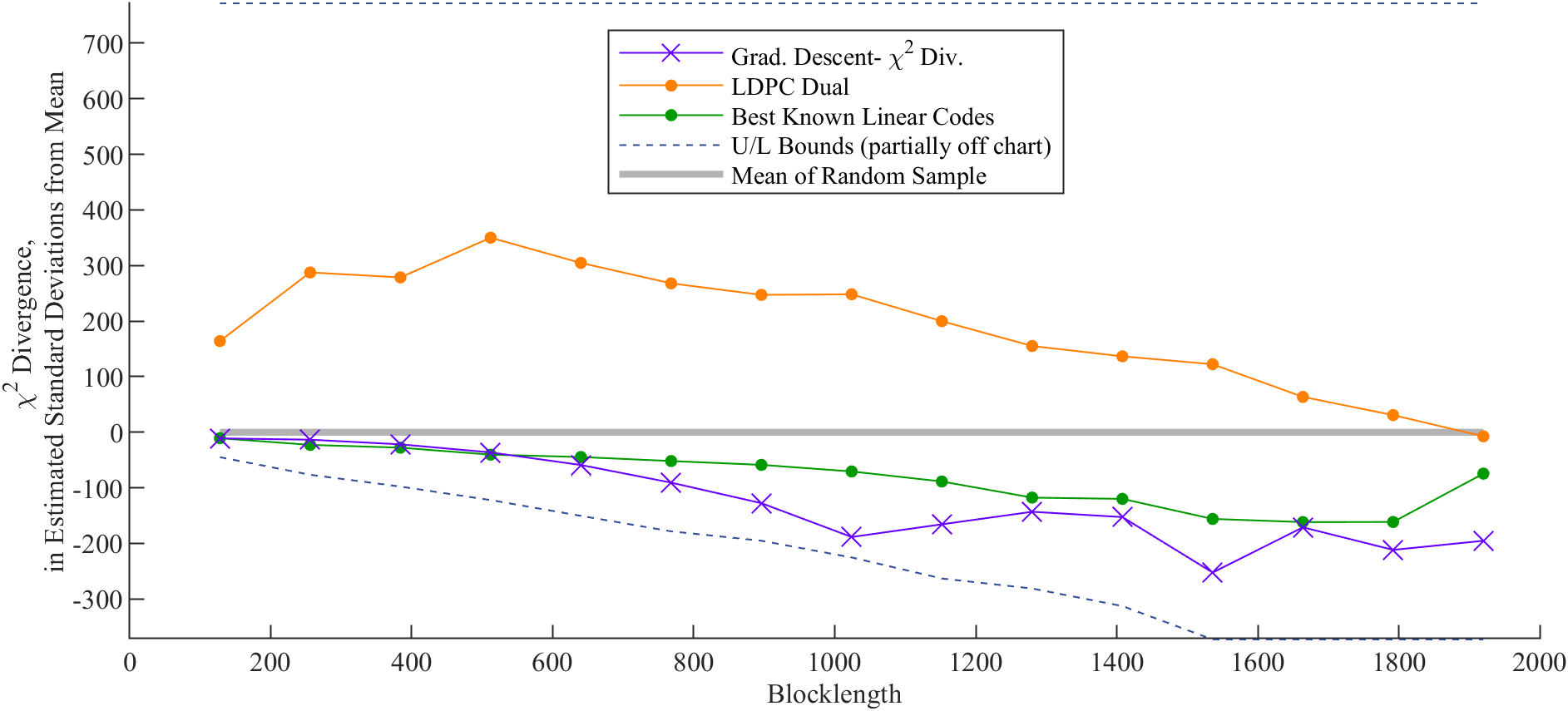}
    \caption{$\chi^2$ Divergence for Codes of Dimension $\kappa=11$, Relative to Random Sample Performance.}
    \label{fig:x2Scaled11}
\end{figure}

\begin{figure}
    \centering
    \includegraphics[trim= 0 0.25cm 0 0.25cm, scale=1.0]{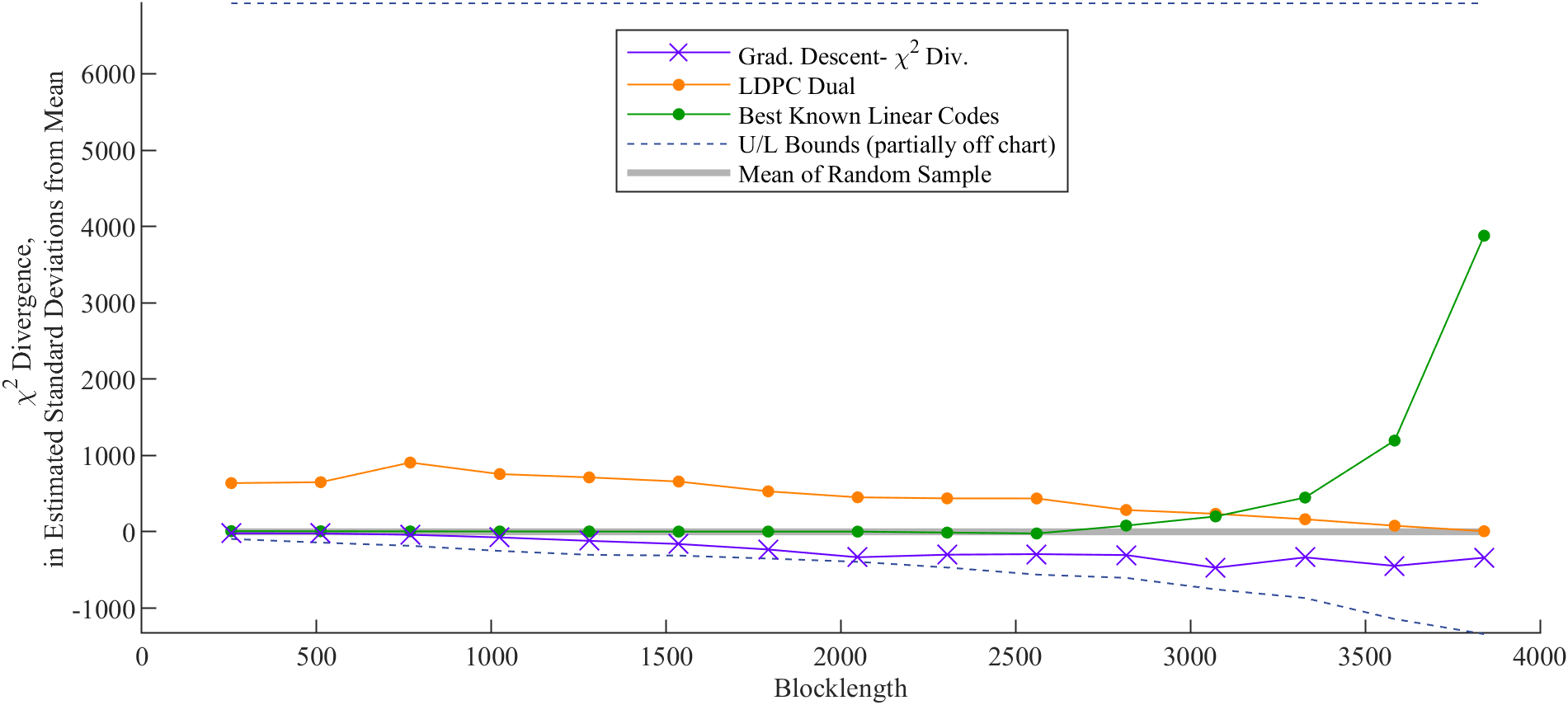}
    \caption{$\chi^2$ Divergence for Codes of Dimension $\kappa=12$, Relative to Random Sample Performance.}
    \label{fig:x2Scaled12}
\end{figure}

The results of gradient descent optimization in terms of $\chi^2$ divergence, as well as limits and comparison codes, are shown for dimensions $\kappa=8$ through $\kappa=12$ in Figs. \ref{fig:x2Nonscaled8} through \ref{fig:x2Nonscaled12}, respectively. In Figs. \ref{fig:x2Scaled8} through \ref{fig:x2Scaled12}, the same results are shown in terms of the estimated standard deviation of the random samples.  

\newpage

\subsection{Observations}
Upon initial inspection of the gradient descent paths shown in Fig. \ref{fig:pathEqLoss_8_64} and \ref{fig:pathX2_8_64}, one of the first revelations is the remarkable complexity of the paths. The path appears chaotic at several points, but after each chaotic period, a new order emerges until the path is eventually forced into a point that represents a realizable code. It is also interesting to note that after some of the elements of $q$ reach the maximum or minimum limit, they later re-emerge into the intermediate region and may cross over to the opposite limit. This suggests that an approach that sequentially selects or excludes columns of $G$ may be sub-optimal, as columns that are included at short code blocklength may need to be excluded at longer blocklength.   
    
From the equivocation loss results shown in Figs. \ref{fig:eqLossNonscaled8} through \ref{fig:x2Scaled12}, it is clear that gradient descent optimization is effective in generating good coset codes. Particularly at larger blocklength, gradient descent methods markedly outperformed all others. It is also remarkable to note that codes produced by gradient descent based on equivocation loss performed remarkably well in terms of $\chi^2$ divergence, and vice versa. This suggests that finding codes using $\chi^2$ divergence (which is practical at larger code dimensions) is a viable method to find codes that have low equivocation loss. 

Another observation from Figs. \ref{fig:eqLossNonscaled8} through \ref{fig:x2Scaled12} (and seen more clearly in Tables \ref{tbl:eqLossResults_8} through \ref{tbl:x2Results_12}) is that bounds based on total variation distance (those defined in \eqref{eqn:finalLimit_EqLoss} and \eqref{eqn:finalLimit_x2}) are very loose, especially for large $n$. In some ways this is unsurprising, as these bounds require a scaling factor which increases with increasing $k$ to be applied to equivocation loss or $\chi^2$ divergence. On the other hand, the $\chi^2$ divergence-based bound defined in \eqref{eqn:boundDirectX2} was much closer to the performance of actual coset codes. 

Another somewhat surprising observation is that the code constructions based on the duals of LDPC codes, though they are proven to achieve strong secrecy, not only perform worse than short-blocklength constructions, but also perform significantly worse than randomly generated coset codes. 

\section{Conclusion}
In this paper, we show that the continuous and differentiable nature of the functions used in subspace decomposition make gradient descent a realistic choice for designing good coset codes, even for blocklengths up to several thousand. To our knowledge, the codes generated using these techniques outperform any other published secrecy codes. 

The results obtained from this work also provide a number of other insights into the design of coset codes. Most strikingly, coset codes that are designed to asymptotically approach the secrecy capacity of a channel perform very poorly at smaller blocklengths. The convoluted path traced by the gradient descent optimization also sheds light on the great complexity of the problem of optimizing secrecy codes, both in terms of equivocation loss and ML decoding by an eavesdropper. 

This work also clearly demonstrates the usefulness of subspace decomposition in tackling practical problems in secrecy coding. The code sizes and performance levels reached here are beyond anything that could be achieved using previously known techniques. Going forward, we expect that these techniques may be applied to many other areas of coding theory to solve problems that are beyond the reach of conventional methods.

\appendices
\section{Equivocation Loss Results in Tabular Form}
\label{apx:EqLossResultsTabular}
The equivocation loss values for each of the codes generated by Algorithm \ref{alg:gradDesc}, as well as the code constructions and limits described in \ref{sec:comparisonMethods}, are given in Tables \ref{tbl:eqLossResults_8} through \ref{tbl:eqLossResults_10}. These values are listed for each of the code sizes examined, and they are given both in terms of raw equivocation loss and in terms of the estimated standard deviation of the sample of randomly-generated codes. 


\bgroup
\setlength\tabcolsep{0.15cm}
\begin{table}[htpb]
\caption{Equivocation loss of codes with dimension $\kappa=8$.}
\centering
\begin{tabular} {|c|c|c|c|c|c|c|c|c|}
    \hline
    & \multicolumn{2}{c|}{Random Sample:} & \multicolumn{4}{c|}{\thead{Equivocation Loss: Absolute and \\(Relative to $\hat{\mu}$ and $\hat{\sigma}$)}} & \multicolumn{2}{c|}{\thead{Bounds: Absolute and \\ (Relative to $\hat{\mu}$ and $\hat{\sigma}$)}}\\
    \hline
    $n$ & \thead{Mean \\ ($\hat{\mu}$)} & \thead{St. Dev. \\ ($\hat{\sigma})$} & \thead{Gradient Descent \\ (Equiv. Loss)} & \thead{Gradient Descent \\ ($\chi^2$ Divergence)} & \thead{Best Known \\ Linear Code} & \thead{LDPC Dual} & \thead{Upper Bound} & \thead{Lower Bound} \\
    \hline
    16 & 1.1283 & 0.08988 & 0.9528\;\,(-1.953\,$\hat{\sigma}$) & 0.9487\;\,(-1.999\,$\hat{\sigma}$) & 0.9452\;\,(-2.037\,$\hat{\sigma}$) & 1.840\;\,(+7.9220\,$\hat{\sigma}$) & 4.016\;\,(+32.125\,$\hat{\sigma}$) & 0.5472\;\,(-6.465\,$\hat{\sigma}$)\\
    \hline
    32 & 1.2602 & 0.019288 & 1.181\;\,(-4.0878\,$\hat{\sigma}$) & 1.194\;\,(-3.4479\,$\hat{\sigma}$) & 1.174\;\,(-4.4564\,$\hat{\sigma}$) & 1.530\;\,(+13.982\,$\hat{\sigma}$) & 12.00\;\,(+556.80\,$\hat{\sigma}$) & 0.5896\;\,(-34.77\,$\hat{\sigma}$)\\
    \hline
    48 & 1.3017 & 0.0087199 & 1.261\;\,(-4.7105\,$\hat{\sigma}$) & 1.261\;\,(-4.6948\,$\hat{\sigma}$) & 1.241\;\,(-6.9415\,$\hat{\sigma}$) & 1.365\;\,(+7.2336\,$\hat{\sigma}$) & 20.0000\;\,(+2144\,$\hat{\sigma}$) & 0.5993\;\,(-80.56\,$\hat{\sigma}$)\\
    \hline
    64 & 1.321 & 0.0044871 & 1.286\;\,(-7.7359\,$\hat{\sigma}$) & 1.286\;\,(-7.7361\,$\hat{\sigma}$) & 1.298\;\,(-5.1253\,$\hat{\sigma}$) & 1.369\;\,(+10.775\,$\hat{\sigma}$) & 28.0000\;\,(+5946\,$\hat{\sigma}$) & 0.6035\;\,(-159.9\,$\hat{\sigma}$)\\
    \hline
    80 & 1.3327 & 0.0027215 & 1.300\;\,(-11.916\,$\hat{\sigma}$) & 1.300\;\,(-11.922\,$\hat{\sigma}$) & 1.304\;\,(-10.534\,$\hat{\sigma}$) & 1.368\;\,(+13.083\,$\hat{\sigma}$) & 36.000\;\,(+12738\,$\hat{\sigma}$) & 0.6059\;\,(-267.1\,$\hat{\sigma}$)\\
    \hline
    96 & 1.3409 & 0.0018573 & 1.309\;\,(-17.181\,$\hat{\sigma}$) & 1.309\;\,(-17.163\,$\hat{\sigma}$) & 1.321\;\,(-10.704\,$\hat{\sigma}$) & 1.362\;\,(+11.445\,$\hat{\sigma}$) & 44.000\;\,(+22969\,$\hat{\sigma}$) & 0.6075\;\,(-394.9\,$\hat{\sigma}$)\\
    \hline
    112 & 1.3465 & 0.0012131 & 1.315\;\,(-25.967\,$\hat{\sigma}$) & 1.315\;\,(-25.967\,$\hat{\sigma}$) & 1.328\;\,(-15.507\,$\hat{\sigma}$) & 1.360\;\,(+11.189\,$\hat{\sigma}$) & 52.000\;\,(+41754\,$\hat{\sigma}$) & 0.6086\;\,(-608.3\,$\hat{\sigma}$)\\
    \hline
    128 & 1.3506 & 0.00085981 & 1.319\;\,(-36.436\,$\hat{\sigma}$) & 1.319\;\,(-36.436\,$\hat{\sigma}$) & 1.319\;\,(-36.436\,$\hat{\sigma}$) & 1.360\;\,(+10.930\,$\hat{\sigma}$) & 60.000\;\,(+68212\,$\hat{\sigma}$) & 0.6094\;\,(-862.1\,$\hat{\sigma}$)\\
    \hline
    144 & 1.3539 & 0.00058821 & 1.336\;\,(-29.808\,$\hat{\sigma}$) & 1.336\;\,(-29.808\,$\hat{\sigma}$) & 1.336\;\,(-29.831\,$\hat{\sigma}$) & 1.361\;\,(+11.568\,$\hat{\sigma}$) & 68.00\;\,(+113303\,$\hat{\sigma}$) & 0.60997\;\,(-1265\,$\hat{\sigma}$)\\
    \hline
    160 & 1.3564 & 0.00039378 & 1.346\;\,(-25.862\,$\hat{\sigma}$) & 1.346\;\,(-25.865\,$\hat{\sigma}$) & 1.355\;\,(-2.5046\,$\hat{\sigma}$) & 1.361\;\,(+10.951\,$\hat{\sigma}$) & 76.00\;\,(+189555\,$\hat{\sigma}$) & 0.61045\;\,(-1894\,$\hat{\sigma}$)\\
    \hline
    176 & 1.3586 & 0.00025298 & 1.352\;\,(-27.219\,$\hat{\sigma}$) & 1.352\;\,(-27.219\,$\hat{\sigma}$) & 1.354\;\,(-19.018\,$\hat{\sigma}$) & 1.360\;\,(+4.6297\,$\hat{\sigma}$) & 84.00\;\,(+326678\,$\hat{\sigma}$) & 0.61084\;\,(-2956\,$\hat{\sigma}$)\\
    \hline
    192 & 1.3603 & 0.00016056 & 1.354\;\,(-37.664\,$\hat{\sigma}$) & 1.354\;\,(-37.664\,$\hat{\sigma}$) & 1.354\;\,(-37.664\,$\hat{\sigma}$) & 1.361\;\,(+1.6883\,$\hat{\sigma}$) & 92.00\;\,(+564521\,$\hat{\sigma}$) & 0.61116\;\,(-4666\,$\hat{\sigma}$)\\
    \hline
    208 & 1.3618 & 8.9468e-05 & 1.359\;\,(-25.907\,$\hat{\sigma}$) & 1.359\;\,(-25.907\,$\hat{\sigma}$) & 1.360\;\,(-18.253\,$\hat{\sigma}$) & 1.362\;\,(-0.5935\,$\hat{\sigma}$) & 100.0\;\,(+1.102e6\,$\hat{\sigma}$) & 0.61143\;\,(-8387\,$\hat{\sigma}$)\\
    \hline
    224 & 1.363 & 4.1696e-05 & 1.362\;\,(-31.710\,$\hat{\sigma}$) & 1.362\;\,(-31.710\,$\hat{\sigma}$) & 1.362\;\,(-31.710\,$\hat{\sigma}$) & 1.363\;\,(-4.7438\,$\hat{\sigma}$) & 108.0\;\,(+2.557e6\,$\hat{\sigma}$) & 0.6117\;\,(-18020\,$\hat{\sigma}$)\\
    \hline
    240 & 1.3641 & 1.1694e-05 & 1.364\;\,(-24.648\,$\hat{\sigma}$) & 1.364\;\,(-24.648\,$\hat{\sigma}$) & 1.364\;\,(-24.648\,$\hat{\sigma}$) & 1.364\;\,(-3.2339\,$\hat{\sigma}$) & 116.0\;\,(+9.803e6\,$\hat{\sigma}$) & 0.6119\;\,(-64329\,$\hat{\sigma}$)\\
    \hline
    
\end{tabular}
\label{tbl:eqLossResults_8}
\end{table}
\egroup

\bgroup
\setlength\tabcolsep{0.15cm}
\begin{table}
\caption{Equivocation loss of codes with dimension $\kappa=9$.}
\centering
\begin{tabular} {|c|c|c|c|c|c|c|c|c|}
    \hline
    & \multicolumn{2}{c|}{Random Sample:} & \multicolumn{4}{c|}{\thead{Equivocation Loss: Absolute and \\(Relative to $\hat{\mu}$ and $\hat{\sigma}$)}} & \multicolumn{2}{c|}{\thead{Bounds: Absolute and \\ (Relative to $\hat{\mu}$ and $\hat{\sigma}$)}}\\
    \hline
    $n$ & \thead{Mean \\ ($\hat{\mu}$)} & \thead{St. Dev. \\ ($\hat{\sigma})$} & \thead{Gradient Descent \\ (Equiv. Loss)} & \thead{Gradient Descent \\ ($\chi^2$ Divergence)} & \thead{Best Known \\ Linear Code} & \thead{LDPC Dual} & \thead{Upper Bound} & \thead{Lower Bound} \\
    \hline
    32 & 1.3116 & 0.02181 & 1.220\;\,(-4.2045\,$\hat{\sigma}$) & 1.220\;\,(-4.2191\,$\hat{\sigma}$) & 1.492\;\,(+8.2677\,$\hat{\sigma}$) & 2.102\;\,(+36.238\,$\hat{\sigma}$) & 11.50\;\,(+467.14\,$\hat{\sigma}$) & 0.6021\;\,(-32.53\,$\hat{\sigma}$)\\
    \hline
    64 & 1.3872 & 0.0063792 & 1.352\;\,(-5.5471\,$\hat{\sigma}$) & 1.351\;\,(-5.6190\,$\hat{\sigma}$) & 1.425\;\,(+5.9731\,$\hat{\sigma}$) & 1.496\;\,(+17.115\,$\hat{\sigma}$) & 27.5000\;\,(+4093\,$\hat{\sigma}$) & 0.6183\;\,(-120.5\,$\hat{\sigma}$)\\
    \hline
    96 & 1.4113 & 0.0028854 & 1.386\;\,(-8.7788\,$\hat{\sigma}$) & 1.386\;\,(-8.7087\,$\hat{\sigma}$) & 1.432\;\,(+7.1088\,$\hat{\sigma}$) & 1.488\;\,(+26.648\,$\hat{\sigma}$) & 43.500\;\,(+14587\,$\hat{\sigma}$) & 0.6228\;\,(-273.3\,$\hat{\sigma}$)\\
    \hline
    128 & 1.4227 & 0.0016071 & 1.400\;\,(-13.950\,$\hat{\sigma}$) & 1.400\;\,(-13.980\,$\hat{\sigma}$) & 1.471\;\,(+30.131\,$\hat{\sigma}$) & 1.476\;\,(+32.883\,$\hat{\sigma}$) & 59.500\;\,(+36138\,$\hat{\sigma}$) & 0.6249\;\,(-496.4\,$\hat{\sigma}$)\\
    \hline
    160 & 1.4295 & 0.0010028 & 1.408\;\,(-21.351\,$\hat{\sigma}$) & 1.408\;\,(-21.372\,$\hat{\sigma}$) & 1.416\;\,(-13.825\,$\hat{\sigma}$) & 1.461\;\,(+31.418\,$\hat{\sigma}$) & 75.500\;\,(+73861\,$\hat{\sigma}$) & 0.6261\;\,(-801.1\,$\hat{\sigma}$)\\
    \hline
    192 & 1.4342 & 0.00059657 & 1.413\;\,(-35.302\,$\hat{\sigma}$) & 1.413\;\,(-35.300\,$\hat{\sigma}$) & 1.424\;\,(-17.424\,$\hat{\sigma}$) & 1.455\;\,(+35.135\,$\hat{\sigma}$) & 91.50\;\,(+150974\,$\hat{\sigma}$) & 0.62695\;\,(-1353\,$\hat{\sigma}$)\\
    \hline
    224 & 1.4375 & 0.00041835 & 1.417\;\,(-50.051\,$\hat{\sigma}$) & 1.417\;\,(-50.054\,$\hat{\sigma}$) & 1.430\;\,(-17.801\,$\hat{\sigma}$) & 1.454\;\,(+39.664\,$\hat{\sigma}$) & 107.5\;\,(+253525\,$\hat{\sigma}$) & 0.62751\;\,(-1936\,$\hat{\sigma}$)\\
    \hline
    256 & 1.44 & 0.00030176 & 1.419\;\,(-69.285\,$\hat{\sigma}$) & 1.419\;\,(-69.285\,$\hat{\sigma}$) & 1.419\;\,(-69.285\,$\hat{\sigma}$) & 1.452\;\,(+39.434\,$\hat{\sigma}$) & 123.5\;\,(+404500\,$\hat{\sigma}$) & 0.62793\;\,(-2691\,$\hat{\sigma}$)\\
    \hline
    288 & 1.442 & 0.00018387 & 1.430\;\,(-64.413\,$\hat{\sigma}$) & 1.430\;\,(-64.414\,$\hat{\sigma}$) & 1.430\;\,(-64.496\,$\hat{\sigma}$) & 1.448\;\,(+31.577\,$\hat{\sigma}$) & 139.5\;\,(+750865\,$\hat{\sigma}$) & 0.62825\;\,(-4426\,$\hat{\sigma}$)\\
    \hline
    320 & 1.4435 & 0.00014142 & 1.437\;\,(-49.464\,$\hat{\sigma}$) & 1.437\;\,(-49.470\,$\hat{\sigma}$) & 1.437\;\,(-48.931\,$\hat{\sigma}$) & 1.447\;\,(+23.300\,$\hat{\sigma}$) & 155.5\;\,(+1.089e6\,$\hat{\sigma}$) & 0.62850\;\,(-5763\,$\hat{\sigma}$)\\
    \hline
    352 & 1.4448 & 8.9699e-05 & 1.440\;\,(-53.655\,$\hat{\sigma}$) & 1.440\;\,(-53.652\,$\hat{\sigma}$) & 1.440\;\,(-48.364\,$\hat{\sigma}$) & 1.447\;\,(+20.342\,$\hat{\sigma}$) & 171.5\;\,(+1.896e6\,$\hat{\sigma}$) & 0.62871\;\,(-9098\,$\hat{\sigma}$)\\
    \hline
    384 & 1.4458 & 5.8082e-05 & 1.442\;\,(-73.449\,$\hat{\sigma}$) & 1.442\;\,(-73.449\,$\hat{\sigma}$) & 1.442\;\,(-58.916\,$\hat{\sigma}$) & 1.447\;\,(+13.522\,$\hat{\sigma}$) & 187.5\;\,(+3.203e6\,$\hat{\sigma}$) & 0.6289\;\,(-14065\,$\hat{\sigma}$)\\
    \hline
    416 & 1.4467 & 3.0926e-05 & 1.445\;\,(-54.259\,$\hat{\sigma}$) & 1.445\;\,(-54.255\,$\hat{\sigma}$) & 1.445\;\,(-53.798\,$\hat{\sigma}$) & 1.447\;\,(+4.6114\,$\hat{\sigma}$) & 203.5\;\,(+6.533e6\,$\hat{\sigma}$) & 0.6290\;\,(-26439\,$\hat{\sigma}$)\\
    \hline
    448 & 1.4474 & 1.5296e-05 & 1.446\;\,(-63.873\,$\hat{\sigma}$) & 1.446\;\,(-63.873\,$\hat{\sigma}$) & 1.447\;\,(-52.047\,$\hat{\sigma}$) & 1.447\;\,(-2.8114\,$\hat{\sigma}$) & 219.5\;\,(+1.426e7\,$\hat{\sigma}$) & 0.6292\;\,(-53496\,$\hat{\sigma}$)\\
    \hline
    480 & 1.4481 & 4.8365e-06 & 1.448\;\,(-46.817\,$\hat{\sigma}$) & 1.448\;\,(-46.817\,$\hat{\sigma}$) & 1.448\;\,(-42.153\,$\hat{\sigma}$) & 1.448\;\,(-8.2139\,$\hat{\sigma}$) & 235.5\;\,(+4.839e7\,$\hat{\sigma}$) & 0.629\;\,(-169305\,$\hat{\sigma}$)\\
    \hline
    
\end{tabular}
\label{tbl:eqLossResults_9}
\end{table}
\egroup

\bgroup
\setlength\tabcolsep{0.15cm}
\begin{table}
\caption{Equivocation loss of codes with dimension $\kappa=10$.}
\centering
\begin{tabular} {|c|c|c|c|c|c|c|c|c|}
    \hline
    & \multicolumn{2}{c|}{Random Sample:} & \multicolumn{4}{c|}{\thead{Equivocation Loss: Absolute and \\(Relative to $\hat{\mu}$ and $\hat{\sigma}$)}} & \multicolumn{2}{c|}{\thead{Bounds: Absolute and \\ (Relative to $\hat{\mu}$ and $\hat{\sigma}$)}}\\
    \hline
    $n$ & \thead{Mean \\ ($\hat{\mu}$)} & \thead{St. Dev. \\ ($\hat{\sigma})$} & \thead{Gradient Descent \\ (Equiv. Loss)} & \thead{Gradient Descent \\ ($\chi^2$ Divergence)} & \thead{Best Known \\ Linear Code} & \thead{LDPC Dual} & \thead{Upper Bound} & \thead{Lower Bound} \\
    \hline
    64 & 1.4416 & 0.0065313 & 1.392\;\,(-7.6260\,$\hat{\sigma}$) & 1.392\;\,(-7.5275\,$\hat{\sigma}$) & 1.401\;\,(-6.2187\,$\hat{\sigma}$) & 1.670\;\,(+34.905\,$\hat{\sigma}$) & 27.0000\;\,(+3913\,$\hat{\sigma}$) & 0.6311\;\,(-124.1\,$\hat{\sigma}$)\\
    \hline
    128 & 1.4839 & 0.0019717 & 1.461\;\,(-11.494\,$\hat{\sigma}$) & 1.460\;\,(-12.021\,$\hat{\sigma}$) & 1.454\;\,(-15.384\,$\hat{\sigma}$) & 1.597\;\,(+57.325\,$\hat{\sigma}$) & 59.000\;\,(+29171\,$\hat{\sigma}$) & 0.6385\;\,(-428.7\,$\hat{\sigma}$)\\
    \hline
    192 & 1.4977 & 0.00094485 & 1.483\;\,(-15.790\,$\hat{\sigma}$) & 1.483\;\,(-15.806\,$\hat{\sigma}$) & 1.481\;\,(-17.755\,$\hat{\sigma}$) & 1.568\;\,(+74.522\,$\hat{\sigma}$) & 91.000\;\,(+94727\,$\hat{\sigma}$) & 0.6408\;\,(-907.0\,$\hat{\sigma}$)\\
    \hline
    256 & 1.5045 & 0.00051625 & 1.490\;\,(-27.208\,$\hat{\sigma}$) & 1.490\;\,(-27.224\,$\hat{\sigma}$) & 1.498\;\,(-12.955\,$\hat{\sigma}$) & 1.546\;\,(+80.650\,$\hat{\sigma}$) & 123.0\;\,(+235341\,$\hat{\sigma}$) & 0.64184\;\,(-1671\,$\hat{\sigma}$)\\
    \hline
    320 & 1.5087 & 0.00038317 & 1.495\;\,(-35.837\,$\hat{\sigma}$) & 1.495\;\,(-35.844\,$\hat{\sigma}$) & 1.504\;\,(-12.586\,$\hat{\sigma}$) & 1.539\;\,(+79.051\,$\hat{\sigma}$) & 155.0\;\,(+400588\,$\hat{\sigma}$) & 0.64248\;\,(-2261\,$\hat{\sigma}$)\\
    \hline
    384 & 1.5114 & 0.00022513 & 1.498\;\,(-60.293\,$\hat{\sigma}$) & 1.498\;\,(-60.296\,$\hat{\sigma}$) & 1.507\;\,(-18.437\,$\hat{\sigma}$) & 1.535\;\,(+105.99\,$\hat{\sigma}$) & 187.0\;\,(+823925\,$\hat{\sigma}$) & 0.64290\;\,(-3858\,$\hat{\sigma}$)\\
    \hline
    448 & 1.5133 & 0.00016278 & 1.500\;\,(-82.983\,$\hat{\sigma}$) & 1.500\;\,(-82.998\,$\hat{\sigma}$) & 1.510\;\,(-21.203\,$\hat{\sigma}$) & 1.526\;\,(+76.478\,$\hat{\sigma}$) & 219.0\;\,(+1.336e6\,$\hat{\sigma}$) & 0.64319\;\,(-5345\,$\hat{\sigma}$)\\
    \hline
    512 & 1.5148 & 9.1177e-05 & 1.501\;\,(-148.26\,$\hat{\sigma}$) & 1.501\;\,(-148.26\,$\hat{\sigma}$) & 1.512\;\,(-33.182\,$\hat{\sigma}$) & 1.522\;\,(+83.123\,$\hat{\sigma}$) & 251.0\;\,(+2.736e6\,$\hat{\sigma}$) & 0.64341\;\,(-9557\,$\hat{\sigma}$)\\
    \hline
    576 & 1.5159 & 7.2897e-05 & 1.508\;\,(-105.76\,$\hat{\sigma}$) & 1.508\;\,(-105.76\,$\hat{\sigma}$) & 1.514\;\,(-29.763\,$\hat{\sigma}$) & 1.521\;\,(+68.798\,$\hat{\sigma}$) & 283.0\;\,(+3.861e6\,$\hat{\sigma}$) & 0.6436\;\,(-11966\,$\hat{\sigma}$)\\
    \hline
    640 & 1.5168 & 4.7978e-05 & 1.512\;\,(-96.539\,$\hat{\sigma}$) & 1.512\;\,(-96.547\,$\hat{\sigma}$) & 1.515\;\,(-40.200\,$\hat{\sigma}$) & 1.520\;\,(+65.942\,$\hat{\sigma}$) & 315.0\;\,(+6.534e6\,$\hat{\sigma}$) & 0.6437\;\,(-18197\,$\hat{\sigma}$)\\
    \hline
    704 & 1.5175 & 3.2874e-05 & 1.514\;\,(-98.395\,$\hat{\sigma}$) & 1.514\;\,(-98.399\,$\hat{\sigma}$) & 1.516\;\,(-59.266\,$\hat{\sigma}$) & 1.519\;\,(+48.212\,$\hat{\sigma}$) & 347.0\;\,(+1.051e7\,$\hat{\sigma}$) & 0.6438\;\,(-26577\,$\hat{\sigma}$)\\
    \hline
    768 & 1.5182 & 2.0208e-05 & 1.515\;\,(-142.21\,$\hat{\sigma}$) & 1.515\;\,(-142.21\,$\hat{\sigma}$) & 1.517\;\,(-49.139\,$\hat{\sigma}$) & 1.519\;\,(+36.107\,$\hat{\sigma}$) & 379.0\;\,(+1.868e7\,$\hat{\sigma}$) & 0.6439\;\,(-43261\,$\hat{\sigma}$)\\
    \hline
    832 & 1.5187 & 1.1328e-05 & 1.518\;\,(-102.03\,$\hat{\sigma}$) & 1.518\;\,(-102.03\,$\hat{\sigma}$) & 1.518\;\,(-64.636\,$\hat{\sigma}$) & 1.519\;\,(+30.072\,$\hat{\sigma}$) & 411.0\;\,(+3.615e7\,$\hat{\sigma}$) & 0.6440\;\,(-77212\,$\hat{\sigma}$)\\
    \hline
    896 & 1.5191 & 5.6451e-06 & 1.518\;\,(-121.02\,$\hat{\sigma}$) & 1.518\;\,(-121.02\,$\hat{\sigma}$) & 1.519\;\,(-0.9083\,$\hat{\sigma}$) & 1.519\;\,(+4.9110\,$\hat{\sigma}$) & 443.0\;\,(+7.821e7\,$\hat{\sigma}$) & 0.644\;\,(-155009\,$\hat{\sigma}$)\\
    \hline
    960 & 1.5195 & 1.6503e-06 & 1.519\;\,(-99.518\,$\hat{\sigma}$) & 1.519\;\,(-99.518\,$\hat{\sigma}$) & 1.520\;\,(+185.19\,$\hat{\sigma}$) & 1.519\;\,(-8.9889\,$\hat{\sigma}$) & 475.0\;\,(+2.869e8\,$\hat{\sigma}$) & 0.644\;\,(-530449\,$\hat{\sigma}$)\\
    \hline

\end{tabular}
\label{tbl:eqLossResults_10}
\end{table}
\egroup

\newpage

\section{$\chi^2$ Divergence Results in Tabular Form}
\label{apx:X2ResultsTabular}
The $\chi^2$ divergence values for each of the codes generated by Algorithm \ref{alg:gradDesc}, as well as the code constructions and limits described in \ref{sec:comparisonMethods}, are given in Tables \ref{tbl:x2Results_8} through \ref{tbl:x2Results_12}. These values are listed for each of the code sizes examined, and they are given both in terms of raw $\chi^2$ divergence and in terms of the estimated standard deviation of the sample of randomly-generated codes. \vspace{10cm}

\bgroup
\setlength\tabcolsep{0.15cm}
\begin{table}
\caption{$\chi^2$ divergence of codes with dimension $\kappa=8$.}
\centering
\begin{tabular} {|c|c|c|c|c|c|c|c|c|}
    \hline
    & \multicolumn{2}{c|}{Random Sample:} & \multicolumn{4}{c|}{\thead{$\chi^2$ Divergence: Absolute and \\(Relative to $\hat{\mu}$ and $\hat{\sigma}$)}} & \multicolumn{2}{c|}{\thead{Bounds: Absolute and \\ (Relative to $\hat{\mu}$ and $\hat{\sigma}$)}}\\
    \hline
    $n$ & \thead{Mean \\ ($\hat{\mu}$)} & \thead{St. Dev. \\ ($\hat{\sigma})$} & \thead{Gradient Descent \\ (Equiv. Loss)} & \thead{Gradient Descent \\ ($\chi^2$ Divergence)} & \thead{Best Known \\ Linear Code} & \thead{LDPC Dual} & \thead{Upper Bound} & \thead{Lower Bound} \\
    \hline
    16 & 2.4298 & 0.2254 & 2.059\;\,(-1.6468\,$\hat{\sigma}$) & 2.052\;\,(-1.6760\,$\hat{\sigma}$) & 2.045\;\,(-1.7053\,$\hat{\sigma}$) & 5.476\;\,(+13.517\,$\hat{\sigma}$) & 1.662\;\,(-3.4058\,$\hat{\sigma}$) & 128.0\;\,(+557.11\,$\hat{\sigma}$)\\
    \hline
    32 & 4.781 & 0.061235 & 4.586\;\,(-3.1911\,$\hat{\sigma}$) & 4.610\;\,(-2.7917\,$\hat{\sigma}$) & 4.570\;\,(-3.4491\,$\hat{\sigma}$) & 5.925\;\,(+18.684\,$\hat{\sigma}$) & 4.274\;\,(-8.2816\,$\hat{\sigma}$) & 8.39e6\;\,(+1.370e8\,$\hat{\sigma}$)\\
    \hline
    48 & 6.2394 & 0.028168 & 6.130\;\,(-3.8772\,$\hat{\sigma}$) & 6.131\;\,(-3.8603\,$\hat{\sigma}$) & 6.079\;\,(-5.7078\,$\hat{\sigma}$) & 6.469\;\,(+8.1428\,$\hat{\sigma}$) & 5.876\;\,(-12.886\,$\hat{\sigma}$) & 5.5e11\;\,(+1.95e13\,$\hat{\sigma}$)\\
    \hline
    64 & 7.19 & 0.01662 & 7.099\;\,(-5.4819\,$\hat{\sigma}$) & 7.099\;\,(-5.4825\,$\hat{\sigma}$) & 7.115\;\,(-4.5391\,$\hat{\sigma}$) & 7.378\;\,(+11.311\,$\hat{\sigma}$) & 6.920\;\,(-16.239\,$\hat{\sigma}$) & 3.6e16\;\,(+2.17e18\,$\hat{\sigma}$)\\
    \hline
    80 & 7.8538 & 0.0093847 & 7.770\;\,(-8.8976\,$\hat{\sigma}$) & 7.770\;\,(-8.9052\,$\hat{\sigma}$) & 7.773\;\,(-8.5869\,$\hat{\sigma}$) & 7.992\;\,(+14.722\,$\hat{\sigma}$) & 7.647\;\,(-22.003\,$\hat{\sigma}$) & 2.4e21\;\,(+2.52e23\,$\hat{\sigma}$)\\
    \hline
    96 & 8.3421 & 0.0068831 & 8.262\;\,(-11.684\,$\hat{\sigma}$) & 8.262\;\,(-11.663\,$\hat{\sigma}$) & 8.284\;\,(-8.4256\,$\hat{\sigma}$) & 8.424\;\,(+11.921\,$\hat{\sigma}$) & 8.181\;\,(-23.384\,$\hat{\sigma}$) & 1.5e26\;\,(+2.25e28\,$\hat{\sigma}$)\\
    \hline
    112 & 8.7165 & 0.0043418 & 8.636\;\,(-18.433\,$\hat{\sigma}$) & 8.636\;\,(-18.433\,$\hat{\sigma}$) & 8.664\;\,(-12.061\,$\hat{\sigma}$) & 8.769\;\,(+11.982\,$\hat{\sigma}$) & 8.589\;\,(-29.346\,$\hat{\sigma}$) & 1.0e31\;\,(+2.34e33\,$\hat{\sigma}$)\\
    \hline
    128 & 9.0112 & 0.0026575 & 8.932\;\,(-29.917\,$\hat{\sigma}$) & 8.932\;\,(-29.917\,$\hat{\sigma}$) & 8.932\;\,(-29.917\,$\hat{\sigma}$) & 9.047\;\,(+13.485\,$\hat{\sigma}$) & 8.911\;\,(-37.797\,$\hat{\sigma}$) & 6.6e35\;\,(+2.50e38\,$\hat{\sigma}$)\\
    \hline
    144 & 9.2496 & 0.0018466 & 9.203\;\,(-24.963\,$\hat{\sigma}$) & 9.203\;\,(-24.963\,$\hat{\sigma}$) & 9.203\;\,(-24.996\,$\hat{\sigma}$) & 9.275\;\,(+13.959\,$\hat{\sigma}$) & 9.171\;\,(-42.705\,$\hat{\sigma}$) & 4.4e40\;\,(+2.36e43\,$\hat{\sigma}$)\\
    \hline
    160 & 9.4466 & 0.0011987 & 9.419\;\,(-23.373\,$\hat{\sigma}$) & 9.419\;\,(-23.377\,$\hat{\sigma}$) & 9.439\;\,(-6.7006\,$\hat{\sigma}$) & 9.463\;\,(+13.555\,$\hat{\sigma}$) & 9.385\;\,(-51.253\,$\hat{\sigma}$) & 2.9e45\;\,(+2.38e48\,$\hat{\sigma}$)\\
    \hline
    176 & 9.6119 & 7.8118e-04 & 9.592\;\,(-25.268\,$\hat{\sigma}$) & 9.592\;\,(-25.268\,$\hat{\sigma}$) & 9.597\;\,(-18.461\,$\hat{\sigma}$) & 9.616\;\,(+5.5500\,$\hat{\sigma}$) & 9.565\;\,(-59.961\,$\hat{\sigma}$) & 1.9e50\;\,(+2.39e53\,$\hat{\sigma}$)\\
    \hline
    192 & 9.7526 & 5.3601e-04 & 9.735\;\,(-33.006\,$\hat{\sigma}$) & 9.735\;\,(-33.006\,$\hat{\sigma}$) & 9.735\;\,(-33.006\,$\hat{\sigma}$) & 9.754\;\,(+1.8849\,$\hat{\sigma}$) & 9.718\;\,(-64.391\,$\hat{\sigma}$) & 1.2e55\;\,(+2.29e58\,$\hat{\sigma}$)\\
    \hline
    208 & 9.8737 & 2.8341e-04 & 9.867\;\,(-24.839\,$\hat{\sigma}$) & 9.867\;\,(-24.839\,$\hat{\sigma}$) & 9.869\;\,(-18.286\,$\hat{\sigma}$) & 9.874\;\,(-0.2792\,$\hat{\sigma}$) & 9.850\;\,(-84.305\,$\hat{\sigma}$) & 8.0e59\;\,(+2.84e63\,$\hat{\sigma}$)\\
    \hline
    224 & 9.9792 & 1.4013e-04 & 9.975\;\,(-29.905\,$\hat{\sigma}$) & 9.975\;\,(-29.905\,$\hat{\sigma}$) & 9.975\;\,(-29.905\,$\hat{\sigma}$) & 9.979\;\,(-4.7202\,$\hat{\sigma}$) & 9.964\;\,(-105.06\,$\hat{\sigma}$) & 5.3e64\;\,(+3.76e68\,$\hat{\sigma}$)\\
    \hline
    240 & 10.0718 & 4.2456e-05 & 10.07\;\,(-22.301\,$\hat{\sigma}$) & 10.07\;\,(-22.301\,$\hat{\sigma}$) & 10.07\;\,(-22.301\,$\hat{\sigma}$) & 10.07\;\,(-2.8111\,$\hat{\sigma}$) & 10.07\;\,(-157.30\,$\hat{\sigma}$) & 3.5e69\;\,(+8.13e73\,$\hat{\sigma}$)\\
    \hline

\end{tabular}
\label{tbl:x2Results_8}
\end{table}
\egroup

\bgroup
\setlength\tabcolsep{0.15cm}
\begin{table}
\caption{$\chi^2$ divergence of codes with dimension $\kappa=9$.}
\centering
\begin{tabular} {|c|c|c|c|c|c|c|c|c|}
    \hline
    & \multicolumn{2}{c|}{Random Sample:} & \multicolumn{4}{c|}{\thead{$\chi^2$ Divergence: Absolute and \\(Relative to $\hat{\mu}$ and $\hat{\sigma}$)}} & \multicolumn{2}{c|}{\thead{Bounds: Absolute and \\ (Relative to $\hat{\mu}$ and $\hat{\sigma}$)}}\\
    \hline
    $n$ & \thead{Mean \\ ($\hat{\mu}$)} & \thead{St. Dev. \\ ($\hat{\sigma})$} & \thead{Gradient Descent \\ (Equiv. Loss)} & \thead{Gradient Descent \\ ($\chi^2$ Divergence)} & \thead{Best Known \\ Linear Code} & \thead{LDPC Dual} & \thead{Upper Bound} & \thead{Lower Bound} \\
    \hline
    32 & 5.342 & 0.080072 & 5.097\;\,(-3.0538\,$\hat{\sigma}$) & 5.097\;\,(-3.0625\,$\hat{\sigma}$) & 5.572\;\,(+2.8705\,$\hat{\sigma}$) & 12.04\;\,(+83.651\,$\hat{\sigma}$) & 4.695\;\,(-8.0817\,$\hat{\sigma}$) & 4.19e6\;\,(+5.238e7\,$\hat{\sigma}$)\\
    \hline
    64 & 8.775 & 0.02223 & 8.673\;\,(-4.5942\,$\hat{\sigma}$) & 8.671\;\,(-4.6718\,$\hat{\sigma}$) & 8.805\;\,(+1.3535\,$\hat{\sigma}$) & 9.279\;\,(+22.670\,$\hat{\sigma}$) & 8.386\;\,(-17.480\,$\hat{\sigma}$) & 1.8e16\;\,(+8.10e17\,$\hat{\sigma}$)\\
    \hline
    96 & 10.5441 & 0.010121 & 10.47\;\,(-6.9320\,$\hat{\sigma}$) & 10.47\;\,(-6.8527\,$\hat{\sigma}$) & 10.56\;\,(+1.1032\,$\hat{\sigma}$) & 10.92\;\,(+36.906\,$\hat{\sigma}$) & 10.28\;\,(-26.093\,$\hat{\sigma}$) & 7.7e25\;\,(+7.64e27\,$\hat{\sigma}$)\\
    \hline
    128 & 11.6033 & 0.0057707 & 11.54\;\,(-10.515\,$\hat{\sigma}$) & 11.54\;\,(-10.551\,$\hat{\sigma}$) & 11.68\;\,(+12.819\,$\hat{\sigma}$) & 11.86\;\,(+43.751\,$\hat{\sigma}$) & 11.41\;\,(-33.262\,$\hat{\sigma}$) & 3.3e35\;\,(+5.76e37\,$\hat{\sigma}$)\\
    \hline
    160 & 12.3058 & 0.0035703 & 12.25\;\,(-16.182\,$\hat{\sigma}$) & 12.25\;\,(-16.208\,$\hat{\sigma}$) & 12.26\;\,(-12.348\,$\hat{\sigma}$) & 12.45\;\,(+40.269\,$\hat{\sigma}$) & 12.16\;\,(-40.676\,$\hat{\sigma}$) & 1.4e45\;\,(+4.00e47\,$\hat{\sigma}$)\\
    \hline
    192 & 12.8045 & 0.0021581 & 12.75\;\,(-26.045\,$\hat{\sigma}$) & 12.75\;\,(-26.044\,$\hat{\sigma}$) & 12.77\;\,(-15.510\,$\hat{\sigma}$) & 12.90\;\,(+43.082\,$\hat{\sigma}$) & 12.69\;\,(-51.934\,$\hat{\sigma}$) & 6.1e54\;\,(+2.84e57\,$\hat{\sigma}$)\\
    \hline
    224 & 13.1768 & 0.0015691 & 13.12\;\,(-35.489\,$\hat{\sigma}$) & 13.12\;\,(-35.492\,$\hat{\sigma}$) & 13.15\;\,(-15.355\,$\hat{\sigma}$) & 13.25\;\,(+45.822\,$\hat{\sigma}$) & 13.09\;\,(-55.839\,$\hat{\sigma}$) & 2.6e64\;\,(+1.68e67\,$\hat{\sigma}$)\\
    \hline
    256 & 13.4654 & 0.0010279 & 13.41\;\,(-54.193\,$\hat{\sigma}$) & 13.41\;\,(-54.193\,$\hat{\sigma}$) & 13.41\;\,(-54.193\,$\hat{\sigma}$) & 13.52\;\,(+49.229\,$\hat{\sigma}$) & 13.40\;\,(-67.047\,$\hat{\sigma}$) & 1.1e74\;\,(+1.10e77\,$\hat{\sigma}$)\\
    \hline
    288 & 13.6955 & 6.7747e-04 & 13.66\;\,(-47.901\,$\hat{\sigma}$) & 13.66\;\,(-47.902\,$\hat{\sigma}$) & 13.66\;\,(-48.003\,$\hat{\sigma}$) & 13.72\;\,(+35.260\,$\hat{\sigma}$) & 13.64\;\,(-79.673\,$\hat{\sigma}$) & 4.9e83\;\,(+7.17e86\,$\hat{\sigma}$)\\
    \hline
    320 & 13.8833 & 5.1058e-04 & 13.86\;\,(-39.129\,$\hat{\sigma}$) & 13.86\;\,(-39.136\,$\hat{\sigma}$) & 13.86\;\,(-38.821\,$\hat{\sigma}$) & 13.90\;\,(+25.843\,$\hat{\sigma}$) & 13.84\;\,(-82.150\,$\hat{\sigma}$) & 2.1e93\;\,(+4.09e96\,$\hat{\sigma}$)\\
    \hline
    352 & 14.0394 & 3.0897e-04 & 14.03\;\,(-46.252\,$\hat{\sigma}$) & 14.03\;\,(-46.248\,$\hat{\sigma}$) & 14.03\;\,(-41.913\,$\hat{\sigma}$) & 14.05\;\,(+23.041\,$\hat{\sigma}$) & 14.01\;\,(-103.37\,$\hat{\sigma}$) & 9e102\;\,(+2.9e106\,$\hat{\sigma}$)\\
    \hline
    384 & 14.1712 & 1.9632e-04 & 14.16\;\,(-65.284\,$\hat{\sigma}$) & 14.16\;\,(-65.284\,$\hat{\sigma}$) & 14.16\;\,(-52.825\,$\hat{\sigma}$) & 14.17\;\,(+15.626\,$\hat{\sigma}$) & 14.15\;\,(-119.64\,$\hat{\sigma}$) & 4e112\;\,(+2.0e116\,$\hat{\sigma}$)\\
    \hline
    416 & 14.2839 & 1.1309e-04 & 14.28\;\,(-46.492\,$\hat{\sigma}$) & 14.28\;\,(-46.487\,$\hat{\sigma}$) & 14.28\;\,(-46.123\,$\hat{\sigma}$) & 14.28\;\,(+5.0862\,$\hat{\sigma}$) & 14.27\;\,(-143.95\,$\hat{\sigma}$) & 2e122\;\,(+1.5e126\,$\hat{\sigma}$)\\
    \hline
    448 & 14.3815 & 5.3694e-05 & 14.38\;\,(-59.283\,$\hat{\sigma}$) & 14.38\;\,(-59.283\,$\hat{\sigma}$) & 14.38\;\,(-48.560\,$\hat{\sigma}$) & 14.38\;\,(-2.6165\,$\hat{\sigma}$) & 14.37\;\,(-187.46\,$\hat{\sigma}$) & 7e131\;\,(+1.3e136\,$\hat{\sigma}$)\\
    \hline
    480 & 14.4668 & 1.5994e-05 & 14.47\;\,(-48.066\,$\hat{\sigma}$) & 14.47\;\,(-48.066\,$\hat{\sigma}$) & 14.47\;\,(-43.340\,$\hat{\sigma}$) & 14.47\;\,(-8.6858\,$\hat{\sigma}$) & 14.46\;\,(-289.80\,$\hat{\sigma}$) & 3e141\;\,(+1.9e146\,$\hat{\sigma}$)\\
    \hline

\end{tabular}
\label{tbl:x2Results_9}
\end{table}
\egroup

\bgroup
\setlength\tabcolsep{0.15cm}
\begin{table}
\caption{$\chi^2$ divergence of codes with dimension $\kappa=10$.}
\centering
\begin{tabular} {|c|c|c|c|c|c|c|c|c|}
    \hline
    & \multicolumn{2}{c|}{Random Sample:} & \multicolumn{4}{c|}{\thead{$\chi^2$ Divergence: Absolute and \\(Relative to $\hat{\mu}$ and $\hat{\sigma}$)}} & \multicolumn{2}{c|}{\thead{Bounds: Absolute and \\ (Relative to $\hat{\mu}$ and $\hat{\sigma}$)}}\\
    \hline
    $n$ & \thead{Mean \\ ($\hat{\mu}$)} & \thead{St. Dev. \\ ($\hat{\sigma})$} & \thead{Gradient Descent \\ (Equiv. Loss)} & \thead{Gradient Descent \\ ($\chi^2$ Divergence)} & \thead{Best Known \\ Linear Code} & \thead{LDPC Dual} & \thead{Upper Bound} & \thead{Lower Bound} \\
    \hline
    64 & 10.5369 & 0.02892 & 10.39\;\,(-5.1573\,$\hat{\sigma}$) & 10.39\;\,(-5.0935\,$\hat{\sigma}$) & 10.40\;\,(-4.5813\,$\hat{\sigma}$) & 11.75\;\,(+41.979\,$\hat{\sigma}$) & 10.04\;\,(-17.136\,$\hat{\sigma}$) & 9.0e15\;\,(+3.11e17\,$\hat{\sigma}$)\\
    \hline
    128 & 14.7802 & 0.0072069 & 14.71\;\,(-9.4931\,$\hat{\sigma}$) & 14.71\;\,(-10.012\,$\hat{\sigma}$) & 14.69\;\,(-12.983\,$\hat{\sigma}$) & 15.45\;\,(+93.116\,$\hat{\sigma}$) & 14.51\;\,(-38.101\,$\hat{\sigma}$) & 1.7e35\;\,(+2.31e37\,$\hat{\sigma}$)\\
    \hline
    192 & 16.6656 & 0.0039461 & 16.62\;\,(-10.738\,$\hat{\sigma}$) & 16.62\;\,(-10.755\,$\hat{\sigma}$) & 16.61\;\,(-13.531\,$\hat{\sigma}$) & 17.06\;\,(+100.86\,$\hat{\sigma}$) & 16.49\;\,(-45.578\,$\hat{\sigma}$) & 3.1e54\;\,(+7.77e56\,$\hat{\sigma}$)\\
    \hline
    256 & 17.7237 & 0.0019406 & 17.68\;\,(-20.111\,$\hat{\sigma}$) & 17.68\;\,(-20.130\,$\hat{\sigma}$) & 17.70\;\,(-13.371\,$\hat{\sigma}$) & 17.94\;\,(+113.88\,$\hat{\sigma}$) & 17.60\;\,(-66.249\,$\hat{\sigma}$) & 5.7e73\;\,(+2.91e76\,$\hat{\sigma}$)\\
    \hline
    320 & 18.3996 & 0.0012211 & 18.36\;\,(-30.978\,$\hat{\sigma}$) & 18.36\;\,(-30.988\,$\hat{\sigma}$) & 18.38\;\,(-14.701\,$\hat{\sigma}$) & 18.55\;\,(+126.09\,$\hat{\sigma}$) & 18.30\;\,(-78.773\,$\hat{\sigma}$) & 1.0e93\;\,(+8.54e95\,$\hat{\sigma}$)\\
    \hline
    384 & 18.8682 & 7.9699e-04 & 18.83\;\,(-46.634\,$\hat{\sigma}$) & 18.83\;\,(-46.638\,$\hat{\sigma}$) & 18.85\;\,(-18.472\,$\hat{\sigma}$) & 18.99\;\,(+146.90\,$\hat{\sigma}$) & 18.79\;\,(-92.444\,$\hat{\sigma}$) & 2e112\;\,(+2.4e115\,$\hat{\sigma}$)\\
    \hline
    448 & 19.2122 & 4.9484e-04 & 19.18\;\,(-74.705\,$\hat{\sigma}$) & 19.18\;\,(-74.729\,$\hat{\sigma}$) & 19.20\;\,(-24.099\,$\hat{\sigma}$) & 19.27\;\,(+116.18\,$\hat{\sigma}$) & 19.15\;\,(-115.84\,$\hat{\sigma}$) & 4e131\;\,(+7.2e134\,$\hat{\sigma}$)\\
    \hline
    512 & 19.4755 & 3.699e-04 & 19.44\;\,(-99.854\,$\hat{\sigma}$) & 19.44\;\,(-99.854\,$\hat{\sigma}$) & 19.47\;\,(-26.795\,$\hat{\sigma}$) & 19.51\;\,(+90.957\,$\hat{\sigma}$) & 19.43\;\,(-121.29\,$\hat{\sigma}$) & 7e150\;\,(+1.8e154\,$\hat{\sigma}$)\\
    \hline
    576 & 19.6835 & 24971e-04 & 19.66\;\,(-86.627\,$\hat{\sigma}$) & 19.66\;\,(-86.627\,$\hat{\sigma}$) & 19.68\;\,(-29.212\,$\hat{\sigma}$) & 19.71\;\,(+86.499\,$\hat{\sigma}$) & 19.65\;\,(-140.53\,$\hat{\sigma}$) & 1e170\;\,(+4.8e173\,$\hat{\sigma}$)\\
    \hline
    640 & 19.8519 & 1.6336e-04 & 19.84\;\,(-82.184\,$\hat{\sigma}$) & 19.84\;\,(-82.196\,$\hat{\sigma}$) & 19.85\;\,(-38.138\,$\hat{\sigma}$) & 19.87\;\,(+81.911\,$\hat{\sigma}$) & 19.82\;\,(-166.46\,$\hat{\sigma}$) & 2e189\;\,(+1.4e193\,$\hat{\sigma}$)\\
    \hline
    704 & 19.991 & 1.1301e-04 & 19.98\;\,(-85.740\,$\hat{\sigma}$) & 19.98\;\,(-85.745\,$\hat{\sigma}$) & 19.98\;\,(-53.936\,$\hat{\sigma}$) & 20.00\;\,(+56.858\,$\hat{\sigma}$) & 19.97\;\,(-182.81\,$\hat{\sigma}$) & 4e208\;\,(+3.6e212\,$\hat{\sigma}$)\\
    \hline
    768 & 20.1078 & 7.33e-05 & 20.10\;\,(-118.92\,$\hat{\sigma}$) & 20.10\;\,(-118.92\,$\hat{\sigma}$) & 20.10\;\,(-46.246\,$\hat{\sigma}$) & 20.11\;\,(+39.506\,$\hat{\sigma}$) & 20.09\;\,(-207.04\,$\hat{\sigma}$) & 8e227\;\,(+1.0e232\,$\hat{\sigma}$)\\
    \hline
    832 & 20.2074 & 3.8533e-05 & 20.20\;\,(-94.817\,$\hat{\sigma}$) & 20.20\;\,(-94.815\,$\hat{\sigma}$) & 20.20\;\,(-67.345\,$\hat{\sigma}$) & 20.21\;\,(+34.728\,$\hat{\sigma}$) & 20.20\;\,(-273.15\,$\hat{\sigma}$) & 1e247\;\,(+3.6e251\,$\hat{\sigma}$)\\
    \hline
    896 & 20.2932 & 1.9771e-05 & 20.29\;\,(-113.08\,$\hat{\sigma}$) & 20.29\;\,(-113.08\,$\hat{\sigma}$) & 20.29\;\,(-17.930\,$\hat{\sigma}$) & 20.29\;\,(+5.8336\,$\hat{\sigma}$) & 20.29\;\,(-329.32\,$\hat{\sigma}$) & 3e266\;\,(+1.3e271\,$\hat{\sigma}$)\\
    \hline
    960 & 20.368 & 6.4665e-06 & 20.37\;\,(-86.777\,$\hat{\sigma}$) & 20.37\;\,(-86.777\,$\hat{\sigma}$) & 20.37\;\,(+106.79\,$\hat{\sigma}$) & 20.37\;\,(-7.8304\,$\hat{\sigma}$) & 20.36\;\,(-467.05\,$\hat{\sigma}$) & 5e285\;\,(+7.4e290\,$\hat{\sigma}$)\\
    \hline
    
\end{tabular}
\label{tbl:x2Results_10}
\end{table}
\egroup

\clearpage

\bgroup
\setlength\tabcolsep{0.15cm}
\begin{table}
\caption{$\chi^2$ divergence of codes with dimension $\kappa=11$.}
\centering
\begin{tabular} {|c|c|c|c|c|c|c|c|}
    \hline
    & \multicolumn{2}{c|}{Random Sample:} & \multicolumn{3}{c|}{\thead{$\chi^2$ Divergence: Absolute and \\(Relative to $\hat{\mu}$ and $\hat{\sigma}$)}} & \multicolumn{2}{c|}{\thead{Bounds: Absolute and \\ (Relative to $\hat{\mu}$ and $\hat{\sigma}$)}}\\
    \hline
    $n$ & \thead{Mean \\ ($\hat{\mu}$)} & \thead{St. Dev. \\ ($\hat{\sigma})$} & \thead{Gradient Descent \\ ($\chi^2$ Divergence)} & \thead{Best Known \\ Linear Code} & \thead{LDPC Dual} & \thead{Upper Bound} & \thead{Lower Bound} \\
    \hline
    128 & 18.6585 & 0.0077281 & 18.57\;\,(-11.025\,$\hat{\sigma}$) & 18.58\;\,(-10.710\,$\hat{\sigma}$) & 19.93\;\,(+164.06\,$\hat{\sigma}$) & 18.31\;\,(-44.837\,$\hat{\sigma}$) & 8.31e34\;\,(+1.075e37\,$\hat{\sigma}$)\\
    \hline
    256 & 23.199 & 0.0023604 & 23.17\;\,(-13.357\,$\hat{\sigma}$) & 23.15\;\,(-22.715\,$\hat{\sigma}$) & 23.88\;\,(+287.54\,$\hat{\sigma}$) & 23.02\;\,(-76.437\,$\hat{\sigma}$) & 2.83e73\;\,(+1.198e76\,$\hat{\sigma}$)\\
    \hline
    384 & 25.0128 & 0.0011877 & 24.99\;\,(-21.738\,$\hat{\sigma}$) & 24.98\;\,(-27.696\,$\hat{\sigma}$) & 25.34\;\,(+278.65\,$\hat{\sigma}$) & 24.90\;\,(-97.719\,$\hat{\sigma}$) & 9.62e111\;\,(+8.099e114\,$\hat{\sigma}$)\\
    \hline
    512 & 25.9857 & 0.00067129 & 25.96\;\,(-36.245\,$\hat{\sigma}$) & 25.96\;\,(-40.301\,$\hat{\sigma}$) & 26.22\;\,(+350.05\,$\hat{\sigma}$) & 25.90\;\,(-122.03\,$\hat{\sigma}$) & 3.27e150\;\,(+4.876e153\,$\hat{\sigma}$)\\
    \hline
    640 & 26.5921 & 0.00040403 & 26.57\;\,(-58.907\,$\hat{\sigma}$) & 26.57\;\,(-44.518\,$\hat{\sigma}$) & 26.72\;\,(+304.67\,$\hat{\sigma}$) & 26.53\;\,(-150.41\,$\hat{\sigma}$) & 1.11e189\;\,(+2.757e192\,$\hat{\sigma}$)\\
    \hline
    768 & 27.0061 & 0.00026008 & 26.98\;\,(-90.635\,$\hat{\sigma}$) & 26.99\;\,(-51.796\,$\hat{\sigma}$) & 27.08\;\,(+267.89\,$\hat{\sigma}$) & 26.96\;\,(-178.29\,$\hat{\sigma}$) & 3.79e227\;\,(+1.457e231\,$\hat{\sigma}$)\\
    \hline
    896 & 27.3067 & 0.00018435 & 27.28\;\,(-127.55\,$\hat{\sigma}$) & 27.30\;\,(-58.641\,$\hat{\sigma}$) & 27.35\;\,(+247.22\,$\hat{\sigma}$) & 27.27\;\,(-195.15\,$\hat{\sigma}$) & 1.29e266\;\,(+6.996e269\,$\hat{\sigma}$)\\
    \hline
    1024 & 27.535 & 0.00012474 & 27.51\;\,(-188.48\,$\hat{\sigma}$) & 27.53\;\,(-70.494\,$\hat{\sigma}$) & 27.57\;\,(+248.23\,$\hat{\sigma}$) & 27.51\;\,(-225.27\,$\hat{\sigma}$) & 4.39e304\;\,(+6.996e269\,$\hat{\sigma}$)\\
    \hline
    1152 & 27.7141 & 8.3304e-05 & 27.70\;\,(-165.58\,$\hat{\sigma}$) & 27.71\;\,(-88.574\,$\hat{\sigma}$) & 27.73\;\,(+199.85\,$\hat{\sigma}$) & 27.69\;\,(-263.11\,$\hat{\sigma}$) & 4.39e304\;\,(+6.996e269\,$\hat{\sigma}$)\\
    \hline
    1280 & 27.8585 & 6.0324e-05 & 27.85\;\,(-143.04\,$\hat{\sigma}$) & 27.85\;\,(-117.58\,$\hat{\sigma}$) & 27.87\;\,(+155.20\,$\hat{\sigma}$) & 27.84\;\,(-281.14\,$\hat{\sigma}$) & 4.39e304\;\,(+6.996e269\,$\hat{\sigma}$)\\
    \hline
    1408 & 27.9773 & 4.1148e-05 & 27.97\;\,(-152.59\,$\hat{\sigma}$) & 27.97\;\,(-119.85\,$\hat{\sigma}$) & 27.98\;\,(+136.59\,$\hat{\sigma}$) & 27.96\;\,(-312.67\,$\hat{\sigma}$) & 4.39e304\;\,(+6.996e269\,$\hat{\sigma}$)\\
    \hline
    1536 & 28.0767 & 2.2466e-05 & 28.07\;\,(-252.34\,$\hat{\sigma}$) & 28.07\;\,(-155.92\,$\hat{\sigma}$) & 28.08\;\,(+122.41\,$\hat{\sigma}$) & 28.07\;\,(-420.48\,$\hat{\sigma}$) & 4.39e304\;\,(+6.996e269\,$\hat{\sigma}$)\\
    \hline
    1664 & 28.1613 & 1.4087e-05 & 28.16\;\,(-170.96\,$\hat{\sigma}$) & 28.16\;\,(-161.62\,$\hat{\sigma}$) & 28.16\;\,(+63.644\,$\hat{\sigma}$) & 28.15\;\,(-464.77\,$\hat{\sigma}$) & 4.39e304\;\,(+6.996e269\,$\hat{\sigma}$)\\
    \hline
    1792 & 28.234 & 7.053e-06 & 28.23\;\,(-211.60\,$\hat{\sigma}$) & 28.23\;\,(-161.41\,$\hat{\sigma}$) & 28.23\;\,(+30.868\,$\hat{\sigma}$) & 28.23\;\,(-574.69\,$\hat{\sigma}$) & 4.39e304\;\,(+6.996e269\,$\hat{\sigma}$)\\
    \hline
    1920 & 28.2972 & 1.966e-06 & 28.30\;\,(-195.26\,$\hat{\sigma}$) & 28.30\;\,(-74.396\,$\hat{\sigma}$) & 28.30\;\,(-7.2788\,$\hat{\sigma}$) & 28.30\;\,(-959.41\,$\hat{\sigma}$) & 4.39e304\;\,(+6.996e269\,$\hat{\sigma}$)\\
    \hline

\end{tabular}
\label{tbl:x2Results_11}
\end{table}
\egroup

\bgroup
\setlength\tabcolsep{0.15cm}
\begin{table}
\caption{$\chi^2$ divergence of codes with dimension $\kappa=12$.}
\centering
\begin{tabular} {|c|c|c|c|c|c|c|c|}
    \hline
    & \multicolumn{2}{c|}{Random Sample:} & \multicolumn{3}{c|}{\thead{$\chi^2$ Divergence: Absolute and \\(Relative to $\hat{\mu}$ and $\hat{\sigma}$)}} & \multicolumn{2}{c|}{\thead{Bounds: Absolute and \\ (Relative to $\hat{\mu}$ and $\hat{\sigma}$)}}\\
    \hline
    $n$ & \thead{Mean \\ ($\hat{\mu}$)} & \thead{St. Dev. \\ ($\hat{\sigma})$} & \thead{Gradient Descent \\ ($\chi^2$ Divergence)} & \thead{Best Known \\ Linear Code} & \thead{LDPC Dual} & \thead{Upper Bound} & \thead{Lower Bound} \\
    \hline
    256 & 30.2289 & 0.0024229 & 30.18\;\,(-20.119\,$\hat{\sigma}$) & 30.25\;\,(+8.2976\,$\hat{\sigma}$) & 31.77\;\,(+637.65\,$\hat{\sigma}$) & 30.00\;\,(-93.810\,$\hat{\sigma}$) & 1.41e73\;\,(+5.834e75\,$\hat{\sigma}$)\\
    \hline
    512 & 34.5765 & 0.00080042 & 34.56\;\,(-21.268\,$\hat{\sigma}$) & 34.58\;\,(+7.1777\,$\hat{\sigma}$) & 35.10\;\,(+650.02\,$\hat{\sigma}$) & 34.46\;\,(-141.75\,$\hat{\sigma}$) & 1.64e150\;\,(+2.045e153\,$\hat{\sigma}$)\\
    \hline
    768 & 36.1936 & 0.0003914 & 36.18\;\,(-39.035\,$\hat{\sigma}$) & 36.20\;\,(+5.1998\,$\hat{\sigma}$) & 36.55\;\,(+907.41\,$\hat{\sigma}$) & 36.12\;\,(-183.56\,$\hat{\sigma}$) & 1.90e227\;\,(+4.842e230\,$\hat{\sigma}$)\\
    \hline
    1024 & 37.0368 & 0.00020093 & 37.02\;\,(-73.567\,$\hat{\sigma}$) & 37.04\;\,(+4.6663\,$\hat{\sigma}$) & 37.19\;\,(+756.15\,$\hat{\sigma}$) & 36.99\;\,(-250.50\,$\hat{\sigma}$) & 2.19e304\;\,(+1.092e308\,$\hat{\sigma}$)\\
    \hline
    1280 & 37.5542 & 0.00012319 & 37.54\;\,(-118.37\,$\hat{\sigma}$) & 37.55\;\,(+3.7315\,$\hat{\sigma}$) & 37.64\;\,(+712.79\,$\hat{\sigma}$) & 37.52\;\,(-301.66\,$\hat{\sigma}$) & 2.19e304\;\,(+1.781e308\,$\hat{\sigma}$)\\
    \hline
    1536 & 37.9041 & 9.0776e-05 & 37.89\;\,(-159.78\,$\hat{\sigma}$) & 37.90\;\,(+2.6903\,$\hat{\sigma}$) & 37.96\;\,(+658.30\,$\hat{\sigma}$) & 37.88\;\,(-311.53\,$\hat{\sigma}$) & 2.19e304\;\,(+1.781e308\,$\hat{\sigma}$)\\
    \hline
    1792 & 38.1565 & 6.2458e-05 & 38.14\;\,(-231.92\,$\hat{\sigma}$) & 38.16\;\,(+3.1084\,$\hat{\sigma}$) & 38.19\;\,(+529.89\,$\hat{\sigma}$) & 38.13\;\,(-350.54\,$\hat{\sigma}$) & 2.19e304\;\,(+1.781e308\,$\hat{\sigma}$)\\
    \hline
    2048 & 38.3471 & 4.3443e-05 & 38.33\;\,(-333.37\,$\hat{\sigma}$) & 38.35\;\,(+2.3980\,$\hat{\sigma}$) & 38.37\;\,(+451.59\,$\hat{\sigma}$) & 38.33\;\,(-392.95\,$\hat{\sigma}$) & 2.19e304\;\,(+1.781e308\,$\hat{\sigma}$)\\
    \hline
    2304 & 38.4962 & 2.841e-05 & 38.49\;\,(-299.83\,$\hat{\sigma}$) & 38.50\;\,(-10.227\,$\hat{\sigma}$) & 38.51\;\,(+437.98\,$\hat{\sigma}$) & 38.48\;\,(-467.97\,$\hat{\sigma}$) & 2.19e304\;\,(+1.781e308\,$\hat{\sigma}$)\\
    \hline
    2560 & 38.6159 & 1.8322e-05 & 38.61\;\,(-292.60\,$\hat{\sigma}$) & 38.62\;\,(-21.907\,$\hat{\sigma}$) & 38.62\;\,(+436.79\,$\hat{\sigma}$) & 38.61\;\,(-560.74\,$\hat{\sigma}$) & 2.19e304\;\,(+1.781e308\,$\hat{\sigma}$)\\
    \hline
    2816 & 38.7142 & 1.2908e-05 & 38.71\;\,(-304.87\,$\hat{\sigma}$) & 38.72\;\,(+80.984\,$\hat{\sigma}$) & 38.72\;\,(+285.46\,$\hat{\sigma}$) & 38.71\;\,(-603.68\,$\hat{\sigma}$) & 2.19e304\;\,(+1.781e308\,$\hat{\sigma}$)\\
    \hline
    3072 & 38.7964 & 7.5824e-06 & 38.79\;\,(-470.24\,$\hat{\sigma}$) & 38.80\;\,(+201.20\,$\hat{\sigma}$) & 38.80\;\,(+235.61\,$\hat{\sigma}$) & 38.79\;\,(-754.49\,$\hat{\sigma}$) & 2.19e304\;\,(+1.781e308\,$\hat{\sigma}$)\\
    \hline
    3328 & 38.8661 & 4.566e-06 & 38.86\;\,(-335.36\,$\hat{\sigma}$) & 38.87\;\,(+448.64\,$\hat{\sigma}$) & 38.87\;\,(+164.09\,$\hat{\sigma}$) & 38.86\;\,(-867.59\,$\hat{\sigma}$) & 2.19e304\;\,(+1.781e308\,$\hat{\sigma}$)\\
    \hline
    3584 & 38.926 & 2.147e-06 & 38.93\;\,(-446.69\,$\hat{\sigma}$) & 38.9285\;\,(+1196\,$\hat{\sigma}$) & 38.93\;\,(+79.678\,$\hat{\sigma}$) & 38.9235\;\,(-1142\,$\hat{\sigma}$) & 2.19e304\;\,(+1.781e308\,$\hat{\sigma}$)\\
    \hline
    3840 & 38.9779 & 7.4085e-07 & 38.98\;\,(-339.26\,$\hat{\sigma}$) & 38.9808\;\,(+3879\,$\hat{\sigma}$) & 38.98\;\,(+7.4914\,$\hat{\sigma}$) & 38.9768\;\,(-1543\,$\hat{\sigma}$) & 2.19e304\;\,(+1.781e308\,$\hat{\sigma}$)\\
    \hline

\end{tabular}
\label{tbl:x2Results_12}
\end{table}
\egroup

\bibliographystyle{IEEEtran}
\bibliography{./SDCC}

\end{document}